\documentclass[a4paper,10pt,notitlepage]{article}
\usepackage{amsfonts,amssymb,amsmath,amsthm,hyperref,colonequals}
\usepackage{caption}
\usepackage{cite}
\usepackage{color}
\usepackage{tabularx}
\usepackage{graphicx}
\usepackage{bbm}
\usepackage{shuffle}
\usepackage[nice]{nicefrac} 
\usepackage{url}
\usepackage{mathtools}
\theoremstyle{definition}
\newcommand{\bea}{\begin{eqnarray}}
\newcommand{\eea}{\end{eqnarray}}
\newcommand{\be}{\begin{equation}}
\newcommand{\ee}{\end{equation}}

\DeclareMathOperator{\Tr}{Tr}

\newcommand{\vac}[1]{\ensuremath{\left< \, #1\, \right>}}

\begin{document}

\thispagestyle{empty}
\setcounter{page}{0}
\begin{flushright}\footnotesize
\vspace{0.5cm}
\end{flushright}
\setcounter{footnote}{0}

\begin{center}
{\huge{
\textbf{Scalar Field Dark Matter in Hybrid Approach} 
}}

{\sc 
Pavel Friedrich$^*$, Tomislav Prokopec$^\dagger$ }\\[5mm]

{\it Institute for Theoretical Physics, Spinoza Institute and the
Center for Extreme Matter and Emergent Phenomena (EMME$\Phi$),\\ 
Utrecht University, Buys Ballot Building,
Princetonplein 5, 3584 CC Utrecht, The Netherlands
}\\[5mm]
\let\thefootnote\relax\footnotetext{\\\texttt{$^*$Electronic address:p.friedrich@uu.nl}\\$^\dagger$
Electronic address:\texttt{t.prokopec@uu.nl}
}

\textbf{Abstract}\\[2mm]
\end{center}
We develop a hybrid formalism suitable for modeling scalar field dark matter, in which the phase-space distribution associated to the real scalar field is modeled by statistical equal-time two-point functions and gravity is treated by two stochastic gravitational fields in the longitudinal gauge (in this work we neglect vector and tensor gravitational perturbations).
  Inspired by the commonly used Newtonian Vlasov-Poisson system, we firstly identify a suitable combination of equal-time two-point functions that defines the phase-space distribution associated to the scalar field and then derive both a kinetic equation that contains relativistic scalar matter corrections as well as linear gravitational scalar field equations whose sources can be expressed in terms of a momentum integral over the phase-space distribution function. Our treatment generalizes the commonly used classical scalar field formalism, in that it allows for modeling of (dynamically generated) vorticity and perturbations in anisotropic stresses of the scalar field. It also allows for a systematic inclusion of relativistic and higher order corrections that may be used to distinguish different dark matter scenarios. We also provide initial conditions for the statistical equal-time two-point functions of the matter scalar field in terms of gravitational potentials and the scale factor.


\newpage
\setcounter{page}{1}


\tableofcontents
\addtolength{\baselineskip}{5pt}

\section{Introduction and Overview}
The standard model of cosmology attributes roughly one third of the universes energy to dark matter, a particle or field whose nature is mostly unknown except for the effect that it interacts with gravity \cite{Ade:2015xua}. There has been success in studying large-scale structures of the universe by modeling dark matter as non-relativistic particles that can be described by a pressureless fluid. Linear perturbation theory can be used up to the scale of non-linearity $k>k_{nl} \sim 0.3~{\rm Mpc}^{-1}$ to predict the distribution of galaxy clusters and the perturbation theory may be used to study higher-order effects \cite{Bernardeau:2001qr}.
On the other hand, interest has recently \cite{Marsh:2015daa}\cite{Hui:2016ltb} been shown in the study of axion or fuzzy dark matter \cite{Turner:1983he} \cite{Sin:1992bg} \cite{Lee:1995af} \cite{Hu:2000ke} \cite{Goodman:2000tg} \cite{Peebles:2000yy} which in the end is a real scalar field with a certain mass range minimally coupled to gravity with self-interaction terms playing a minor role. Common to most minimal scalar field dark models is that the mass is much bigger than the Hubble rate. It has been studied in linear perturbation theory in different gauges \cite{Ratra:1990me} \cite{Hwang:1996xd}. 
The non-relativistic limit of the Klein-Gordon equation of a classical scalar field yields the Schr\"odinger equation.
By defining energy-density and fluid velocity via the so-called Madelung transformation, one can reproduce non-relativistic, non-linear hydrodynamic equations in FLRW-space-time for real \cite{Marsh:2015daa} and complex \cite{Suarez:2015fga} classical scalar field theories. From a quantum field theory point of view, we think of the classical fields entering these models as Bose condensates that are obtained by coherent quantum states whose one-point function defines the classical field. In view of the semi-classical Einstein equations, it is natural to extent the analysis to the statistical limit of the full two-point functions where the additional degrees of freedom can account for all features of a fluid of massive collisionless particles in the classical limit. In fact, it is the expectation value of squares of (non-composite) field operators at equal times that couples to the Einstein tensor in semi-classical gravity. Thus, we should think of these two-point functions as building blocks of the fluid. In the classical limit these equal-time two-point functions reduce to the statistical or Hadamard two-point function. It is a priori not clear why they should reduce only to the product of classical fields, i.e. expectation values of one field operator insertion. Despite that one has to argue on how such condensates are generated in a quadratic potential in late-time cosmology, working only with classical fields cuts down degrees of freedom that might be important for cold dark matter models. 
We underpin the later point by deriving that statistical two-point functions are in a gradient approximation related to phase-space densities whose position and momentum dependence is initially generic by means of the connected piece of the two-point function, i.e. the part which does not reduced to a product of expectation values. This makes it clear that they contain more features of the scalar field fluid than the one-point functions or classical fields are able to describe.
From the perspective of a classical particle that is coupled to gravity, the studies of phase-space dynamics are inevitable when a single-stream fluid {\it Ansatz} breaks down due to what is called shell-crossing. The kinetic theory underlying dark matter is summarized in the Vlasov equation \cite{lrr-2011-4}, which represents a phase-space description that does not break down in the non-linear regime since there is no shell-crossing in phase-space. 
Phase-space densities and the corresponding Vlasov equation have previously been derived by using the Wigner transformation of the non-relativistic Schr\"odinger-Poisson system \cite{Widrow:1993qq} \cite{Davies:1996kp}\cite{Uhlemann:2014npa} and also for a relativistic scalar field \cite{Widrow:1996eq}. Once again, only one-point functions have been considered and the richness of the connected part of the statistical two-point function is lost.
\par
In this paper, we want to put forward the discussion about real scalar field dark matter from the perspective of phase-space dynamics which is according to us still incomplete at the moment. We show that instead of using classical fields, the more natural objects are statistical two-point functions which via the additional space-time-dependence can be used to derive a momentum-dependence as it occurs in kinetic theory. Integrating out this momentum dependence still leaves us with a non-homogeneous space-time dependence that is induced by the stochastic gravitational fields. Furthermore, two-point functions naturally arise in quantum field theory, whose broad apparatus might even be used to simplify non-linear calculations once a mapping to observables in cosmology is established as we do in this paper. Defining phase-space densities from two-point functions is a known business in Minkowki-space \cite{ trove.nla.gov.au/work/9783845}, it is a generalization of non-relativistic Wigner transformation \cite{Hillery:1983ms} to special relativity. The idea is to change the coordinates to a collective and difference coordinate and to Fourier-transform with respect to the difference coordinate to obtain a momentum dependence.
However, there are few publications on a generalization of this idea to curved space-times. Two independent works in \cite{Winter:1986da} and in \cite{Calzetta:1987bw} postulate off-shell curved space-time Wigner transformations in a mathematical complicated expansion by using geodesics and Riemann normal coordinates, respectively. The transformation is done with respect to a coordinate-independent physical distance between the two points on the space-time manifold. A similar approach has again been discussed by \cite{Fonarev:1993ht}.
However, in this paper we make use of a simpler transformation that allows us to write down exact equations in a one-step transformation. The idea is to think of the two-point function as an object that depends on one point of the space-time manifold and on another point that belongs to the tangent space over that point on the space-time manifold. Consequently, the momentum is a variable of the cotangent space over that point on the space-time manifold. This approach was used in \cite{Garbrecht:2002pd} to define particle densities in an unperturbed FLRW-univserse where the authors started with off-shell equations and projected them onto on-shell quantities via integration. A similar approach was already proposed in \cite{Antonsen:1997dc} for general space-times and its implication were studied for Fermionic systems, however, no on-shell projection was discussed.
As far as we know, none of the previous works makes an attempt to clearly derive a set of equation that reduce in the classical limit to the Newtonian on-shell Vlasov-Poisson system that is used in kinetic theory of dark matter. We consider this as an important gap in the theory of scalar field or axion dark matter and it is the task of our paper to close it. Once we have a clear pictures on how the dynamics of dark matter is embedded in quantum field theory on curved space-time, we might discover more ways to calculate  cosmological quantities in the non-linear regime.
\par
We call the approach in this paper a hybrid approach for the reason that we start in principle from a quantum field theory for the real scalar field but do not properly integrate out the gravitational constraint. Thus, we approximate the self-interaction terms that would be generated by this procedure in terms of the gravitational potentials treated as external sources which are by means of the semi-classical Einstein equations related to the statistical two-point functions themselves. The source of stochasticity is in part in the quantum origin of scalar field fluctuations
and in part in the fact that making an initial Gaussian state {\it Ansatz} neglects interactions of dark matter 
with other matter fields and with gravity, which in general will create higher order (non-Gaussian) correlations that are 
neglected (coarse-grained) in our formalism.
\par
The paper is structured as follows. We start by deriving a dynamical system of on-shell two-point functions that is converted from a pure space-time dependence to a dependence on phase-space variables. We specialize to a scalar linearized longitudinal gauge without vector perturbations and without gravitons. But we keep the gravitational slip (defined as the difference between the two gravitational potentials), which induces high-order corrections in the fluid dynamics of scalar field dark matter that have not been captured so far in the one-point function approach. We derive Einstein's equations in that gauge and rewrite the energy-momentum tensor as momentum integrals over two-point functions.
This allows us in turn to define scalar hydrodynamic variables like energy density, rest-mass density and pressure. However, hydrodynamic variables containing the four-velocity can only be defined as composite operators leaving space for anisotropy. We then consider a gradient expansion by introducing a variety of perturbation parameters on top of the linearization in the gravitational potentials and show that we indeed recover the generalization of the continuity and Euler equation in the FLRW-space-time. We also use the energy-momentum to identify even and odd phase-space densities which brings us finally to the derivation of the on-shell Vlasov equation by making use of the unintegrated dynamical equations for the statistical two-point functions.
\par
In the hybrid approach that we put forward in this paper we have two types of two-point functions involved. We have a statistical two-point function (also called Hadamard function) which consists of the expectation value of anti-commutators of the scalar field operators evaluated with respect to some initial density matrix. In our hybrid approach, this initial density matrix is  taken to depend on the stochastic gravitational potentials as they appear in the context of cosmological perturbation theory. This dependence makes the Hadamard two-point function (which is a c-number) itself a stochastic quantity arising from the initial conditions. Thus, we can now take the expectation value of the product of Hadamard two-point functions with respect to the stochastic initial conditions and integrated versions thereof will correspond for example to density-density correlators in cosmological perturbation theory. Unless, we do not make a clear distinction, we mean statistical or Hadamard two-point functions whenever we speak generically of two-point functions.
\pagebreak

\section{Phase-Space Distribution From 2-Point Functions \label{phaseFrom2}}
\subsection{Microscopic Theory in the Operator Formalism}
Let us start by writing down the microscopic theory that captures the fundamental dynamics.
It is a real scalar quantum field theory that indirectly self-interacts via a minimal and semi-classical coupling to gravity.
\par We work in units where $c=1$ and write down the action for the system
\be
S\left[ \phi, g_{\mu \nu} \right] = S_{g} \left[ g_{\mu \nu} \right] + S_{m} \left[ \phi, g_{\mu \nu} \right]\, ,
\ee
where
\be
 S_{m} \left[ \phi, g_{\mu \nu} \right] = -  \frac{1}{2} \int d^D x \sqrt{-g} \left[ g^{\mu \nu} \partial_{\mu} \phi \partial_{\nu} \phi + \frac{m^2}{\hbar^2} \phi^2 \right]\, , \label{sMatter}
\ee
is the matter action and
\be
S_{g} \left[ g_{\mu \nu} \right] = \frac{M_P^2}{2 \hbar} \int d^D x \sqrt{-g} R  \, \label{SemiClassGravityAction}
\ee
is the classical gravity action with $R$ being the Ricci scalar.
We will work with a metric for linearized scalar perturbations in Newtonian (longitudinal) gauge which is specified by the two gravitational potentials $\Phi_G$ and $\Psi_G$,
\be
g_{00} (\eta, x^i ) = - a^{2}(\eta ) \left[ 1 + 2 \Phi_G (\eta, x^i ) \right] \,, \,  g_{ij}(\eta, x^i ) = a^2 (\eta ) \delta_{ij} \left[ 1- 2 \Psi_G(\eta, x^i ) \right] \, .	\label{longMetric}
\ee
We drop all quadratic terms $\Phi_G^2\, , \Psi_G^2 \, ,\Phi_G \cdot \Psi_G$ as higher-order corrections from the very beginning. We also drop vector and tensorial perturbations for simplicity although in general we expect them to be generated due to non-linear evolution.
Inflation generates gravitational potentials that can be to a good approximation treated as classical stochastic fields that are at large redshifts approximated by a Gaussian distribution. 
We note that the metric \eqref{longMetric} is particularly useful to study the Newtonian limit of general relativity. It generalizes the longitudinal metric that has been used in the classical real scalar field theory approach to dark matter in\cite{Marsh:2015daa} by allowing for a non-zero gravitational slip that we define in $D$ space-time dimensions as
\be
\text{gravitational slip} := \Phi_G - (D-3) \Psi_G\, .
\ee
\par
The quantum theory in the operator formalism is specified by the time-evolution or Hamilton operator $\hat{H}$ which is a functional of the field operator $\hat{\phi}$ and its canonical momentum field operator $\hat{\Pi}$. We work in the Heisenberg picture and  the canonical momentum operator evaluates to
\bea
\hat{\Pi} (x) = a^{(D-2)}(\eta) \Big[1 - \Phi_G(\eta , x^i) -(D-1)\Psi_G(\eta , x^i)  \Big]  \hat{\phi}^{\prime}(\eta , x^i)\, ,
\eea
where
\be
x := \big(\eta, x^i \big)\, \qquad \text{and} \qquad \Big( \, . \,  \Big)^{\prime} := \frac{\partial}{\partial \eta} \Big( \, . \,  \Big)\, .
\ee
The field operators obey equal-time commutation relations
\be
\Big[ \hat{\phi} (\eta, x^i)\, , \hat{\Pi}(\eta, \widetilde{x}^i) \Big] = i \hbar \delta^{D-1} (x^i - \widetilde{x}^i)\, , \, \Big[ \hat{\phi}(\eta, x^i)\, , \hat{\phi}(\eta, \widetilde{x}^i) \Big] = 0\, , \, \Big[ \hat{\Pi}(\eta, x^i)\, , \hat{\Pi}(\eta, \widetilde{x}^i) \Big] = 0\, . \label{commRel}
\ee
Since we are working in semi-classical gravity the Hamiltonian $\hat{H}$ additionally depends on the gravitational potentials $\Phi_G\, ,\Psi_G$ that act as external, stochastic fields,
\bea
\hat{H} (\hat{\phi} , \hat{\Pi}, g_{\mu \nu}) &\equiv & \int d^{D-1} x\, \hat{\Pi} \,   \hat{\phi}^{\prime} - \int d^{D-1} x \, \hat{\mathcal{L}}_m \Big[ \hat{\phi}, \hat{\phi}^{\prime},g_{\mu \nu} \Big] \nonumber \\&=& - \frac{1}{2 }\int d^{D-1} x \frac{\hat{\Pi}^2 }{\sqrt{-g}g^{00}} +\frac{1}{2} \int d^{D-1} x \sqrt{-g} \left[ g^{ij} \partial_{i} \hat{\phi} \partial_{j} \hat{\phi} + \frac{ m^2}{\hbar^2} \hat{\phi}^2 \right]\, . 
\eea
Using the Heisenberg equations we find the following time evolution of the canonical operators,
\bea
\hat{\phi}^{\prime} (x) &=&  a^{-(D-2)}(\eta) \Big[1 + \Phi_G(x) +(D-1)\Psi_G(x)  \Big] \hat{\Pi}(x)\, , \label{phiPrime} \\
\hat{\Pi}^{\prime} (x) &=& a^{D-2} (\eta)  \delta^{ij} \partial_i \Bigg[ \Big[1+ \Phi_G(x) - (D-3) \Psi_G(x) \Big] \partial_j \hat{\phi}(x)   \Bigg]\nonumber \\ && \qquad \qquad  - \frac{ m^2}{\hbar^2} a^D (\eta) \Big[1+ \Phi_G(x) - (D-1) \Psi_G(x) \Big] \hat{\phi}(x)  \, .\label{piPrime} 
\eea
We stress that  the constraint fields $\Phi_G$, $\Psi_G$ do not evolve independently. Thus, we are not fully fixing the gauge by integrating out the constraint fields. We do this in order to make the connection to the Einstein-Vlasov system clearer. This means, at the same time that we are approximating scalar interactions that are induced via gravity by stochastic two-point functions. However, examining the fully gauged fixed theory at the quantum level is planned for the future. 

\subsection{Why a Quantum Field Formalism for a Classical Problem? \label{whyQuantum}}

One might object that the quantum field theory framework we presented so far is a completely exaggerated tool to describe effects that arise in the classical treatment of late-time cosmology.
However, this description has on the one hand the advantage of being based on a fundamental theory which permits a Lagrangian description 
and in which in our simple model contains one parameter, the scalar field mass $m$. On the other hand, it is related to the typical classical non-relativistic particle description by imposing conditions that approximate a classical stochastic rather than a quantum description as well as a gradient expansion that contains relativistic corrections. We note that quantum path integrals generalize classical stochastic path integrals where the quantum commutators \eqref{commRel} are replaced by Poisson brackets \cite{Berges:2015kfa}. For us it is important to inherit the stochastic correlations of two-point functions from the quantum field theory framework since late-time cosmology is a classical stochastic theory whose stochastic seeds are given by the gravitational potentials. That quantum effects are potentially present in this approach is a completely negligible add-on rather than a crucial ingredient.
Still, this perspective has the advantage that we always keep the bridge to non-equilibrium quantum field theoretic techniques as for example the Schwinger-Keldysh formalism \cite{Schwinger:1960qe} \cite{Keldysh:1964ud} that might be useful when studying non-linear evolution. 
Our formalism is a hybrid formalism in the sense that we use a mixture of 2-PI formalism \cite{jackiw} for the scalar field matter and 1-PI formalism for gravity and do not fully fix gravitational gauge in the sense that we do not fully solve (gravitational) constraints of the theory, but instead we leave the gravitational potentials as external stochastic sources (keeping in mind that they are eventually fixed by the linear Einstein equations).

The condition of being in the classical stochastic regime of a quantum field theory rather than in the quantum regime can be formulated as follows: the (classical) correlators
 that contain anti-commutators
 (and no time ordering) are much larger than the quantum correlators defined in terms of anti-commutators with or without time ordering
 (example of which include the causal or spectral two-point function $\langle[\hat{\phi}(x),\hat{\phi}(\widetilde{x})]\rangle$, retarded and advanced
 propagators, etc.). We therefore assume that our two-point functions obey,
 \begin{equation}
   2F(x;\tilde x)\equiv \langle\{\hat\phi(x),\hat\phi(\widetilde{x})\}\rangle\gg \big| \langle[\hat\phi(x),\hat\phi(\widetilde{x})]\rangle \big|
   \,,
\label{classicality condition}
\end{equation}
where $\{\hat\phi(x),\hat\phi(\widetilde{x})\}$ and $[\hat\phi(x),\hat\phi(\widetilde{x})]$ denote anti-commutator and commutator operation, respectively. Rigorously speaking, the classicality condition~(\ref{classicality condition}) is never satisfied for all space-time points.
By assuming~(\ref{classicality condition}) we are saying that we restrict ourselves to those space-time points where~(\ref{classicality condition})
is amply satisfied. Rather than rigorously going through a procedure that would achieve that in practice, here we just sketch how such a procedure 
can be exacted. In the case of interest for dark matter, the condition~(\ref{classicality condition}) will be met for sufficiently large spatial separations.
One can make use of a suitable window (smearing) function, which projects out of the full two-point function its classical part. When the 
complementary (`quantum')  part of the two-point function is integrated out, one will generate local geometric divergent contributions (that can be renormalized by adding suitable local geometric counterterms). Apart from renormalizing the Newton and cosmological constant to its observable values, 
the remaining geometric terms will have a negligible effect on the evolution of late time two-point functions, and we shall neglect them here. The remaining infrared parts of the two-point functions will satisfy the classicality condition~(\ref{classicality condition}).

To get a better feeling on what classicality really means, it is helpful to assume adiabaticity with respect to gradient expansion
(discussed in more detail below), in which case one can perform a Wigner transform with respect to the relative spatial coordinate,
$x^i-\tilde{x}^i$, resulting in the statistical two-point function,  $F(X^i,p_i,t,t')$, $X^i\equiv (x^i+\tilde{x}^i)/2$. 
When $F(X^i,p_i,t,t')[\partial_t\partial_{t'}F(X^i,p_i,t,t')]_{t'=t}\gg (\hbar/2)^2$
is satisfied, then one is in the classical regime. \footnote{An alternative (and related) 
criterion for classicality of a state is given by the von Neumann entropy of the Gaussian part of the density matrix being much larger than one \cite{Koksma:2011dy}\cite{Koksma:2011fx}.}
More concretely, in the case at study we expect the classicality condition~(\ref{classicality condition}) to be satisfied for two-point functions 
that are smeared on distances larger than the co-moving distance corresponding to the end of inflation,
which is of the order $\sim 1{\rm m} $. Since the two-point functions we use to describe dark matter
on the scales of large-scale structures, they are in a deeply classical regime and the condition \eqref{classicality condition} is royally satisfied.

\medskip

  The second important condition to get into the regime of non-relativistic particles is related to validity of the gradient expansion. Roughly speaking,
 the expansion is valid when the following two conditions are met,
\begin{equation}
 \|\hbar\,\partial_{\vec X}\cdot \partial_{\vec p}\| \ll 1
 \,,\quad
  \|\hbar\, \partial_{\eta}\, \partial_E \|\ll 1
 \,,
 \label{gradient expansion validity}
\end{equation}
where the norm is to be understood in the sense that one derivative acts on one test function (such as a two-point function)
and the other on another object (such as a gravitational potential).
\footnote{The validity of this expansion depends on the initial density matrix which ought to be classical enough. The initial density matrix can for example be taken to be Gaussian, containing initial one-point functions and connected two-point functions. In particular, without any coarse-graining only the connected part of the two-point function can satisfy the conditions of gradient expansion.} 
Assuming that two-point functions vary on scales of momentum
(energy) given by the momentum (energy), i.e. $\hbar\partial_E\sim \hbar/E\sim \hbar/(mc^2)\sim \lambda_C/c$, and
$\hbar\partial_{\vec p}\sim \hbar/\|\vec p\|\sim \hbar/(mv)\sim \lambda_{dB}$, where $\lambda_C$ and
$\lambda_{dB}$ denote the Compton and de Broglie wavelength, respectively,
 the conditions
\eqref{gradient expansion validity} can be recast as,
\begin{equation}
      L \gg \lambda_{dB}
 \,,\quad
        T  \gg \frac{\lambda_{C}}{c}
 \,,
 \label{gradient expansion validity:2}
\end{equation}
where $L$ and $T$ represent the characteristic length and time scales over which gravitational potentials
or two-point functions vary.
$L$ can be as small as the smallest large-scale structures we are interested in (which is of the order $\sim {\rm kpc}$)
which implies that $T$ must be much larger than the time light crosses about one mega-parsec
which is about a million years or so (this estimate follows from
the observation that $\lambda_{dB}\sim 10^3\lambda_C$ as $v\sim (10^{-3}-10^{-2})c$).
To get a feeling on how good that approximation is, note that the inequalities are amply satisfied for a dark matter 
whose mass is of the electroweak scale $\sim 10^2~{\rm GeV}$. However, when one considers ultra-light scalar 
such as in references \cite{Marsh:2015daa} \cite{Hui:2016ltb}, the scalar mass is of the order $m\sim 10^{-22}-10^{-24}~{\rm eV}$,
the Compton and de Broglie wave lengths are $\lambda_C\sim 10^{-4}-10^{-2}~{\rm kpc}$, $\lambda_{dB}\sim 10^{-2}-10~{\rm kpc}$,
the quantities in \eqref{gradient expansion validity:2} can become comparable for smallest scales of interest, and hence one expects
significant higher order gradient corrections. 
One is typically interested in modeling dark matter at an accuracy better than $1\%$ (as it will be tested by upcoming observations), which then defines the order in gradient expansion that one ought to keep.
The corrections of the gradient expansion can be subsumed by the following perturbation parameters
\bea
\varepsilon_{\hbar} &\sim& \Bigg\lbrace\varepsilon_{\text{k}} \sim \hbar\frac{ \partial_X}{ma}\,,  \label{pertParam1}\,
\varepsilon_{\text{k/p}} \sim \hbar \frac{ \partial_X}{p} \sim \hbar { \partial_X}{\partial p}\,, \,  
\varepsilon_{\text{{H}}} \sim \hbar \frac{ \mathcal{H}}{m a } \,,\,   
\varepsilon_{\partial \eta} \sim   {\hbar } \frac{\partial_{\eta}}{ma }\Bigg\rbrace \, .
\eea

Next, there are relativistic corrections due to the relativistic nature of dark matter.
The fact that the energy is not equal to the rest energy we can fully capture in our formalism as long as we keep
on-shell energy roughly speaking equal to its quasi-particle value, $E=\sqrt{m^2+p_{ph}^2}$,
where $p_{ph}^2= g_{ij}p^ip^j$ denotes the physical momentum squared (for simplicity in here
we do not include all of the metric corrections). To study these corrections one can systematically
include them order by order if,
\begin{equation}
 \varepsilon_{\text{p}} \sim \frac{p_{\rm com}}{ma}\ll 1
\label{relativistic condition}
\end{equation}
where $p_{com}$ denotes the comoving momentum today.
These corrections occur e.g. as relativistic corrections to the energy-momentum tensor whose most important components
are energy density and pressure which source the generalized Poisson-like equations for the gravitational potentials.

Furthermore,  there are relativistic corrections induced by the relativistic nature of the scalar field Klein-Gordon equations,
and these appear as higher order time derivatives in the Vlasov (or collisionless Boltzmann) equation. These corrections are small if the second condition
in equations \eqref{gradient expansion validity} and \eqref{gradient expansion validity:2} is met and if
$\|\partial_{\eta}\|\ll \mathcal{H}$, where $\mathcal{H}$ denotes the Hubble expansion rate in conformal time.

Next, there are relativistic corrections arising from general relativity being different from
Newton's gravity. These corrections occur as higher time derivative corrections to gravitational potentials 
and as the corrections induced by the Universe expansion. The latter are suppressed by the Hubble rate $H$
and they are small if,
\begin{equation}
 T\ll \frac{1}{a\mathcal{H}}
\,.
\label{Hubble scale correction}
\end{equation}

Furthermore, since general relativity has more degrees of freedom than Newton's gravity,
there are general relativistic corrections expressed as a non-vanishing gravitational slip.
Finally, we expect that as a result of non-linear interactions between matter and gravity, gravitational vector and
tensor perturbations will be (dynamically) generated (even if they are not present at the initial time).
In this paper we neglect these
type of perturbations, but they can be easily included in our formalism by including them in the 
{\it Ansatz} for the metric tensor.

Of course, there are also higher order gravitational perturbations.
However, since on the large scales we are interested in gravitational potentials
do not grow much beyond their initial value,
\be
\varepsilon_{\text{g}}^2 \sim \Phi_G\, , \, \Psi_G\,  \, \sim 10^{-5} \ll 1  \, ,\label{pertGrav}
\ee
we can safely neglect terms of the form $\Phi_G^2$ and $\Psi_G^2$;
higher order vector and tensor perturbations can be also neglected since vectors and tensors remain smaller than gravitational scalars
throughout the evolution.
We will also encounter the parameter
\be
\varepsilon_{\text{{H}/k}} \sim  \mathcal{H} \partial_X^{-1} \,\label{pertParamn} ,
\ee
that controls whether we are on sub- or superhorizon scales. 

\subsection{Phase-Space Distributions from Wigner Transformation} \label{phasefromWigner}
The concept of the Wigner transformation (see for example \cite{trove.nla.gov.au/work/9783845} or \cite{calzetta2008nonequilibrium}) was introduced to extract phase-space distributions and its Boltzmann equation from particle wavefunctions, then generalized to field theory in Minkowski space-time and even later to field theory in curved space-time to yield a covariant Boltzmann or Vlasov equation. However, the generalization of the Wigner transformation to arbitrary curved space-times must still be considered as an active research field since there are as far as we know merely four major contributions to this area \cite{Winter:1986da} \cite{Calzetta:1987bw} \cite{Fonarev:1993ht} \cite{Antonsen:1997dc} which agree only in the limit where $\hbar$ goes to zero. Apart from the  proposal by \cite{Antonsen:1997dc}, all approaches are based on perturbative expressions. On the other hand, all of these papers cover almost entirely off-shell phase-space distribution $f(X^{\mu}, p_{\nu}) $ in the sense that the momentum conjugate to the time difference $\Delta \eta$ in the two-point function, $p_0$, is still an independent variable that needs to be put on shell by integrating it out since the starting point for the Wigner transformation is a non-equal-time two-point function, $\Delta \eta \neq 0$. As it was to our knowledge first pointed out by \cite{PhysRevD.57.6525} for electrodynamics in Minkowski space-times, going on-shell requires more than one moment in $p_0$ space, i.e. $\int \frac{dp_0}{2 \pi \hbar}  p_0^{n} f(X^{\mu}, p_{\nu})$. This approach of taking several moments and relating them has been applied to homogeneous cosmological backgrounds by \cite{Garbrecht:2002pd} to define a particle number density. Our goal is to obtain candidates for on-shell phase-space distributions $f(X^{\mu}, p_{i}) $  and their dynamics for the non-homogeneous linearized longitudinal metric \eqref{longMetric} we provided in the beginning. We seek them by computing the dynamics of two-point functions on-shell or at equal-times in the operator formalism and then performing a $(D-1)$-dimensional Wigner transformation of these  equal-time two-point functions.  Apart from our interest in scalar field dark matter, we want to use the linearized longitudinal metric as a guideline to gain some intuition for on-shell Wigner transformation in curved space-time and generalize it in future work to arbitrary non-perturbative metrics.
\par Let us start by looking at the four equal-time two-point functions whose combination might provide candidates for phase-space distributions after a $(D-1)$-dimensional  Wigner transformation
\bea
 {F}_{00} \big(\eta, x^i, \widetilde{x}^i \big) &:=& \vac{  {\hat{\phi} \big(\eta, x^i \big)} {\hat{\phi}\big(\eta, \widetilde{x}^i \big)}} = \Tr \Big[ \hat{\rho}_{\text{ini}}\big[\hat{\phi},\hat{\Pi},  \Phi_G, \Psi_G \big] \,   {\hat{\phi} \big(\eta, x^i \big)} {\hat{\phi}\big(\eta, \widetilde{x}^i \big)} \Big]   \label{F00}   \, , \\
F_{10} \big(\eta, x^i, \widetilde{x}^i \big)  &:=& \vac{\hat{\Pi}\big(\eta, x^i \big) {\hat{\phi}\big(\eta, \widetilde{x}^i \big)}} = \Tr \Big[  \hat{\rho}_{\text{ini}}\big[\hat{\phi},\hat{\Pi},  \Phi_G, \Psi_G \big]\,  \hat{\Pi}\big(\eta, x^i \big) {\hat{\phi}\big(\eta, \widetilde{x}^i \big)} \Big] \, , \label{F10} \\
 F_{01} \big(\eta, x^i, \widetilde{x}^i \big)  &:=& \vac{{\hat{\phi}\big(\eta, x^i \big)} \hat{\Pi}\big(\eta, \widetilde{x}^i \big)} = \Tr \Big[  \hat{\rho}_{\text{ini}}\big[\hat{\phi},\hat{\Pi},  \Phi_G, \Psi_G \big]\, {\hat{\phi}\big(\eta, x^i \big)} \hat{\Pi}\big(\eta, \widetilde{x}^i \big) \Big]\, ,  \label{F01} \\
 F_{11} \big(\eta, x^i, \widetilde{x}^i \big)  &:=& \vac{{\hat{\Pi}\big(\eta, x^i \big) } {\hat{\Pi}\big(\eta, \widetilde{x}^i \big)} }= \Tr \Big[ \hat{\rho}_{\text{ini}}\big[\hat{\phi},\hat{\Pi},  \Phi_G, \Psi_G \big]\, {\hat{\Pi}\big(\eta, x^i \big) } {\hat{\Pi}\big(\eta, \widetilde{x}^i \big)} \Big] \,  \label{F11} ,
\eea
where the expectation values are taken with respect to some initial density matrix $ \hat{\rho}_{\text{ini}}$ which functionally depends on the operators $\hat{\phi}$ and $\hat{\Pi}$ at some initial time $\eta_{\text{ini}}$. Note that in the context of cosmology the initial density matrix $ \hat{\rho}_{\text{ini}}$ depends also functionally on the stochastic gravitational potentials $\Phi_G$, $\Psi_G$ at this initial time $\eta_{\text{ini}}$,
and in its general form it allows for implementation of effects of 
coherent states, squeezing and state mixing.
In this way, two-point functions can be stochastic quantities, 
\begin{multline}
F_{00} (\eta ,x^i, \widetilde{x}^i ) \neq \vac{F_{00} }_{(\Phi_G , \Psi_G)}(\eta ,x^i, \widetilde{x}^i )  := \vac{  \vac{{\hat{\phi} \big(\eta, x^i \big)} {\hat{\phi}\big(\eta, \widetilde{x}^i \big)} }_{ \hat{\rho}}}_{(\Phi_G , \Psi_G)} \\ :=  \int \mathcal{D}\Phi_G \mathcal{D}\Psi_G \mathcal{P}\big[\Phi_G, \Psi_G \big] \Tr \Big[ \hat{\rho}_{\text{ini}}\big[\hat{\phi},\hat{\Pi},\Phi_G, \Psi_G \big]\,   {\hat{\phi} \big(\eta, x^i \big)} {\hat{\phi}\big(\eta, \widetilde{x}^i \big)} \Big]   \, ,
\end{multline}
where $\mathcal{P}$ is a probability distribution for the gravitational potentials. The reason to introduce this formalism lies in its application to cosmology in the sense that we want to bridge a gap from semi-classical quantum field theory to cosmological perturbation theory.\footnote{We remark, that this might be closely related to the stochastic gravity framework proposed in \cite{Hu:2008rga} although we did not investigate this further. For us, the stochasticity of two-point functions is more an ad-hoc {\it Ansatz} that turns out to be very convenient in relation to cosmological perturbation theory.} Thus, $\Phi_G$ and $\Psi_G$ are stochastic, homogeneously distributed fields that evolve into non-Gaussian fields due to the evolution of large-scale structures. We want to think of this model more as a conceptional test case on how to relate full quantum microscopic theories to models in cosmological perturbation theory as for example the cold dark matter model. Once we find that this is a fruitful {\it Ansatz}, we will provide generalizations for arbitrary metrics and even wave the semi-classical approach by integrating out the gravitational constraint fields $\Phi_G\, ,\Psi_G$, which at the moment act as approximate self-interactions of the scalar field theory since the Einstein equations constrain them to be related to the scalar field's two-point functions. We also want to point out that the reducible and connected pieces of the two-point functions entering \eqref{F00} to \eqref{F11} do not decouple in general which is due to the gravitational fields since the semi-classical Einstein equation will constrain them to be related to both one-point functions as well as the connected matter field two-point function. On the other hand, this makes clear that setting initially either the one-point functions of the matter field or connected parts of the matter field two-point function to zero removes them from \eqref{F00} to \eqref{F11}  for all times.
Before we turn to the Wigner transformation itself, we write down the dynamics of the equal-time two-point functions \eqref{F00} to \eqref{F11}. Let us therefore define the differential operator
\begin{multline}
\mathcal{D} (\eta, x^i) :=   a^{D-2}(\eta) \delta^{ij}\Bigg[  \Big[\partial_i  \Phi_G \Big](\eta, x^i) - (D-3)  \Big[ \partial_i \Psi_G \Big](\eta, x^i) \Bigg] \frac{\partial}{\partial x^j} \\  + a^{D-2}(\eta) \Big[ 1 + \Phi_G(\eta, x^i) -(D-3) \Psi_G(\eta, x^i) \Big] \Delta^x  - \frac{ m^2}{\hbar^2}  a^{D}(\eta) \Big[1+ \Phi_G(\eta, x^i) - (D-1) \Psi_G(\eta, x^i)\Big] \, ,
\end{multline}
where we used the Laplace operator $\Delta$ which on conformally flat cosmological spaces equals $\delta^{ij} \partial_i \partial_j$.
We also define the following function as a shorthand
\be
h(\eta, x^i) :=a^{-(D-2)}(\eta) \Big[ 1 + \Phi_G(\eta, x^i)  + (D+1) \Psi_G(\eta, x^i) \Big] \, .
\ee
Then, based on the Hamilton's equation for the canonical operators \eqref{phiPrime} and \eqref{piPrime}, we get the following system of equations
\bea
F_{00}^{\prime}(\eta, x^i, \widetilde{x}^i )&=&  h(\eta, x^i)   F_{10} (\eta, x^i, \widetilde{x}^i )   + F_{01} (\eta, x^i, \widetilde{x}^i ) h(\eta , \widetilde{x}^i)\, , \label{firstDerF00}  \\
F_{10}^{\prime}(\eta, x^i, \widetilde{x}^i ) &=&  \mathcal{D}(\eta, x^i) \, F_{00}(\eta, x^i, \widetilde{x}^i )  +     F_{11}  (\eta, x^i, \widetilde{x}^i )h(\eta , \widetilde{x}^i)  \, \label{firstDer10} , \\
F_{01}^{\prime}(\eta, x^i, \widetilde{x}^i )  &=&  \mathcal{D}(\eta, \widetilde{x}^i) \,  F_{00} (\eta, x^i, \widetilde{x}^i )  +h(\eta, x^i)  F_{11}  (\eta, x^i, \widetilde{x}^i )  \, \label{firstDer01} , \\
 F_{11}^{\prime}(\eta, x^i, \widetilde{x}^i ) &=&   \mathcal{D}(\eta, x^i)  \, F_{01}(\eta, x^i, \widetilde{x}^i )  + \mathcal{D}(\eta, \widetilde{x}^i) \, F_{10}(\eta, x^i, \widetilde{x}^i )   \, \label{firstDer11} . 
\eea
This is a system of four first-order differential equations with four independent initial conditions. However, we have to keep in mind that these two-point functions obey certain symmetry properties and we realize by combining $F_{10}$ and $F_{01}$ that we can specify three symmetric functions and one anti-symmetric function as initial conditions.
We remark, that this system of equations closes in the sense that we need no information about higher n-point functions.
This is due to the following reasons: firstly, we neglected manifest self-interactions of the scalar field (e.g. $\sim \lambda \phi^4$) that are not due to gravity, secondly, we approximate the self-interactions that are induced via gravity. These interactions are non-local in space but local in time via an inversion of the generalized Poisson equation. It means that the scalar field couples in this approximation only to its two-point functions since the gravitational potentials $\Phi_G$, $\Psi_G$ are -  via the semi-classical Einstein equations - entirely expressible in terms of the scalar field two-point functions.  We call this the hybrid approach. Thirdly, the scalar field does not interact with dynamical part of gravity, the gravitons, since we put them to zero by hand as an assumed negligible effect. \par
Let us continue to manipulate the system of equations \eqref{firstDerF00} to \eqref{firstDer11} by switching to collective and difference coordinates for the spatial parts,
\be
X^i := \frac{ x^i + \widetilde{x}^i}{2}\, , \qquad r^i :=   x^i - \widetilde{x}^i\, . \label{defCollandDiff}
\ee
We define the Wigner transform with respect to covariant momenta $p_{i}$ and its zeroth moment denoted by a bar as
\bea
F_{*} (\eta, X^i,p_i) &:=&  \int {d^{D-1} r}\, e^{- \frac{i}{\hbar} p_i  r^i}  F_{*} (\eta, X^i,r^i)\, , \\
\overline{F}_{*} (\eta, X^i) &:=& \int \frac{d^{D-1} p}{(2 \pi \hbar)^{D-1}} F_{*} (\eta, X^i,p_i)\,.
\eea
This definition is our equal-time, thus on-shell, version of the several curved space-time generalizations of the Minkowski space-time Wigner transformation of course for the our specific choice of a longitudinal linearized metric without gravitons. It is a direct generalization of the {\it Ansatz} in \cite{Garbrecht:2002pd} from the homogeneous FLRW-space-time to its non-homogeneous perturbed form. This definition identifies the time coordinate $\eta$ and spatial collective coordinates $X^i$ as a single point on the curved space-time manifold, whereas the spatial difference coordinates $r^i$ and the momenta $p_i$ belong to the tangent and cotangent space, respectively, that is associated to that point. We neither make use of any geodesic expansion nor do we use Riemannian coordinates. This implies also that our equations are exact apart from the linearization in the gravitational potentials. We also would like to mention that the on-shell operator formalism for curved space-times we are using resolves the problem of perturbatively solving the off-shell constraint equation which always accompanies the off-shell Vlasov equation by providing a manifest closure
for on-shell correlators \cite{Winter:1986da} \cite{Calzetta:1987bw} \cite{Fonarev:1993ht} \cite{Antonsen:1997dc}.   
\par
It will turn out that upon Wigner transforming the equations \eqref{firstDerF00} to \eqref{firstDer11}, two other correlators are much more useful than $F_{01}$ and $F_{10}$. Thus, we define the following combination of equal-time two-point functions
\bea
 F_{+} (\eta, x^i, \widetilde{x}^i) &:=&  \frac{1}{2} \Big[ F_{10} (\eta, x^i, \widetilde{x}^i) + F_{01} (\eta, x^i, \widetilde{x}^i)\Big]\, \label{defFPlus}\\ & =&  \frac{1}{4}  \vac{\Big\lbrace \hat{\Pi}(\eta, x^i)\, , {\hat{\phi}(\eta, \widetilde{x}^i)} \Big\rbrace  +\Big\lbrace \hat{\phi}(\eta, x^i)\, , {\hat{\Pi}( \eta, \widetilde{x}^i)}\Big\rbrace }  \, ,\nonumber  \\
  F_{-} (\eta, x^i, \widetilde{x}^i) &:=&  \label{defFMinus} \frac{i}{2} \Big[ F_{10} (\eta, x^i, \widetilde{x}^i) - F_{01} (\eta, x^i, \widetilde{x}^i)\Big] - \frac{\hbar}{2} \, \delta^{D-1} (x^i - \widetilde{x}^i)\\  &=& \frac{i}{4}  \vac{\Big\lbrace \hat{\Pi}(\eta, x^i)\, , {\hat{\phi}(\eta, \widetilde{x}^i)} \Big\rbrace  - \Big\lbrace \hat{\Pi}(\eta, \widetilde{x}^i) \, , {\hat{\phi}(\eta, x^i)}\Big\rbrace } \, , \nonumber  
\eea
 where $\big\lbrace\, . \,, \, . \,  \big\rbrace$ denotes the anti-commutator.
The calculation of transforming the two-point function dynamics into Wigner space is shown in appendix \ref{appWignerDyn}.
We remark that this calculation is exact up to the linearization in the gravitational potentials.
Since we want to identify phase-space distributions we have to treat the problem by utilizing the gradient approximation .
Therefore, we consider again the perturbation parameters in \eqref{pertParam1} to \eqref{pertGrav} and solve perturbatively for $F_{+}$ and $F_{11}$ which are determined in terms of $F_{00}$ and $F_{-}$,
\begin{multline}
  F_+
=  \Bigg\lbrace \frac{1}{2} \Big[ 1 - \Phi_G  - (D-1)  \Psi_G \Big]  a^{D-2} F_{00}^{\prime}    \\ 
- \frac{\hbar}{2}  \frac{\partial}{\partial X^i} \Big[ \Phi_G  + (D-1)  \Psi_G \Big] \frac{\partial}{\partial p_i}  F_-  \Bigg\rbrace \times \Bigg\lbrace 1+\mathcal{O} \big(\varepsilon_{\hbar}^2 \cdot \varepsilon_{\text{g}}^2 \big)+ \mathcal{O} \big(\varepsilon_{\hbar}^4 \big) \Bigg\rbrace \, , \label{F+IntermsOfFMinusandF00}
\end{multline}
\begin{multline}
 F_{11}  =    a^{D-2}\Bigg\lbrace \frac{1}{2} \Big[ a^{D-2} F_{00}^{\prime} \Big]^{\prime}    
\\ - \Bigg[\frac{\hbar}{2}  \frac{\partial}{\partial X^i} \Big[ \Phi_G  + (D-1)  \Psi_G \Big] \frac{\partial}{\partial p_i}  F_- \Bigg]^{\prime}  -  \frac{\Delta_X}{4} \Big[ a^{D-2} F_{00} \Big] 
   +\frac{p^2}{\hbar^2} \Big[1 - 2(D-2) \Psi_G \Big]   \Big[ a^{D-2} F_{00} \Big]  
 \\ + \frac{ m^2}{\hbar^2} a^2 \Big[1 - 2(D-1) \Psi_G \Big] \Big[ a^{D-2} F_{00} \Big] \Bigg\rbrace \times \Bigg\lbrace 1+\mathcal{O} \big(\varepsilon_{\hbar}^2 \cdot \varepsilon_{\text{g}}^2 \big)+ \mathcal{O} \big(\varepsilon_{\hbar}^4 \big) \Bigg\rbrace \, , \label{F11IntermsOfFMinusandF00}
\end{multline}
where
\be
p^2 := \delta^{ij} p_i p_j\, .
\ee
The dynamics of $F_{00}$ and $F_{-}$ is given by the following coupled equations,
\begin{multline}
 F_{-}^{\prime}  = \Bigg\lbrace
 \frac{1}{ \hbar}  \frac{\partial}{\partial X^k} \Big[ \Phi_G +  \Psi_G \Big] \frac{\partial}{\partial p_k}
 \Bigg[ {p^2} \Big[ a^{D-2} F_{00} \Big] \Bigg]
 - \frac{p}{ \hbar} \cdot \partial_X \Big[ \Phi_G - (D-3) \Psi_G \Big] \Big[ a^{D-2} F_{00} \Big]
 \\ - \Big[1+ \Phi_G - (D-3) \Psi_G \Big] \frac{p_k }{\hbar} \frac{\partial}{\partial X^k}   \Big[ a^{D-2} F_{00} \Big] 
 \\  +  \frac{ m^2 a^2  }{\hbar}  \frac{\partial}{\partial X^i} \Phi_G  \frac{\partial}{\partial p_i} \Big[ a^{D-2} F_{00} \Big] \Bigg\rbrace \times \Bigg\lbrace 1+\mathcal{O} \big(\varepsilon_{\hbar}^2 \cdot \varepsilon_{\text{g}}^2 \big)+ \mathcal{O} \big(\varepsilon_{\hbar}^4 \big) \Bigg\rbrace\, , \label{FMinDynhleader}
\end{multline}
\begin{multline}
\Bigg\lbrace\frac{1}{2} \Bigg[ a^{D-2} F_{00}^{\prime} \Bigg]^{\prime\prime}   + (D-2) \frac{\mathcal{H}}{2} \Bigg[ a^{D-2} F_{00}^{\prime}  \Bigg]^{\prime}       +\frac{p^2}{\hbar^2} \Big[1 - 2(D-2) \Psi_G \Big]^{\prime}   \Big[ a^{D-2} F_{00} \Big] \\ 
 + \Bigg[\frac{ m^2}{\hbar^2} a^2 \Big[1 - 2(D-1) \Psi_G \Big]\Bigg]^{\prime}  \Big[ a^{D-2} F_{00} \Big]      +  2\frac{ m^2}{\hbar^2} a^2   \Big[ 1  - 2(D-1)  \Psi_G \Big] \Big[ a^{D-2} F_{00} \Big]^{\prime}  \\ 
-  \frac{\Delta_X}{2}   \Big[a^{D-2} F_{00}\Big]^{\prime}  
+  2 \frac{p^2}{\hbar^2} \Big[ 1 - 2(D-2)  \Psi_G \Big] \Big[ a^{D-2} F_{00} \Big]^{\prime}  
-2 \frac{{p^2} }{\hbar} \frac{\partial}{\partial X^k}   \Big[ \Phi_G +  \Psi_G \Big] \frac{\partial}{\partial p_k}    F_{-} \\
 + 2 \Big[1+ \Phi_G - (D-3) \Psi_G \Big] \frac{p}{\hbar }\cdot \partial_X   F_{-} 
- 2\frac{ m^2a^2}{\hbar}    \frac{\partial}{\partial X^i} \Phi_G   \frac{\partial}{\partial p_i}  F_-\Bigg\rbrace \times \Bigg\lbrace 1+\mathcal{O} \big(\varepsilon_{\hbar}^2 \cdot \varepsilon_{\text{g}}^2 \big)+ \mathcal{O} \big(\varepsilon_{\hbar}^4 \big) \Bigg\rbrace =0 \, .\label{F00Dynhleader}
\end{multline}
We conclude that to this order in the gradient approximation, we still keep all degrees of freedom that were contained on the original first-order system \eqref{firstDerF00} to \eqref{firstDer11}. Dropping the third-order time derivative would leave us with the degrees of freedom of a symmetric and an anti-symmetric function. 
 Before we try to recover a Vlasov equation from these equations let us pause a bit and make it clear how this equation reflects the difference between products of one-point functions and connected two-point functions. We will also realize that the higher-order time-derivatives correspond to oscillatory degrees of freedom. For simplicity, we set $D=4$ and focus on the homogeneous part of equation \eqref{F00Dynhleader} in the large mass limit ($p \ll m$), where we denote the homogeneous approximation of $F_{00}$ by ${F}_{00}^{\text{hom}}$,
\be
\frac{1}{2} \Big[ a^{2} \big[F_{00}^{\text{hom}} \big]^{\prime} \Big]^{\prime\prime}   +  {\mathcal{H}} \Big[ a^{2} \big[F_{00}^{\text{hom}} \big]^{\prime}   \Big]^{\prime}       
 +2 \mathcal{H}\frac{ m^2}{\hbar^2} a^2  \Big[ a^{2} F_{00}^{\text{hom}} \Big]      +  2\frac{ m^2}{\hbar^2} a^2    \Big[ a^{2} F_{00}^{\text{hom}} \Big]^{\prime}  \approx 0 \, . 
\ee
Expanding this equation, we arrive at
\begin{multline}
\Big[a^3  F_{00}^{\text{hom}} \Big]^{\prime \prime \prime}-3 \mathcal{H}\Big[a^3  F_{00}^{\text{hom}}  \Big]^{\prime \prime}+\Big[4 \frac{m^2}{\hbar^2}a^2 - \mathcal{H}^2 - 7 \mathcal{H}^{\prime}\Big]\Big[a^3  F_{00}^{\text{hom}}  \Big]^{\prime } \\+3 \Big[\mathcal{H}^3 + \mathcal{H}\mathcal{H}^{\prime} - \mathcal{H}^{\prime \prime} \Big] \Big[a^3  F_{00}^{\text{hom}} \Big] \approx 0 \, . \label{hom1}
\end{multline}
In order to make progress, we also have to provide the Einstein equations, which we derive in appendix \ref{appEinstein} in equations \eqref{Einstein1} to \eqref{Einstein2}. The spatially homogeneous equations with neglected momenta $p \ll m$ read in $D=4$ dimensions (all terms $p^2 m^{-2}$ are dropped),
\bea
- 2 \mathcal{H}^{\prime} - \mathcal{H}^2 
   &\approx&  \frac{\hbar}{2  M_P^2} \Bigg\lbrace  a^{-4} F_{11}^{\text{hom}}
  - \frac{ m^2 a^2}{\hbar^2} F_{00}^{\text{hom}} \Bigg\rbrace \,, \label{hom2}  \\ 
3 \mathcal{H}^2  &\approx& \frac{\hbar}{2 M_P^2} \Bigg\lbrace    a^{-4} F_{11}^{\text{hom}}
+ {a^2} \frac{m^2}{\hbar^2}  F_{00}^{\text{hom}} \Bigg\rbrace  \,, \label{hom3}
\eea
and from \eqref{F11IntermsOfFMinusandF00}, we have
\be
 F_{11}^{\text{hom}} \approx    \frac{a^{2}}{2} \Big[ a^{2} \big[F_{00}^{\text{hom}} \big]^{\prime}\Big]^{\prime}    
 + \frac{ m^2}{\hbar^2} a^6    F_{00}^{\text{hom}} \, .\label{hom4}
\ee
The equations \eqref{hom1} to \eqref{hom4} admit three independent solutions. We can guess them quickly by noting once more that $F_{00}^{\text{hom}}$ is constructed out of one-point functions and a connected piece
\be
F_{00}^{\text{hom}} = {\langle \hat{\phi} \rangle}^{\text{hom}} {\langle \hat{\phi} \rangle}^{\text{hom}} + {\langle \hat{\phi} \hat{\phi} \rangle}_\text{connected}^{\text{hom}} \, .
\ee
In the limit $\mathcal{H} \ll \hbar^{-1} ma$, the solutions for the one-point functions are through the Klein-Gordon equations approximately given by 
\be
{\langle \hat{\phi} \rangle}^{\text{hom, sol 1}} \approx a^{-3/2} \cos \Big( \int d \eta \,  m a \Big) \,, \quad {\langle \hat{\phi}\rangle}^{\text{hom, sol 2}} \approx a^{-3/2}\sin \Big( \int d \eta \,  m a \Big)\, .
\ee
Had we used only one-point functions $\langle \hat{\phi} \rangle$ to construct $F_{00}^{\text{hom}}$, our analysis would be complete at this stage since we can only impose two initial conditions for $\langle \hat{\phi} \rangle$ and they would completely determine $F_{00}^{\text{hom}}$ which in this case has always an oscillatory contribution.
However, let us forget about the one-point functions and focus on the connected part of $F_{00}^{\text{hom}}$. We see that the following functions are two independent solutions to \eqref{hom1},\footnote{Beeing ignorant about any $p$ dependence for the moment, the initial density matrix for homogeneous two-point functions contains five initial conditions (see e.g. \cite{Berges:2015kfa}): the one-point functions $\langle \hat{\phi} \rangle$ and $\langle \hat{\Pi} \rangle$, and the connected parts of the two-point functions $F_{00}$, $F_{11}$ and $F_{+}$ or equivalently $F_{00}$, $F_{00}^{\prime}$ and $F_{00}^{\prime \prime}$. The fourth condition for the connected part of $F_{-}$ is trivially satisfied in the homogeneous case. } 
\be
F_{00}^{\text{hom, sol 1}} \approx a^{-3} \cos^2 \Big( \int d \eta \,  m a \Big) \,, \quad F_{00}^{\text{hom, sol 2}} \approx a^{-3} \sin^2 \Big( \int d \eta \,  m a \Big)\, ,
\ee
\be
F_{11}^{\text{hom, sol 1}} \approx \frac{m^2}{\hbar^2} a^{3} \sin^2 \Big( \int d \eta \,  m a \Big) \,, \quad F_{11}^{\text{hom, sol 2}} \approx  \frac{m^2}{\hbar^2}a^{3} \cos^2 \Big( \int d \eta \,  m a \Big)\, .
\ee
The, corresponding solutions for the Hubble rate are given by the following leading order terms
\bea
 \Big[ \mathcal{H}^2 a \Big]^{\text{sol 1,2}} & \approx & \text{const}    \,, \\
 \Big[ 2 \mathcal{H}^{\prime} + \mathcal{H}^2 \Big]^{\text{sol 1,2}}
   &\approx &  \pm 3 \mathcal{H}^2  \times \cos \Big( 2 \int d \eta \,  m a \Big)  \,,  \\
    \Big[   \mathcal{H}^{\prime \prime} +  \mathcal{H} \mathcal{H}^{\prime} \Big]^{\text{sol 1,2}}
   &\approx &  \mp 3 m a \mathcal{H}^2   \sin \Big( 2 \int d \eta \,  m a \Big)  \,.
\eea
We than choose a linear combination of these solutions which is not oscillatory and can only be provided by means of the connected part of $F_{00}^{\text{hom}}$. It is to leading order simply given by
\be
F_{00}^{\text{hom, non-osc}} = \frac{1}{2} \big[F_{00}^{\text{hom, sol 1}}+ F_{00}^{\text{hom, sol 2}} \big]\approx a^{-3} \,, \quad F_{11}^{\text{hom, non-osc}} \approx \frac{m^2}{\hbar^2} a^{3} \,.
\ee
The, corresponding solutions for the Hubble rate are given by the following leading order terms
\bea
 \Big[ \mathcal{H}^2 a \Big]^{\text{non-osc}} & \approx & \text{const}    \,, \\
 \Big[ 2 \mathcal{H}^{\prime} + \mathcal{H}^2 \Big]^{\text{non-osc}}
   &\approx &  0  \, .
\eea
We of course get this solution by dropping all higher-order time derivatives on $F_{00}^{\text{hom}}$ in the limit $\mathcal{H} \ll \hbar^{-1} ma$.
To summarize, we have shown that -- by means of the connected part -- the two-point function formalism allows  to overcome the oscillatory behavior of time derivatives of the Hubble rate, and equivalently of pressure, without any averaging procedure. It is in this respect significantly richer in comparison to the approach based on classical real scalar fields.

\section{Generalized On-Shell Vlasov Equation}
\subsection{Identification of Scalar Field On-Shell Phase-Space Distributions}
We now compare the real scalar field energy-momentum tensor expressed in terms of scalar field two-point functions (see \eqref{T00PureF00} to \eqref{Tij} in appendix \ref{appEinstein} ) with a general energy-momentum tensor in kinetic theory. This will allow us to identify phase-space distributions based on the scalar field. The phase-space distribution $f_{\text{cl}}$ of classical collisionless particles in general relativity obeys the Vlasov equation \cite{Debbasch20091079}
\be
\Bigg[\frac{\partial}{\partial \eta}  + \frac{{p}^{i}_{cl}}{{p}^0_{cl}} \frac{\partial}{\partial X^{i}}  + \Gamma^{\alpha}_{\, i \beta } \frac{p_{\alpha}^{cl} {p}^{\beta}_{cl}}{{p}^0_{cl}} \frac{\partial}{\partial {p}^{i}_{cl}} \Bigg] f_{\text{cl}}(\eta, X^j, p_k^{cl}) =0. \label{VlasovClassic}
\ee
The energy-momentum tensor in kinetic theory is then given by 
\be
T_{\mu \nu}^{\text{kin}} (\eta, X^i) =   \int { d^{D-1}p^{cl}} \Big[\gamma^{-1/2} \frac{p_{\mu}^{cl} p_{\nu}^{cl}}{E_{cl} } \Big] (\eta, X^i, p_i^{cl}) f_{\text{cl}}(\eta, X^i, p_i^{cl}) \, . \label{kinEnergyMom}
\ee
Here, the quantity $\gamma$ is the determinant of the spatial metric.
The particle energy $E_{cl} $ and the temporal momentum ${p}_0^{cl}$ are related to the on-shell condition 
\be
p_{\mu}^{cl} p^{\mu}_{cl} = - m_{cl}^2\, , 
\ee
which gives in longitudinal gauge
\bea
 p_{0}^{cl}  &=& g_{00} \sqrt{ m_{cl}^2 + g^{ij} p_{i}^{cl} p_{j}^{cl}}\, , \quad g_{0i} =0\, ,\\
E_{cl} &=&  - |g^{00}|^{1/2} p_{0}^{cl} \, .
\eea
Of course we want to identify similar quantities through the covariant Wigner momenta $p_i$. By using a tilde from now on, we want to clearly distinguish between the covariant Wigner momentum $p_i$, which is an integration variable and derived quantities that are related to it via the metric,
\bea
E(\eta,  X^i, p_i) &=&   \Big[m^2  + \frac{p^2}{a^2(\eta)} \Big]^{1/2} \Big[1 +\frac{p^2 }{m^2 a^2(\eta) + p^2 }\Psi_G(\eta,  X^i)   \Big]\, , \\
\widetilde{p}^0(\eta,  X^i, p_i) &=&  a^{-1}(\eta) \big[ 1-\Phi_G(\eta,  X^i) \big]E(\eta,  X^i, p_i) \, ,\\
\widetilde{p}_0(\eta,  X^i, p_i)  &=&- a^2 (\eta) \big[ 1 + 2\Phi_G(\eta,  X^i) \big]  \widetilde{p}^0 (\eta,  X^i, p_i) \, ,\\
\widetilde{p}^k (\eta,  X^i, p_i) &=& a^{-2}(\eta)\big[ 1+2\Psi_G(\eta,  X^i) \big] \delta^{ki} p_i \, .
\eea
In particular, we emphasize that there is no independent integration variable $p_0$ which is encountered in off-shell Wigner transformations, we have only the on-shell quantity $\widetilde{p}_0(p_i)$.
As we derive in appendix \ref{appEinstein}, the $00$- and $0i$-components of the real scalar field energy-momentum tensor are to leading order in $\hbar$ given by (see equations \eqref{T00PureF00} and \eqref{T0iPureF-})
\begin{multline}
\vac{\hat{T}_{00}}(\eta, X^i) = T_{00}(\eta, X^i) =  \Bigg[\frac{m^2 a^2}{\hbar^2} \Big[1+2 \Phi_G(\eta, X^i) \Big] \overline{F}_{00}  (\eta, X^i) 
\\ +\Big[1+2 \Phi_G(\eta, X^i) +2 \Psi_G(\eta, X^i)  \Big] \int \frac{ d^{D-1}p}{(2\pi \hbar)^{D-1}} \frac{p^2}{\hbar^2}  F_{00}(\eta, X^i,p_i) \Bigg]  \times \Bigg[ 1+\mathcal{O} \big(\varepsilon_{\hbar}^2 \big)  \Bigg] \,, \label{T00leadh}
\end{multline}
and
\begin{multline}
\vac{\hat{T}_{0i}}(\eta, X^i)=T_{0i}(\eta, X^i)= - a^{-(D-2)}    \Big[1 + \Phi_G(\eta, X^i) +  (D-1) \Psi_G(\eta, X^i) \Big]\\ \times  \int \frac{ d^{D-1}p}{(2\pi \hbar)^{D-1}} \frac{p_i}{\hbar} F_- (\eta, X^i,p_i)    \times \Bigg[ 1+\mathcal{O} \big(\varepsilon_{\hbar}^2  \big) \Bigg] \, \label{T0ileadh}.
\end{multline}
We realize that according to equation \eqref{T00leadh} a phase-space density candidate in the classical particle limit would be given by
\begin{multline}
f_{\text{cl}}^{\text{even}}  \longrightarrow   f^{\text{even}}_{\phi}  := \frac{E \gamma^{1/2}}{(2\pi\hbar)^{D-1}} \frac{ F_{00} }{\hbar^2}  
\\= \frac{(m^2 a^2 + p^2 )^{1/2}}{(2\pi\hbar)^{D-1}} \Big[1-(D-1) \Psi_G  +\frac{p^2}{m^2 a^2 + p^2 } \Psi_G \Big] \frac{  a^{D-2} F_{00} }{\hbar^2} . \label{feven}
\end{multline}
However, when looking at the $T_{0i}$ equation \eqref{T0ileadh}, we would rather come to the conclusion that the classical phase-space-density should be given by
\be
f_{\text{cl}}^{\text{odd}} \; \longrightarrow \;  f^{\text{odd}}_{\phi} := \frac{1}{(2\pi\hbar)^{D-1}} \frac{F_-}{\hbar}  \,. \label{fodd}
\ee
In equations \eqref{F00IsEven} and \eqref{FMinusIsOdd} in appendix \ref{appWignerDyn} we summarize that according to their fundamental definitions the two-point function $F_{00}$ is of even parity in $p_i$ whereas the two-point function $F_{-}$ is an odd parity function in $p_i$. Thus, up to a rescaling and corrections in our perturbation parameters, the quantity $F_{00}$ seems to play the role of the phase-space-density for even moments and $F_-$ seems to play the role of the phase-space-density for odd moments in $p_i$.
\subsection{Dynamics of On-Shell Phase-Space Distributions}
Based on the identifications in the previous paragraph, we rewrite \eqref{FMinDynhleader} and \eqref{F00Dynhleader} in terms of the definitions \eqref{feven} and \eqref{fodd}. We find
\begin{multline}
 \Bigg[ \widetilde{p}^0 \frac{\partial }{\partial \eta} f^{\text{odd}}_{\phi} + \widetilde{p}^k \frac{\partial}{\partial X^k}f^{\text{even}}_{\phi} 
 - \widetilde{p}^i p_i   \frac{\partial}{\partial X^k} \Big[ \Phi_G +  \Psi_G \Big] \frac{\partial}{\partial p_k} f^{\text{even}}_{\phi} 
  -  { m^2   } \frac{\partial}{\partial X^k} \Phi_G   \frac{\partial}{\partial p_k}f^{\text{even}}_{\phi}\Bigg]  \\ \times \Bigg[ 1+\mathcal{O} \big(\varepsilon_{\hbar}^2 \cdot \varepsilon_{\text{g}}^2 \big)+ \mathcal{O} \big(\varepsilon_{\hbar}^4 \big) \Bigg]=0  \, ,
\end{multline}
\begin{multline}
  \Bigg[   \widetilde{p}^0 \frac{\partial }{\partial \eta} f^{\text{even}}_{\phi}  + \widetilde{p}^k \frac{\partial}{\partial X^k}  f^{\text{odd}}_{\phi}
-  \widetilde{p}^i p_i  \frac{\partial}{\partial X^k}   \Big[ \Phi_G + \Psi_G \Big] \frac{\partial}{\partial p_k}    f^{\text{odd}}_{\phi}
- { m^2}  \frac{\partial}{\partial X^k} \Phi_G    \frac{\partial}{\partial p_k}  f^{\text{odd}}_{\phi}\\
+ \frac{\hbar^2}{4 } a^{-2(D-1)}  \Bigg[a^{D-2} \frac{\partial}{\partial \eta} \Bigg]^3 \Bigg[a^{-(D-2)} (m^2 a^2 + p^2 )^{-1/2} f^{\text{even}}_{\phi}\Bigg]  \\-  \hbar^2 a^{-2}  \frac{\Delta_X}{4}   \frac{\partial}{\partial \eta}\Big[(m^2 a^2 + p^2 )^{-1/2} f^{\text{even}}_{\phi}\Big]\Bigg]  \times \Bigg[ 1+\mathcal{O} \big(\varepsilon_{\hbar}^2 \cdot \varepsilon_{\text{g}}^2 \big)+ \mathcal{O} \big(\varepsilon_{\hbar}^4 \big) \Bigg] = 0 \, .
\end{multline}
We emphasize again that $\widetilde{p}^0$ is on-shell.
We set\footnote{Note, that we could also have considered
\be
f_{\phi}^{\text{time-rev}} :=  f^{\text{even}}_{\phi} - f^{\text{odd}}_{\phi}\, , 
\ee 
which yields phase-space dynamics for reversed momenta or equally for reversed times.
However, the equation for the time-reversed density amounts only to a flip of the sign of the momentum $p_i$ and thus yields no new information. }
\be
f_{\phi} := f^{\text{even}}_{\phi} + f^{\text{odd}}_{\phi} =  (2 \pi \hbar)^{-(D-1)}\Big[{E \gamma^{1/2}}\frac{ F_{00} }{\hbar^2}+ \frac{F_-}{\hbar}\Big] \, ,\label{psdens}
\ee
and find
\begin{multline}
  \Bigg[   \Big[  \widetilde{p}^0\frac{\partial }{\partial \eta}    + \widetilde{p}^k \frac{\partial}{\partial X^k}  
 -  \widetilde{p}^i p_i  \frac{\partial}{\partial X^k}   \big[ \Phi_G + \Psi_G \big] \frac{\partial}{\partial p_k}    
- { m^2}  \frac{\partial}{\partial X^k} \Phi_G    \frac{\partial}{\partial p_k} \Big] f_{\phi} \\+ \frac{\hbar^2}{4 } a^{-2(D-1)}  \Bigg[a^{D-2} \frac{\partial}{\partial \eta} \Bigg]^3 \Bigg[a^{-(D-2)} (m^2 a^2 + p^2 )^{-1/2} f^{\text{even}}_{\phi}\Bigg]  \\- \hbar^2 a^{-2}  \frac{\Delta_X}{4}   \frac{\partial}{\partial \eta}\Big[(m^2 a^2 + p^2 )^{-1/2} f^{\text{even}}_{\phi}\Big] \Bigg] \times \Bigg[ 1+\mathcal{O} \big(\varepsilon_{\hbar}^2 \cdot \varepsilon_{\text{g}}^2 \big)+ \mathcal{O} \big(\varepsilon_{\hbar}^4 \big) \Bigg] =0 \, . \label{finalVlasov}
\end{multline}
Equation \eqref{finalVlasov} is the main result of this paper. It tells us that we can obtain a corrected on-shell Vlasov equation from statistical two-point functions of a scalar field theory. It includes a third-order time derivative such that we keep all degrees of freedom from the initial first-order system \eqref{firstDerF00} to \eqref{firstDer11} originating from the relativistic Klein-Gordon equation. However, dropping these third-order time derivatives as small corrections we are left with the degrees of freedom of a one-particle phase-space distribution. We note, that dropping these third-order time derivatives can be justified by either using certain initial conditions in the case that  $F_{00}$ is given only in terms of the connected two-point functions (concretely $F_{+} \sim \mathcal{H} F_{00}$). In the case where $F_{00}$ is only given by a product of oscillatory one-point functions, we can drop those contributions only after an averaging procedure. Apart from the third-order time derivatives the generalized Vlasov equation \eqref{finalVlasov} contains corrections in the gradient expansion. Dropping the later, we find the same form as for the classical, collisionless  on-shell Vlasov equation given in \eqref{VlasovClassic}. We note that since we used a spin zero field we expect corrections due to non-trivial spin in other quantum field theoretical settings. We also note that the degrees of freedom of the system containing all time derivatives might be constrained by conserved quantities. We remark that conserved quantities have been identified in homogeneous settings \cite{Garbrecht:2002pd} and we plan to investigate this  issue for our non-homogeneous case in the future.
\par A few more comments are in order. We acknowledge that an off-shell version of this equation has been derived much earlier by \cite{Winter:1986da} and \cite{Calzetta:1987bw} for arbitrary metrices. As we pointed out above and explicitely derived here, the on-shell Vlasov cannot be derived from a single on-shell two-point function which came apparent from the derivation of real scalar field particle densities by \cite{Garbrecht:2002pd} in homogeneous FLRW-space-time and which was pointed out earlier by \cite{PhysRevD.57.6525} for a QED Vlasov equation in Minkowski space-time. However, we remark that integrating out off-shell energies implies a renormalization procedure that is avoided by using a pure on-shell formulation up to this point. An equation similar to \eqref{finalVlasov} may be derived by employing a quasi-particle approximation that is resticted to positive off-shell energies which are then integrated over \cite{Prokopec:2003pj} \cite{Prokopec:2004ic}. However, this quasi-particle picture does not need to hold and it is a priori not settled why negative off-shell energies should not contribute. Thus, as far as we know, this is the first explicit derivation of the on-shell Vlasov equation for a non-homogeneous metric from a fundamental theory that is not employing a quasi-particle approximation i.e. that does not restrict excitation to be lumped in region of positive off-shell energies. We also acknowledge that equation \eqref{finalVlasov} has been derived on phenomenological grounds from a Schr\"odinger equation based on one-point functions as discussed for example in \cite{Uhlemann:2014npa}, however without any coarse-graining this approach does not allow for independent moments in momentum space since they will be related by spatial derivatives and thus, this one-point function approach cannot model a generic non-perfect fluid with gravitational interactions without any coarse-graining. 
\par
Let us comment a bit more on the physical meaning of this real scalar field phase-space  distribution  $f_{\phi}$ and why we think that it can model a general fluid with gravitational interactions. First, we note that it is no surprise that the definition \eqref{psdens} yields a generalization of the Vlasov equation since it is related to the Wigner transformation of the Schr\"odinger operators $\hat{\psi}$ in the non-relativistic limit. We see this by writing
\be
\hat{\phi}(\eta , x^i) \sim  \hat{\psi} (\eta , x^i) \exp \Big[ {-i \frac{m}{\hbar} \int^{\eta} d \widetilde{\eta} \,  a(\widetilde{\eta})} \Big] +  \hat{\psi}^{\dagger} (\eta , x^i)  \exp \Big[{+i \frac{m}{\hbar} \int^{\eta}  d \widetilde{\eta} \, a(\widetilde{\eta})} \Big] \, , 
\ee 
and after some manipulations we find
\begin{multline}
f_{\phi}(\eta, X^i, p_i)  \sim a^{D-1} \int d^{D-1}r \, e^{-\frac{i}{\hbar} r^i p_i} \vac{\hat{\psi} (\eta, X^k + r^k/2)\hat{\psi}^{\dagger} (\eta, X^k - r^k/2)}\\\times \Bigg[1+ \mathcal{O}\Big( \varepsilon_{g}^2 \Big)+\mathcal{O}\Big( \varepsilon_{ \hbar}^2 \Big)  + \text{oscillatory terms} \Bigg]\, .
\end{multline}
We already remarked in section \ref{phasefromWigner} that the oscillatory terms can be removed by an appropriate initial density matrix.
Thus, we rediscover in the large mass limit the definition of the non-relativistic Wigner quasi-probability distribution based on an initial density matrix \cite{Hillery:1983ms}. This is another strong hint that our on-shell approach to construct a phase-space distribution $f_{\phi}$ has a classical interpretation provided the classically condition and the gradient expansion we discussed in \ref{whyQuantum} apply. This implies a suitable choice for the initial density matrix, in other words the state it represents has to be classical enough.
We can also formulate the approximate equivalence between a classical one-particle phase-space model and the classical limit of the real scalar quantum field theory in the following way: since the dynamics for the scalar field phase-space distribution $f_{\phi}$ and the classical distribution $f_{cl}$ agree on length and time-scales that are associated to the classical limit, the difference between the two quantities is encoded in the possibility to formulate arbitrary initial conditions. 
For any smooth classical phase-space distribution we can write
\begin{multline}
f_{cl} (\eta_{\text{ini}}, X^i, p_i ) =  \int d^{D-1}r \, e^{i r^i p_i} {f}_{cl} (\eta_{\text{ini}}, X^k,  r^k ) \\ =  \int d^{D-1}(x - y) \, e^{i (x - y)^i p_i} \widetilde{f}_{cl} (\eta_{\text{ini}}, x^k, y^k )\, ,
\end{multline}
which in particular means that any moment can be written as 
\begin{multline}
f^{(k_1, ..., k_n)}_{cl} (\eta_{\text{ini}}, X^i) = \int \frac{ d^{D-1}p}{(2\pi \hbar)^{D-1}} p_{k_1} ...  p_{k_n} f_{cl} (\eta_{\text{ini}}, X^i, p_i ) \\ =  \frac{\partial^{n}}{\partial (x-y)^{k_1} ... \partial (x-y)^{k_n}}\widetilde{f}_{cl} (\eta_{\text{ini}}, x^i, y^i )\Big|_{(x-y)^i=0} \, .
\end{multline}
From this we conclude that an arbitrary smooth, classical, one-particle phase-space distribution is initially specified by an arbitrary function in two spatial coordinates.
However, we can always provide such a function based on a Gaussian initial density matrix for the quantum theory which is encoded in the connected part of the two-point function,
\be
\langle{\hat{\psi} (\eta_{\text{ini}}, x^k)\hat{\psi}^* (\eta_{\text{ini}}, y^k)} \rangle = \langle{\hat{\psi}(\eta_{\text{ini}}, x^k)} \rangle  \langle{\hat{\psi}^* (\eta_{\text{ini}}, y^k) } \rangle + \langle{\hat{\psi} (\eta_{\text{ini}}, x^k)\hat{\psi}^* (\eta_{\text{ini}}, y^k)} \rangle _{\text{connected}}\, .
\ee
Note that the product of two one-point functions is not general enough to cover an arbitrary function of two arguments, so we really need the connected term.
We can provide similar arguments for the fully relativistic scalar field theory by splitting the classical distribution into even and odd parts whose arbitrary initial conditions can always be specified by providing the initial connected parts of the two-point functions $F_{00}$ and $F_-$ as well as $F_{+}$ and $F_{11}$ to fix oscillatory behavior. \footnote{It is worth pointing out that the formalism in which one uses a single particle wave function resulting from the
evolution of a certain class of initial states may possess no classical limit in the sense that 
the phase-space distribution (Wigner) function can exhibit rapid (space and/or time) oscillations in the limit when 
$\hbar \rightarrow 0$, thus invalidating the gradient expansion. 
Since the  Schr\"odinger equation is the non-relativistic limit of the Klein-Gordon equation this argument could in principle also apply to our scalar field phase-space distribution $f_{\phi}$. However, the question whether a spatial gradient expansion does apply or not is tied to the specification of the initial density matrix: it ought to be such 
that it yields two-point functions that satisfy the classicality criteria spelled out in section \ref{whyQuantum}.}
 
Finally, let us go to a coarser approximation of equation \eqref{finalVlasov} by pursuing the large mass limit in order to clearly see the relation to the cold dark matter particle picture,
\begin{multline}
\Bigg[\frac{\partial}{\partial \eta} +  \frac{p_i}{ m a(\eta) } \frac{\partial}{\partial X^{i}}  -m a(\eta) \frac{\partial}{\partial X^i} \Phi_G (\eta, X^i )  \frac{\partial}{\partial p_i} \Bigg] f_{\phi}(\eta, X^i ,p_i ) \\ \times \Bigg[1+ \mathcal{O}\Big( \varepsilon_{g}^2 \Big)+\mathcal{O}\Big( \varepsilon_{ \hbar}^2 \Big) + \mathcal{O}\Big( \varepsilon_{\text{p}}^2 \Big) \Bigg] =0 \, . \label{VlasovBigM}
\end{multline}
The last equation is used for a collisionless gas in the context of cold dark matter as for example in \cite{Bernardeau:2001qr} or \cite{Bertschinger:1993xt} and is simply the non-relativistic limit of \eqref{VlasovClassic},
\be
\Bigg[\frac{\partial}{\partial \eta}  + \frac{{p}_{i}^{cl}}{m_{cl} a(\eta)} \frac{\partial}{\partial X^{i}} -m_{cl} a(\eta) \frac{\partial}{\partial X^i} \Phi_G (\eta, X^i )  \frac{\partial}{\partial {p}_{i}^{cl}} \Bigg] f_{\text{cl}}(\eta, X^j, p_k^{cl}) =0. \label{VlasovClassicNr}
\ee
\par
We thus conclude that collisionless dark matter obeying a smooth phase-space distribution can always be mimicked by a real scalar field theory based on scales where the mass dominates. Initial non-trivial moments of the phase-space distribution can be provided by non-trivial initial density matrices for the connected parts of the two-point functions. Taking moments of equation \eqref{VlasovBigM} shows that this generates in principle an infinite hierarchy of moments. 

Apart from the evolution of the phase-space distribution we also have to specify the Einstein equations that determine the gravitational potentials. The Einstein equations take a particularly convenient form if we write them in terms of hydrodynamic quantities. To identify those is the goal of the next section.
\pagebreak
\section{Hydrodynamics Based on Two-Point Functions}
\subsection{Perfect or Non-Perfect Fluid? \label{perfOrNot}}
A perfect fluid description is suitable for early (linear) evolution of cold dark matter. However, it stops being correct in the non-linear regime, in which gravitational slip, vorticity and anisotropic stresses (as well as the corresponding gravitational perturbations) get generated even if they were not present initially.
The question we want to address in this subsection is whether those non-perfect fluid components can generically be modeled with the scalar field two-point function approach we present in this paper.
All two-point functions used to generate different components of the energy-momentum tensor discussed in this section are assumed to arise from 
suitably smeared (classical) 
two-point functions (a more detailed discussion on this important point can be found in section~\ref{whyQuantum}). 
In the above section we worked out phase-space distributions based on real scalar field two-point functions and argued that a non-trivial initial density matrix can give rise to a non-trivial hierarchy of moments.
We now work out how this is reflected on the level of hydrodynamic quantities. 
In particular we argue that although the energy-momentum tensor of the real scalar field theory has the \textit{apparent form} of the energy-momentum tensor of a perfect fluid it \textit{does not} correspond to one\footnote{
It is well known (see for example \cite{Weinberg:1102255}), that the energy-momentum tensor of a classical real scalar field theory has the form of a perfect fluid
\be
T_{\mu \nu}^{\text{cl}} = (e_{\text{cl}}+P_{\text{cl}}) u_{\mu}^{\text{cl}} u_{\nu}^{\text{cl}}  + g_{\mu \nu} P_{\text{cl}}\, , \label{EMTensorClass}
\ee
 with the following classical energy density $e_{\textsc{pf}}$, pressure $P_{\textsc{pf}}$ and four-velocity $ u_{\mu}^{\textsc{pf}}$,
\bea
e_{\text{cl}} &=& -\frac{1}{2} g^{\mu \nu} \partial_{\mu} \phi_{\text{cl}} \partial_{\nu} \phi_{\text{cl}} + \frac{1}{2}m^2 \phi_{\text{cl}}^2 \, , \\  \label{ePF}
P_{\text{cl}} &=& -\frac{1}{2} g^{\mu \nu}\partial_{\mu} \phi_{\text{cl}}\partial_{\nu} \phi_{\text{cl}} - \frac{1}{2}m^2 \phi_{\text{cl}}^2 \, ,\\\label{PPF}
 u^{\mu}_{\text{cl}} &=& - \Big[e_{\text{cl}} +P_{\text{cl}} \Big]^{-1/2} g^{\mu \nu} \partial_{\nu} \phi_{\text{cl}}\, .\label{uPF}
\eea
This classical viewpoint, based on one-point functions,  has been exploited linearly (see e.g. \cite{Hwang:2009js} or \cite{Nambu:1996gf}) in different gauges and non-lineary for real and complex classical fields  in the longitudinal gauge \cite{Marsh:2015daa}\cite{Suarez:2015fga}.}
\begin{multline}
 \vac{\partial_{\mu} \hat{\phi} \partial_{\nu} \hat{\phi}} - \frac{g_{\mu \nu}}{2} \Big[  \vac{\partial^{\alpha} \hat{\phi} \partial_{\alpha} \hat{\phi}} + \frac{m^2}{ \hbar^2} \vac{\hat{\phi}^2} \ \Big] \\ = \vac{T_{\mu \nu}} \neq T_{\mu \nu}^{\textsc{pf}} = \partial_{\mu} {\phi}_{\text{cl}} \partial_{\nu} {\phi}_{\text{cl}} - \frac{g_{\mu \nu}}{2} \Big[  \partial^{\alpha} {\phi}_{\text{cl}} \partial_{\alpha} {\phi}_{\text{cl}} + \frac{m^2}{ \hbar^2} {\phi}_{\text{cl}}^2 \ \Big] \, ,
\end{multline}
where 
\be
\phi_{\text{cl}} := \vac{\hat{\phi}}\, .
\ee
Thus, we will in general \textit{not} speak of a perfect fluid in the two-point function approach.
In our scalar field model the fundamental reason is that the statistical two-point functions split into a reducible and an irreducible or connected piece
\be
\vac{ \hat{\phi}(x)  \hat{\phi}(y)} = \phi_{\text{cl}} (x) \phi_{\text{cl}} (y) + \vac{ \hat{\phi} (x) \hat{\phi} (y) }_{\text{connected}} \, , \label{redandIrr}
\ee
where the connected piece can initially be an arbitrary function of $x^i$ and $y^i$ in position space or equivalently of $X^i$ and $p_i$ in Wigner space. The reducible piece is given by the product of one-point functions. 
The question whether the one-point function or the connected two-point function or both contribute to the whole two-point function 
depends on the dark matter production mechanism which is model dependent. Here we focus our attention mainly on the connected piece when discussing statistical two-point functions of the scalar matter field because
the scalar field in our model~(\ref{sMatter}) does not couple linearly to external sources, such that one-point functions 
are generally absent (unless they are imposed as the initial condition as it is e.g. done for the inflaton). 
 

In order to make contact to a hydrodynamic description, we define the following energy density $e$ and apparent pressure $P$ analogously to the one-point function approach,
\bea
e &:=& -\frac{1}{2} g^{\mu \nu} \vac{\partial_{\mu} \hat{\phi} \partial_{\nu} \hat{\phi}} + \frac{1}{2}m^2 \vac{\hat{\phi}^2} \, , \\ 
P &:=& -\frac{1}{2} g^{\mu \nu} \vac{\partial_{\mu} \hat{\phi} \partial_{\nu} \hat{\phi}} - \frac{1}{2}m^2 \vac{\hat{\phi}^2} \, , \label{apparentPressure}
\eea
as well as a composite quantity containing a notion of four-velocity,
\be
\Big[  (e+P) u_{\mu} u_{\nu} \Big]_{\text{com}} := \vac{\partial_{\mu} \hat{\phi} \partial_{\nu} \hat{\phi}}= \partial_{\mu}  \phi_{\text{cl}}\partial_{\nu} \phi_{\text{cl}} + \vac{\partial_{\mu} \hat{\phi} \partial_{\nu} \hat{\phi}}_{\text{connected}}\, .
\ee
We then have
\be
T_{\mu \nu} = \Big[  (e+P) u_{\mu} u_{\nu} \Big]_{\text{com}}  + g_{\mu \nu} P \, . \label{compPFEMT}
\ee
We stress that up to now, we do not have a definition of an irreducible four-velocity, we only have a definition of a composite operator that will contain it\footnote{
We remark that composite fluid quantities can also arise from genuine perfect fluids by introducing a smoothing scale \cite{Baumann:2010tm}. However, this origin is conceptually  different from the connected two-point function approach we are advertising here.}.
We now make the following identification
\bea
\text{0-th moment} & \hat{=} &  -\Big[  (e+P) u^{0} u_{0} \Big]_{\text{com}} \, ,\\ \label{mom0}
 \text{1-st moment} & \hat{=} & \;\;\,\Big[  (e+P) u^{0} u_{i} \Big]_{\text{com}}
\, , \\\label{mom1}
\text{2-nd moment}& \hat{=} & \;\;\,\Big[  (e+P) u^{i} u_{j} \Big]_{\text{com}} \, ,\label{mom2}
\eea
and define a quantity that we will play the role of the stress tensor
\be
\sigma_{ij} := \delta_{ik} \Big[  (e+P) u^{k} u_{j} \Big]_{\text{com}} - (e+P)^{-1} \Big[  (e+P) u^{0} u_{i} \Big]_{\text{com}}\Big[  (e+P) u^{0} u_{j} \Big]_{\text{com}}\, . \label{anisoStress}
\ee
We will see below that these identification are justified based on the continuity and Euler equation.
We find neglecting contributions of the metric
\bea
 \Big[  (e+P) u^{0} u_{0} \Big]_{\text{com}}(X)& \sim& \vac{\hat{\Pi}^2(X)} \, ,\\ 
\Big[  (e+P) u^{0} u_{i} \Big]_{\text{com}}(X)&\sim& \Big[\Big(\frac{1}{2} \partial_i^X  -\partial_i^r \Big) \vac{\hat{\Pi} ({X+r/2}) \hat{\phi}({X-r/2})  }\Big]_{r=0} 
\, , \\
\Big[  (e+P) u^{i} u_{j} \Big]_{\text{com}} (X) &\sim &  \Big[ \Big(\frac{1}{4}\partial_{i}^X \partial_{j}^X-\partial_{i}^r \partial_{j}^r  \Big) \vac{ \hat{\phi}(X+r/2)\hat{\phi} (X-r/2)  } \Big]_{r=0} \, .
\eea
Since the various two-point functions appearing can be independently specified, we conclude that the composite four-velocity objects that we are referring to as moments are independent. One might object that the equations of motion enforce certain two-point functions to be proportional to other ones in the classical limit where higher-time derivatives or spatial derivatives are small. Indeed this is the case as we find in equation \eqref{F11IntermsOfFMinusandF00}. So to lowest order in the gradient expansion in the parameters \eqref{pertParam1} to \eqref{pertParamn} we find
\bea
 \Big[  (e+P) u^{0} u_{0} \Big]_{\text{com}}(X)& \approx & m^2 \vac{\hat{\phi}^2(X)}  \label{indPos1} \, ,\\ 
\Big[  (e+P) u^{0} u_{i} \Big]_{\text{com}}(X)&\approx &  \partial_i^r \Big[ \vac{\hat{\Pi} ({X+r/2}) \hat{\phi}({X-r/2})  } - (r \rightarrow -r )\Big]_{r=0}   
\, , \\
\Big[  (e+P) u^{i} u_{j} \Big]_{\text{com}} (X) &\approx &   \partial_{i}^r \partial_{j}^r  \Big[\vac{ \hat{\phi}(X+r/2)\hat{\phi} (X-r/2)  }\Big]_{r=0} \, .\label{indPos3}
\eea
or in using the notion of section \ref{phaseFrom2}
\bea
 \Big[  (e+P) u^{0} u_{0} \Big]_{\text{com}}(X)& \approx & m^2 \Big[ F_{00} (X ,r) \Big]_{r=0}  \, ,\\ 
\Big[  (e+P) u^{0} u_{i} \Big]_{\text{com}}(X)&\approx &  \partial_i^r \Big[F_- (X ,r\Big]_{r=0}   
\, , \\
\Big[  (e+P) u^{i} u_{j} \Big]_{\text{com}} (X) &\approx &   \partial_{i}^r \partial_{j}^r  \Big[F_{00} (X,r)\Big]_{r=0} \, .
\eea
Thus, in the classical limit even moments in Wigner momentum space will be related to even numbers of $r^i$-derivatives of the scalar field two-point function $F_{00}(X^i, r^j)$ evaluated at zero whereas for odd moments the same is true for odd numbers of $r^i$-derivatives of the function $F_- (X^i, r^i)$, the two-point functions we already identified as phase-space densities on the previous section. The zeroth and the second moment are the first two non-vanishing Taylor coefficients of the arbitrary function $F_{00}(X^i, r^j)$ and are thus independent. The function $F_{00}(X^i, r^j)$ is arbitrary since its connected part can be freely specified by the initial density matrix.
In other words, the stress tensor $\sigma_{ij}$ is completely generic and not only related to spatial $X^i$-derivatives of the zeroth moments as it would be the case for  one-point functions or classical real scalar fields. In particular, it may contain an addition contribution to the pressure on top of the apparent pressure $P$ we defined in \eqref{apparentPressure}. But not only this, we realize that the first moment shows that we can specify a non-vanishing vorticity even at early times in linear perturbation theory since we have
\be
\overline{\rho}  \epsilon^{ijk} \frac{\partial}{\partial X^j }  \delta v_k (X) \approx \epsilon^{ijk} \frac{\partial}{\partial X^j }\Big[ \frac{\partial}{\partial r^k } F_-(X, r) \Big]_{r=0} \neq 0\, .
\ee
\par
In \eqref{mom0} to \eqref{mom2} we labeled components of the composite energy-momentum tensor as moments such that we are interpreting them as rest-mass-density, momentum-densities and stress tensor of particles in a certain non-relativistic limit. One might worry, that this fluid picture does not hold since we do not have a conserved rest-mass density. However, as we show in appendix \ref{appGenCompPerfFluid}, local energy- and momentum-conservation reduce to the familiar continuity and Euler equation in an expanding universe in the limit of small pressure, weak gravitational fields and small velocity. This holds for every energy-momentum tensor of the form \eqref{compPFEMT}. In particular it holds for the real scalar field energy-momentum tensor. Identifying energy density, apparent pressure and composite velocities in terms of the Wigner transformed two-point functions will allow us to reformulate the aforementioned non-relativistic limits clearly as a large mass limit. 
We find
\begin{multline}
e = \frac{m^2}{\hbar^2}   \overline{F}_{00} - a^{-2} \frac{\partial_X}{4}\Big[ \Phi_G -(D-3) \Psi_G \Big] \partial_X  \overline{F}_{00} 
-{a^{-2}}\Big[1 + 2 \Psi_G \Big] \frac{\Delta_X}{4} \overline{F}_{00} \\ + \frac{1}{2} a^{-D} \Big[1-\Phi_G +(D-1)\Psi_G  \Big]\Bigg[\frac{a^{D-2}}{2} \Big(1-\Phi_G -(D-1)\Psi_G \Big) \overline{F}_{00} ^{\prime} \Bigg]^{\prime} 
 \label{energyDens}  \, .
\end{multline}
For the apparent pressure $P$ we get then 
\begin{multline}
P  =\frac{1}{2} a^{-D} \Big[1-\Phi_G +(D-1)\Psi_G  \Big]\Bigg[\frac{a^{D-2}}{2} \Big(1-\Phi_G -(D-1)\Psi_G \Big) \overline{F}_{00}^{\prime} \Bigg]^{\prime} \\
-a^{-2} \frac{\partial_X}{4}\Big[ \Phi_G -(D-3) \Psi_G \Big] \partial_X  \overline{F}_{00}
-{a^{-2}}\Big[1 + 2 \Psi_G \Big]\frac{\Delta_X}{4}  \overline{F}_{00}   \, , \label{pressure}
\end{multline}
and we realize that the choice of our perturbation parameters in \eqref{pertParam1} to \eqref{pertParamn} amounts to having a small apparent pressure. We note that the apparent pressure is due to the fundamental field theory we started with, i.e. it is build out of wave- rather than particle phenomena. 
We identify $m^2 F_{00} $ as the rest-mass density in the particle picture and  remark that it is consistent with the lowest-order expression we get in the phase-space language,
\begin{multline}
\rho (\eta ,X^i):= m  \Big[  \frac{m}{\hbar^2} \overline{F}_{00}(\eta ,X^i) \Big] =\frac{m^2}{\hbar^2}  \vac{ \hat{\phi}(\eta , X^i)\hat{\phi}(\eta, X^i)} 
\\= m  \int { d^{D-1}p}   f^{\text{even}}_{\phi}(\eta, X^i,p_i) \gamma^{-1/2}(\eta , X^i) \Bigg[ 1+ \mathcal{O} \big( \varepsilon_{\text{p}}^2\big) \Bigg] \, .\label{restMassDef}
\end{multline}
With this definition we have
\be
 e = \rho + P \, ,
\ee
as well as
\begin{multline}
P =\frac{\hbar^2}{2m^2} a^{-D} \Big[1-\Phi_G +(D-1)\Psi_G  \Big]\Bigg[\frac{a^{D-2}}{2} \Big(1-\Phi_G -(D-1)\Psi_G \Big) \rho^{\prime} \Bigg]^{\prime} \\
- \frac{\hbar^2}{4m^2 a^2} \frac{\partial}{\partial X^k}\Big[ \Phi_G -(D-3) \Psi_G \Big] \frac{\partial}{\partial X^k} \rho
-\frac{\hbar^2}{4m^2a^2}\Big[1 + 2 \Psi_G \Big]{\Delta_X} \rho  \, .
\end{multline}
The composite fluid-four velocity quantities evaluate to
\begin{multline}
\Big[ (e+P) u_0 u_i\Big]_{\text{com}} = -a^{-(D-2)}\Big[ 1+ \Phi_G + (D-1) \Psi_G \Big] \int \frac{ d^{D-1}p}{(2\pi \hbar)^{D-1}} \frac{p_i}{\hbar} F_-  \\ + \frac{\hbar^2}{4 m^2} \Big[ 1+ \Phi_G + (D-1) \Psi_G \Big] \frac{\partial}{\partial X^i} \Bigg[ \Big[1- \Phi_G - (D-1) \Psi_G \Big] \rho^{\prime} \Bigg] \, , \label{u0uicomp}
\end{multline}
as well as
\be
\Big[  (e+P) u_i u_j\Big]_{\text{com}} = T_{ij}- g_{ij}P  
 =\frac{\hbar^2}{4 m^2 }\frac{\partial^2 }{\partial X^i \partial X^j} \rho +  \int \frac{ d^{D-1}p}{(2\pi \hbar)^{D-1}}  \frac{p_i p_j}{\hbar^2}  F_{00} \, .
\ee
Note, that we consistently find
\be
\Big[  (e+P) u^0 u_0\Big]_{\text{com}} = T^{0}_{\; 0} - P = -\Bigg[ e+P + \Big[  (e+P) u^{i} u_{i}  \Big]_{\text{com}}\Bigg]\,.
\ee
Late us now return to the scheme we presented \eqref{mom0} to \eqref{mom2} by applying the definition of the scalar field phase-space distribution we worked out in \eqref{psdens}. 
We find in the large mass limit
\bea
- \Big[  (e+P) u^{0} u_{0} \Big]_{\text{com}} & \hat{=} & \text{0-th moment} \, \approx \, m \int { d^{D-1}p}   f_{\phi} \gamma^{-1/2}\, \approx \, \rho\, ,\\ 
\Big[  (e+P) u^{0} u_{i} \Big]_{\text{com}} & \hat{=} & \text{1-st moment} \, \, \approx \, m \int { d^{D-1}p} \frac{ p_i}{m}  f_{\phi} \gamma^{-1/2}\, , \\
\Big[  (e+P) u_{i} u_{j} \Big]_{\text{com}} & \hat{=} & \text{2-nd moment} \approx\,  m \int { d^{D-1}p} \frac{ p_i p_j }{m^2}  f_{\phi} \gamma^{-1/2}  \, ,
\eea
and thus the stress tensor
\begin{multline}
\sigma_{ij} \approx m \int { d^{D-1}p} \delta_{ik} \frac{ \tilde{p}^k p_j }{m^2}  f_{\phi} \gamma^{-1/2} \\  - \frac{m^2}{\rho}\Bigg[ \int { d^{D-1}p} \frac{ p_i}{m}  f_{\phi} \gamma^{-1/2} \Bigg] \Bigg[ \int { d^{D-1}p} \frac{ p_j}{m}  f_{\phi} \gamma^{-1/2} \Bigg] \, ,
\end{multline}
is an arbitrary quantity since the phase-space distribution $f_{\phi}$ can freely be specified via the initial density matrix. This is just a different way of phrasing the independence of moments in position space as we did in \eqref{indPos1} to \eqref{indPos3}. We underpin again that this is even valid on scales where $ma \gg \partial_X$ where there is no quantum pressure term involved yet.
\subsection{Einstein Equations \label{EinsteinEqFluidVar}}
Having identified hydrodynamic variables in the last section, we would like the express the Einstein equations in terms of these variables. We find
\begin{multline}
G_{00}= \frac{1}{2}(D-1)(D-2) \mathcal{H}^2 +(D-2)\Delta \Psi_G - (D-1)(D-2) \mathcal{H}\Psi_G^{\prime} \\=a^2 \frac{\hbar}{M_P^2} \Big[1+2\Phi_G+2\Psi_G \Big] \Bigg[ e + \Big[  (e+P) u_{i} u_{i}  \Big]_{\text{com}}\Bigg]\, ,
\end{multline}
\be
G_{0i}=(D-2) \frac{\partial}{\partial X^i} \Psi_G^{\prime} +  (D-2) \mathcal{H} \frac{\partial}{\partial X^i}  \Phi_G= \frac{\hbar}{M_P^2}  \Big[ (e+P) u_i u_0 \Big]_{\text{com}} \,, 
\ee
\begin{multline}
G_{ii} = - (D-2) \mathcal{H}^{\prime} -\frac{1}{2}(D-2)(D-3) \mathcal{H}^2  + (D-2) \Psi_G^{\prime \prime} + \frac{D-2}{D-1} \Delta \Big[ \Phi_G  - (D - 3)  \Psi_G \Big]\\
   +  (D-2) \Big[ 2\mathcal{H}^{\prime} + (D-3) \mathcal{H}^2 \Big]  ( \Phi_G +\Psi_G)  +(D-2) \mathcal{H} \Big[ \Phi_G^{\prime} +(D-2)\Psi_G^{\prime} \Big] \\  = a^2
\frac{\hbar}{M_P^2} \Bigg[ P + \frac{1}{D-1} \Big[  (e+P) u^{i} u_{i}  \Big]_{\text{com}}  \Bigg] \,,
\end{multline}
\begin{multline}
 \frac{\Delta_X G_{kk}}{D-1} \delta_{ij}  - G_{ij} =\Big[ \frac{\Delta}{D-1} \delta_{ij} - \partial_i \partial_j \Big] \Big[ \Phi_G - (D-3) \Psi_G \Big]\\ = - \frac{\hbar}{M_P^2}\Bigg[ \frac{\Big[ (e+P) u_{k} u_{k}  \Big]_{\text{com}} }{D-1}  \delta_{ij} -  \Big[  (e+P) u_{i} u_{j}  \Big]_{\text{com}} \Bigg]  \,. 
\end{multline}
In particular the last equation shows that the gravitational slip is sourced by terms non-linear in the fluid velocities and should be taken into account for higher-order corrections. We also remind the reader that the spatial Einstein equations with neglected gravitons can be obtained from the temporal equations via the Bianchi identity and the energy-momentum conservation.
We combine this set of four redundant Einstein equations into two independent ones that express the gravitational potentials in terms of the hydrodynamic fields.
For the Hubble rate we find
\be
 \frac{1}{2}(D-1)(D-2) \mathcal{H}^2 = a^2 \frac{\hbar}{M_P^2} \vac{ e + \delta^{ij}  \Big[  (e+P) u_{i} u_{j}  \Big]_{\text{com}} }_{\Phi_G, \Psi_G}  \approx  a^2 \frac{\hbar}{M_P^2} \vac{\rho}_{\Phi_G, \Psi_G}  \, ,
\ee
where we remark that the expectation value here is taken with respect to the stochastic variables $\Phi_G$ and $\Psi_G$ that are taken to be Gaussian. Together with the approximated dynamical matter equations we recover the usual behavior of the scalar factor of the cold dark matter scenario. The gravitational potentials themselves are determined by
\begin{multline}
\Delta_X \Big[ \Delta_X \Psi_G -(D-1)(D-2) \mathcal{H}^2\Psi_G \Big] \\ = \frac{a^2 }{D-2}\frac{\hbar}{M_P^2} \Bigg[ \Delta_X  e + \delta^{ij} \Delta_X\Big[  (e+P) u_{i} u_{j}  \Big]_{\text{com}}+\mathcal{H}(D-1)\frac{\partial}{\partial X^i}  \Big[ (e+P) u_i u_0 \Big]_{\text{com}}\Bigg]\, ,
\end{multline}
\begin{multline}
\Delta^2_X \Big[ \Phi_G - (D-3) \Psi_G \Big] \\ = a^2 \frac{\hbar}{M_P^2}\Bigg[ \delta^{ij} \frac{\Delta_X}{D-2}\Big[ (e+P) u_{i} u_{j}  \Big]_{\text{com}}  - \frac{D-1}{D-2} \frac{\partial^2}{\partial X^i\partial X^j} \Big[  (e+P) u_{i} u_{j}  \Big]_{\text{com}} \Bigg]  \,. 
\end{multline}
One has to be aware that the hydrodynamic fields, if again expressed in terms of two-point functions, still contain the gravitational potentials but only with spatial derivatives. These terms can however be neglected for a leading order approximation in the large mass limit. We clearly see that the gravitational fields enter our calculation as non-dynamical constraint fields which was of course expected.
Again, using our approximation we see that we recover the Poisson equation on subhorizon scales
\be
 \Delta_X \Psi_G \approx \frac{a^2}{D-2}\frac{\hbar}{M_P^2} \Big[ \rho -  \vac{\rho}_{\Phi_G, \Psi_G} \Big]\,.
\ee
We also see that the gravitational slip is of higher order.
\subsection{Continuity and Euler Equation for Real Scalar Field Fluid }
In this section we want to derive non-linear equations for the real scalar field fluid based on our two-point function approach. These equations are identical to local energy and momentum conservation and can be derived in a much more general setting as we show it in appendix \ref{appGenCompPerfFluid}. However, as a consistency check, we want to explicitly use the dynamical  equations for the two-point functions of the first part of the paper. The computation may be found in appendix \ref{appConEul}.
We obtain two differential equations for composite operators that are exact up to the linearization in the gravitational potentials and that correspond to the Euler and continuity equation 
\begin{multline}
 \partial_{\eta} \Bigg[\Big[ (e+P) u^{0} u_0 \Big]_{\text{com}} +P \Bigg] + \Big[1- \Phi_G +(D-1)  \Psi_G  \Big]\partial_k \Bigg[ \Big[1+ \Phi_G -(D-1)  \Psi_G  \Big]  \Big[ (e+P) u^{k} u_0 \Big]_{\text{com}}\Bigg]\\+ (D-1)\Big[\mathcal{H} - \Psi_G^{\prime}\Big] \Big[ (e+P) u^{0} u_0 \Big]_{\text{com}} - \Big[\mathcal{H} - \Psi_G^{\prime}\Big]  \Big[ (e+P) u^{k} u_k \Big]_{\text{com}} =0 \,, \label{contRelMainText}
\end{multline}
\begin{multline}
\partial_{\eta} \Big[ (e+P) u^{0} u_i \Big]_{\text{com}} + \partial_k \Big[ (e+P) u^{k} u_i \Big]_{\text{com}} 
+\partial_i P \\+ \Big[D\mathcal{H}+\Phi_G^{\prime}-(D-1)\Psi_G^{\prime}  \Big]\Big[ (e+P) u^{0} u_i \Big]_{\text{com}}
+\Big[\partial_k \Phi_G -(D-1)\partial_k \Psi_G \Big] \Big[ (e+P) u^{k} u_i \Big]_{\text{com}} \\
- \partial_i \Phi_G \Big[ (e+P) u^{0} u_0 \Big]_{\text{com}} + \partial_i \Psi_G \Big[ (e+P) u^{k} u_k \Big]_{\text{com}} =0 \, . \label{EulerRelMainText}
\end{multline}
We want to see the non-relativistic limit of these equations. With the definition of the rest-mass density in \eqref{restMassDef} which yields ($P \ll e$)
\be
\rho = e - P \approx  e +  P \approx - \Big[ (e+P) u^{0} u_0 \Big]_{\text{com}} -P\, .
\ee
We also need to define the proper fluid velocity
\be
  v^{i} :=  - \rho^{-1} \Big[1+\Psi_G + \Phi_G \Big] \Big[ (e+P) u^{i}u_0\Big]_{\text{com}} \Bigg[1 +  (e+P)^{-1}\Big[ (e+P) u^{k} u_k \Big]_{\text{com}} \Bigg]^{1/2}\, ,
\ee
and expand the composite term including spatial velocities in terms of the stress tensor as we did in \eqref{anisoStress}. We find approximately
\be
 \Big[ (e+P) u^{i}u_k\Big]_{\text{com}}  \approx \delta^{ij} \Big[ \sigma_{jk}  +  \rho \cdot  v_j \cdot v_k +  \frac{\hbar^2}{4m^2a^2} \partial_j \partial_k \rho \Big]  \, .
\ee
We now have all ingredients to approximate equations \eqref{contRelMainText} to \eqref{EulerRelMainText} as the non-relativistic continuity equation in an FLRW-universe
\be
    \partial_{\eta} \rho + (D-1)\mathcal{H}  \rho + \partial_i \Big[ \rho \cdot v^{i} \Big] \approx0 \,, \label{ContApprox}
\ee
and generalized Euler equation 
\begin{multline}
   \partial_{\eta} \Big[ \rho  \cdot v^i \Big] + D \mathcal{H}\Big[ \rho \cdot v^i \Big]  + \rho \partial_i \Phi_G   
 \\ + \partial_k \Bigg[\delta^{ij}\delta^{km} \sigma_{jm}  +  \rho\cdot v^i \cdot v^k  -  \frac{\hbar^2}{4m^2a^2} \Big[  \delta^{ik} (D-2) \mathcal{H} \partial_{\eta} \rho+\delta^{ik}     \partial_{\eta}^2 \rho \Big]   \Bigg]\approx  0 \, . \label{Eulerli}
\end{multline}
The higher time derivatives are the only terms remaining from the apparent pressure P. We remark that equation \eqref{Eulerli} generizalizes the usual form of the classical non-linear scalar field dark matter equations as stated for example in \cite{Marsh:2015daa} or \cite{Hui:2016ltb} by first, an stress tensor that can be specified in accordance with inflationary predictions at initial times and which is not only given in terms of a quantum pressure\footnote{The quantum pressure term for a pure one-point function approach may be recovered from the stress $\sigma_{ij}$.} and second, a fluid velocity which generically allows for non-vanishing vorticity as was pointed out in section \ref{perfOrNot}. As we pointed out earlier, the perfect fluid description is a suitable one at early times for linear evolution. However, non-linear evolution generates non-vanishing stresses and vorticity although they are negligible small initially. Our model is capable of capturing these contributions that get more and more important at late times, i.e. in the non-linear regime.
\section{Initial Conditions}
We derived the correspondence between the approximated dynamics of the scalar field phase-space distribution $f_{\phi}$ and a classical one-particle phase-space distribution in the cold dark matter scenario. Before exploiting this correspondence in a more detail analysis, one would have to specify initial conditions as well. 
A typical assumption within the non-relativistic regime is that only the first two moments of the phase-distribution are present initially, although higher moments will be generated at late times in the non-linear regime
\be
f_{\text{cl}}^{\text{ini}}(X^k, p_i) \approx \frac{\rho_{\text{cl}}^{\text{ini}}(X^k)  }{m  }\gamma^{1/2}(X^k) \delta^{D-1} \Big(p_i - \delta_{ij} m a_{\text{ini}} \big[1-  \Psi_G^{\text{ini}}(X^k) \big] \big[ v^{\text{ini}}_{\text{cl}} \big]^j (X^k) \Big)\, ,
\ee
such that the initial physical rest-mass density is given by
\be
\rho_{\text{cl}}^{\text{ini}} (X^k) \approx m \gamma^{-1/2}_{\text{ini}}(X^k) \int d^{D-1} p f_{\text{cl}}^{\text{ini}}(X^k, p_l)\, ,
\ee
whereas the initial physical momentum density equals
\be
\rho_{\text{cl}}^{\text{ini}} \big[ v^{\text{ini}}_{\text{cl}} \big]^i (X^k) \approx \delta^{ij} a^{-1}_{\text{ini}} \big[1+  \Psi_G^{\text{ini}}(X^k) \big] \gamma^{-1/2}_{\text{ini}}(X^k) \int d^{D-1}p \, p_j f_{\text{cl}}^{\text{ini}}(X^k, p_l)\, .
\ee
Higher-order cumulants and thus the stress tensor are absent initially. 
We transfer this setting to the scalar field phase-space density and make use of the large mass approximation used in the cold dark matter kinetic equation \eqref{VlasovClassicNr},
\be
f_{\phi}^{\text{ini}}(X^k, p_i)  \approx  \rho_{\text{ini}}(X^k)  \frac{\gamma^{1/2}_{\text{ini}}(X^k)}{m}\delta^{D-1} \Bigg(p_i -  m \frac{\big[(e+P) u^0 u_i \big]_{\text{com}}^{\text{ini}}(X^k) }{\rho_{\text{ini}}(X^k)  }    \Bigg)\, .
\ee
We rewrote the initial phase-space density such that we can relate it easier to the gravitational potentials and the scale factor, i.e. the metric constraint fields that contain initially the information from previous regimes in the cosmological evolution due to their relation to other matter that were dominant in those.
Using the Einstein equations from section \ref{EinsteinEqFluidVar}, we have
\begin{multline}
f_{\phi}^{\text{ini}}(X^k, p_i)  \approx \frac{D-2 }{ma^2_{\text{ini}}}\frac{M_P^2}{\hbar}\gamma^{1/2}_{\text{ini}}(X^k)  \Big[\frac{D-1}{2} \mathcal{H}^2_{\text{ini}} + \Delta_X \Psi_G^{\text{ini}} (X^k) \Big] \\\times  \delta^{D-1} \Bigg(p_i - m a_{\text{ini}} \frac{2}{D-1} \frac{\partial_i \Phi_G^{\text{ini}} (X^k) }{\mathcal{H}_{\text{ini}}}   \Bigg)\, , \label{fIni}
\end{multline}
where we are sticking to the typical set-up in which the gravitational potentials are initially constant and the decaying mode is neglected. The gravitational slip is initially also equal to zero within cosmological perturbation theory, however, we keep it here to illustrate that the potential $\Psi_G$ is sourced by the density perturbation whereas the potential $\Phi_G$ is to source to leading order by the velocity perturbation. 
We split \eqref{fIni} into even and odd parts in order to be able to write down initial conditions for the scalar field two-point functions. We find
\begin{multline}
F_{00}^{\text{ini}} (X^k,p_l) \approx {\hbar  } \frac{D-2 }{2}\frac{M_P^2}{m^2 a^2_{\text{ini}}}  \Big[\frac{D-1}{2} \mathcal{H}^2_{\text{ini}} + \Delta_X \Psi_G^{\text{ini}} (X^k) \Big] \\ \times\int {d^{D-1} r} e^{- \frac{i}{\hbar} p_i r^i} \cos \Bigg(  \frac{2}{D-1}\frac{ma_{\text{ini}}}{\hbar}  \frac{r^i \partial_i \Phi_G^{\text{ini}} (X^k) }{\mathcal{H}_{\text{ini}}}   \Bigg)\, ,
\end{multline}
and 
\begin{multline}
F_{-}^{\text{ini}} (X^k,p_l) \approx i {m} \gamma^{1/2}_{\text{ini}}(X^k) \frac{D-2 }{2}\frac{M_P^2}{m^2 a^2_{\text{ini}}} \Big[\frac{D-1}{2} \mathcal{H}^2_{\text{ini}} + \Delta_X \Psi_G^{\text{ini}} (X^k) \Big] \\ \times \int {d^{D-1} r}  e^{- \frac{i}{\hbar} p_i r^i} \sin \Bigg( \frac{2}{D-1}  \frac{ma_{\text{ini}}}{\hbar}  \frac{r^i \partial_i \Phi_G^{\text{ini}} (X^k) }{\mathcal{H}_{\text{ini}}}   \Bigg)\, .
\end{multline}
Carrying out the linearization in the gravitational potentials, these expressions reduce to
\bea
F_{00}^{\text{ini}} (X^k,p_l) &\approx &  {\hbar  } \frac{D-2 }{2}\frac{M_P^2}{m^2 a^2_{\text{ini}}}  \Big[\frac{D-1}{2} \mathcal{H}^2_{\text{ini}} + \Delta_X \Psi_G^{\text{ini}} (X^k) \Big] (2 \pi \hbar)^{D-1} \delta^{D-1} \big( p_l \big) \, , \\
F_{-}^{\text{ini}} (X^k,p_l) &\approx & - {m} a^{D-1}_{\text{ini}} \frac{D-2 }{2}\frac{\mathcal{H}_{\text{ini}} M_P^2}{m a_{\text{ini}}}    \frac{\partial}{\partial X^i} \Phi_G^{\text{ini}} (X^k)  (2 \pi \hbar)^{D-1}  \frac{\partial}{\partial p_i}\delta^{D-1} \big( p_l \big)  \, .
\eea
We perform a Wigner transformation to obtain the correlators in coordinate space
\bea
F_{00}^{\text{ini}} (x^k,y^l) & \approx &  \frac{D-2 }{2}\frac{M_P^2}{\hbar} \frac{\hbar^2}{m^2 a^2_{\text{ini}}} \Big[\frac{D-1}{2} \mathcal{H}^2_{\text{ini}} + \big[\Delta \Psi_G^{\text{ini}} \big] \Big( \frac{(x + y)^k}{2} \Big)   \Big] \, ,\\ 
F_{-}^{\text{ini}} (x^k,y^l) & \approx & i a^{D-2}_{\text{ini}}\mathcal{H}_{\text{ini}}    \frac{D-2 }{2}\frac{M_P^2}{\hbar}  { (x - y)^i \big[ \partial_i \Phi_G^{\text{ini}}  \big]\Big( \frac{(x + y)^k}{2} \Big)  }  \, .
\eea
As we found out earlier, we can express the remaining two-point functions $F_+$ and $F_{11}$ in terms of $F_{00}$ and $F_{-}$ (see \eqref{F+IntermsOfFMinusandF00} to \eqref{F11IntermsOfFMinusandF00}),
\begin{multline}
  F_+^{\text{ini}} (x^k,y^l)
\approx  - a^{D-2}_{\text{ini}}\mathcal{H}_{\text{ini}} \frac{D-2 }{4}\frac{M_P^2}{\hbar} \frac{\hbar^2}{m^2 a^2_{\text{ini}}} \Big[\frac{(D-1)^2}{2} \mathcal{H}^2_{\text{ini}}\\ + \big[(D-1)\Delta \Psi_G^{\text{ini}}+\Delta \Phi_G^{\text{ini}} \big] \Big( \frac{(x + y)^k}{2} \Big)   \Big]  \, , 
\end{multline}
\be
 F_{11}^{\text{ini}}  (x^k,y^l)   \approx   \frac{D-2 }{2}a^{2D-4}_{\text{ini}} \frac{M_P^2}{\hbar}  \Big[\frac{D-1}{2} \mathcal{H}^2_{\text{ini}} + \big[\Delta \Psi_G^{\text{ini}} \big] \Big( \frac{(x + y)^k}{2} \Big)   \Big] \, . 
\ee
We have now specified all ingredients of a Gaussian initial density matrix for the real scalar field theory we started with in the beginning. Note, that this is a very choice that maps on an initially perfect fluid. We could as well have taken into account higher cumulants in momentum space.
\section{Conclusion}

In this work we develop a formalism for the dynamics of dark matter in which we start with a tree-level relativistic 
action for a real scalar field and obtain an effective action description that includes leading order interactions mediated by gravity.  Our formalism is relativistic, in that 
it allows for a systematic inclusion of both relativistic matter field effects as well as relativistic gravitational effects.
The non-relativistic limit of our dynamical equations could be obtained from a second quantized scalar field formalism
akin to the one used in condensed matter literature. However, since we are in particular interested in capturing relativistic
corrections (that can be important in the gravitational sector when one is interested in the scales comparable to
the Hubble scale and in the matter sector e.g. when the scalar field is ultralight), using such a formalism 
would restrict its validity too much. 

Furthermore, we identify a phase-space distribution $f_{\phi}$ based on four on-shell, equal-time real scalar field statistical two-point functions \eqref{psdens}. The statistical two-point functions obey a system of first order differential equations that closes because we first neglected manifest self-interactions of the matter field and the dynamical gravitational fields and second, did not integrate out the gravitational constraint fields. In the language of Feynman diagrams this amounts to approximating loop contributions with external sources whose evolution is determined by the semi-classical Einstein equation.
  The evolution of the phase-space distribution $f_{\phi}$ is determined by a generalized Vlasov equation including relativistic corrections, third-order time derivatives and corrections in a gradient expansion \eqref{finalVlasov}. Dropping the third-order time-derivatives as small corrections reduces the degrees of freedom to those of classical one-particle phase-space distribution. The statistical two-point functions of the scalar matter field entering the definition of $f_{\phi}$ are evaluated with respect to an initial density matrix and thus have generically reducible and connected pieces, in other words they contain a part given by one-point functions or classical fields. Focusing on the connected piece of the statistical matter two-point function makes the major distinction from previous approaches of modeling real scalar field fluids that focused on one-point functions. The reason is that it allows for generic initial conditions in two arguments without coarse-graining, either in position space or in Wigner space which then translates into a hierarchy of non-related moments in momentum space and in particular enables us to model a fluid that generically can include vorticity and anisotropy \eqref{anisoStress}. At the level of hydrodynamics we realized this by facing what we called a composite term that written as fluid quantities reduces into products of velocities and an irreducible piece \eqref{redandIrr} comprising a stress tensor. We derived non-linear imperfect hydrodynamic equations \eqref{ContApprox} \eqref{Eulerli} from the integrated system of statistical matter two-point functions that are exact up the linearized scalar metric we worked with.

 We note that using statistical two-point functions from the beginning allows us to treat gravity on a semi-classical level where we introduced linearized stochastic gravitational potentials that couple to the statistical two-point functions and thereby make the phase-space distribution $f_{\phi}$ stochastic. This is what we call hybrid approach and it bridges  the gap to cosmological perturbation theory since we now can calculate in a second step two-point function with respect to the gravitational potentials. The Einstein equations relating them were derived \ref{EinsteinEqFluidVar}.
 We also provided initial conditions for the statistical matter two-point functions and thus provided all ingredients to treat this problem with non-equilibrium quantum field theory techniques like the Schwinger-Keldysh formalism which we intent to do in the future.
 
We once more remark that - to model dark matter on non-linear scales at late times - it is necessary to go beyond the perfect fluid
approximation. The perfect fluid description is valid only in 
the linear regime which is reflected in the initial conditions and breaks down on scales $k>k_{nl} \sim 0.3~{\rm Mpc}^{-1}$ primarily due to shell-crossing
and generation of other types of perturbations that in the non-linear regime get dynamically generated. These perturbations 
include gravitational slip, vector and tensor metric perturbations as well as vorticity and anisotropic stresses at the matter side. 
Except for vector and tensor metric perturbations, all of these can be consistently treated in our formalism, which models dark matter by utilizing statistical (Hadamard) two-point functions.  In future work we intent to go beyond this hybrid approach, including a full non-linear treatment of gravity that is not restricted to the Newtonian gauge. 

\paragraph{Acknowledgments.}
We are grateful for conversation on this subject with
C. Uhlemann.
This work is part of the research programme of the Foundation for Fundamental Research on Matter (FOM), which is part of the Netherlands Organisation for Scientific Research (NWO). This work is in part supported by the D-ITP consortium, a program of the Netherlands Organization for Scientific Research (NWO) that is funded by the Dutch Ministry of Education, Culture and Science (OCW). 
\appendix
\section{Appendix}
\subsection{Wigner Tranformation of 2-Point Function Dynamics \label{appWignerDyn}}

In this appendix we drop the ubiquitous $\eta$ dependence to save some space.
We define the following operator 
\be
\Big[ f(X^i, p_i) \Big] \overleftarrow{\partial_X} \cdot \overrightarrow{\partial_p} \Big[ g(X^i, p_i)  \Big] := \Bigg[ \frac{\partial f }{\partial X^k}\Bigg] (X^i, p_i)   \Bigg[ \frac{\partial g }{\partial p_k }\Bigg] (X^i, p_i)  \, .
\ee
By using partial integration, one can show that the following relation holds up to boundary terms
\begin{multline}
\int d^{D-1}(x-\widetilde{x}) e^{-ip_i (x^i-\widetilde{x}^i)} \int d^{D-1}z A(x^i,z^i) \, B(z^i, \widetilde{x}^i)\\ =  A(X^i, p_i) e^{i \frac{\hbar}{2} \big( \overleftarrow{\partial_X} \cdot \overrightarrow{\partial_p} -  \overleftarrow{\partial_p}  \cdot \overrightarrow{\partial_X} \big) } B(X^i,p_i) + \text{boundary terms}\, .
\end{multline}
Using the definition
\be
p^2 := \delta^{ij} p_i p_j\, 
\ee 
we rewrite the equations \eqref{firstDerF00} -  \eqref{firstDer11} in Wigner space in the following way
\begin{multline}
a^{D-2}  F_{00}^{\prime}(X^i, p_i)=  \Big[1+ \Phi_G(X^i) + (D-1) \Psi_G(X^i) \Big] e^{i \frac{\hbar}{2}  \overleftarrow{\partial_X} \cdot \overrightarrow{\partial_p} } F_{10} (X^i, p_i) \\  + F_{01} (X^i, p_i)  e^{-i \frac{\hbar}{2} \overleftarrow{\partial_p}  \cdot \overrightarrow{\partial_X}  } \Big[1+ \Phi_G(X^i) + (D-1) \Psi_G(X^i) \Big] \, ,
\end{multline}
\begin{multline}
F_{10}^{\prime}(X^i, p_i )  = \frac{\partial}{\partial X^k} \Big[ \Phi_G (X^i) - (D-3)  \Psi_G(X^i) \Big]  e^{i \frac{\hbar}{2}  \overleftarrow{\partial_X} \cdot \overrightarrow{\partial_p} } \Bigg[  \Big[\frac{1}{2}\frac{\partial}{\partial X^k}  +i \frac{p_k}{\hbar} \Big] \Big[ a^{D-2} F_{00} \Big](X^i, p_i )\Bigg]   \\ + \Bigg[1+ \Phi_G (X^i) - (D-3)  \Psi_G(X^i)\Bigg]  e^{i \frac{\hbar}{2} \overleftarrow{\partial_X} \cdot \overrightarrow{\partial_p} } \Bigg\lbrace   \Big[\frac{\Delta_X}{4}+ i  \frac{p}{\hbar }\cdot \partial_X  - \frac{p^2}{\hbar^2} \Big] \Big[ a^{D-2} F_{00} \Big](X^i, p_i )\Bigg\rbrace    \\ - \frac{ m^2}{\hbar^2} a^2 \Big[1+ \Phi_G(X^i) - (D-1) \Psi_G(X^i) \Big] e^{i \frac{\hbar}{2}  \overleftarrow{\partial_X} \cdot \overrightarrow{\partial_p} } \Big[ a^{D-2} F_{00} \Big](X^i,p_i )\\  +   \Big[ a^{-(D-2)} F_{11} \Big] (X^i,p_i) e^{-i \frac{\hbar}{2} \overleftarrow{\partial_p}  \cdot \overrightarrow{\partial_X}  }  \Big[1+ \Phi_G(X^i) + (D-1) \Psi_G(X^i) \Big]\, ,
\end{multline}
\begin{multline}
F_{01}^{\prime}(X^i, p_i )  = \Bigg\lbrace \Big[\frac{1}{2}\frac{\partial}{\partial X^k} -i \frac{p_k}{\hbar} \Big] \Big[ a^{D-2} F_{00} \Big](X^i, p_i ) \Bigg\rbrace \cdot e^{-i \frac{\hbar}{2} \overleftarrow{\partial_p}  \cdot \overrightarrow{\partial_X} }    \frac{\partial}{\partial X^k}\Big[ \Phi_G (X^i) - (D-3)  \Psi_G(X^i) \Big]  \\ + \Bigg\lbrace  \Big[\frac{\Delta_X}{4}- i \frac{p}{\hbar }\cdot \partial_X   - \frac{p^2}{\hbar^2} \Big] \Big[ a^{D-2} F_{00} \Big](X^i, p_i ) \Bigg\rbrace  e^{-i \frac{\hbar}{2} \overleftarrow{\partial_p} \cdot \overrightarrow{\partial_X} } \Bigg\lbrace   1+ \Phi_G (X^i) - (D-3)  \Psi_G(X^i)\Bigg\rbrace    \\ - \frac{ m^2}{\hbar^2} a^2  \Big[ a^{D-2} F_{00} \Big](X^i,p_i )e^{-i \frac{\hbar}{2}  \overleftarrow{\partial_p} \cdot \overrightarrow{\partial_X} } \Big[1+ \Phi_G(X^i) - (D-1) \Psi_G(X^i) \Big]\\   +   \Big[1+ \Phi_G(X^i) + (D-1) \Psi_G(X^i) \Big]  e^{i \frac{\hbar}{2} \overleftarrow{\partial_X}  \cdot \overrightarrow{\partial_p}  }\Big[ a^{-(D-2)} F_{11} \Big] (X^i,p_i) \, .
\end{multline}
\begin{multline}
 a^{-(D-2)} F_{11}^{\prime}(X^i, p_i ) =   \\
+  \frac{\partial}{\partial X^k} \Big[ \Phi_G (X^i) - (D-3)  \Psi_G(X^i) \Big] \cdot e^{i \frac{\hbar}{2}  \overleftarrow{\partial_X} \cdot \overrightarrow{\partial_p}  } \Bigg[  \Big[\frac{1}{2}\frac{\partial}{\partial X^k}  +i \frac{p_k}{\hbar} \Big] F_{01}(X^i, p_i )\Bigg]   \\ + \Bigg\lbrace 1+ \Phi_G (X^i) - (D-3)  \Psi_G(X^i)\Bigg\rbrace  e^{i \frac{\hbar}{2} \overleftarrow{\partial_X} \cdot \overrightarrow{\partial_p} } \Bigg\lbrace   \Big[\frac{\Delta_X}{4}+ i \frac{p}{\hbar }\cdot \partial_X   - \frac{p^2}{\hbar^2} \Big] F_{01}(X^i, p_i )\Bigg\rbrace    \\ - \frac{ m^2}{\hbar^2} a^2  \Big[1+ \Phi_G(X^i) - (D-1) \Psi_G(X^i) \Big] e^{i \frac{\hbar}{2}  \overleftarrow{\partial_X} \cdot \overrightarrow{\partial_p} } F_{01}(X^i,p_i )\\
+ \Bigg\lbrace \Big[\frac{1}{2}\frac{\partial}{\partial X^k} -i \frac{p_k}{\hbar}\Big] F_{10}(X^i, p_i ) \Bigg\rbrace \cdot e^{-i \frac{\hbar}{2}  \overleftarrow{\partial_p}  \cdot \overrightarrow{\partial_X}  }   \frac{\partial}{\partial X^k} \Big[ \Phi_G (X^i) - (D-3)  \Psi_G(X^i) \Big]  \\ + \Bigg\lbrace  \Big[\frac{\Delta_X}{4}- i \frac{p}{\hbar }\cdot \partial_X   - \frac{p^2}{\hbar^2} \Big] F_{10}(X^i, p_i ) \Bigg\rbrace  e^{-i \frac{\hbar}{2} \overleftarrow{\partial_p} \cdot \overrightarrow{\partial_X} } \Bigg\lbrace   1+ \Phi_G (X^i) - (D-3)  \Psi_G(X^i)\Bigg\rbrace    \\ - \frac{ m^2}{\hbar^2} a^2  F_{10}(X^i,p_i )e^{-i \frac{\hbar}{2}  \overleftarrow{\partial_p} \cdot \overrightarrow{\partial_X} } \Big[1+ \Phi_G(X^i) - (D-1) \Psi_G(X^i) \Big] 
 \, .
\end{multline}
By using the definitions \eqref{defFPlus} and \eqref{defFMinus}, we get the following system of equations 
\begin{multline}
 a^{D-2} F_{00}^{\prime}(X^i, p_i)
=   2 \Big[1+ \Phi_G(X^i) + (D-1) \Psi_G(X^i) \Big] \cos \Big[ \frac{\hbar}{2} \, \overleftarrow{\partial_X} \cdot \overrightarrow{\partial_p} \Big] F_+ (X^i,p_i) \\ + 2 \Big[\Phi_G(X^i) + (D-1) \Psi_G(X^i) \Big]  \sin \Big[ \frac{\hbar}{2} \, \overleftarrow{\partial_X} \cdot \overrightarrow{\partial_p} \Big] F_- (X^i,p_i) \, , \label{F00Fin}
\end{multline}
\begin{multline}
 F_{+}^{\prime}(X^i, p_i )  = \frac{\partial}{\partial X^k}  \Big[ \Phi_G (X^i) - (D-3)  \Psi_G(X^i) \Big]   \cos \Big[  \frac{\hbar}{2}  \overleftarrow{\partial_X} \cdot \overrightarrow{\partial_p}\Big] \frac{1}{2}\frac{\partial}{\partial X^k}  \Big[ a^{D-2} F_{00} \Big](X^i, p_i )  \\
+   \Big[  1+ \Phi_G (X^i) - (D-3)  \Psi_G(X^i) \Big]   \cos \Big[  \frac{\hbar}{2}  \overleftarrow{\partial_X} \cdot \overrightarrow{\partial_p}\Big]  \Bigg\lbrace \Big[  \frac{\Delta_X}{4}- \frac{p^2}{\hbar^2} \Big]  \Big[ a^{D-2} F_{00} \Big](X^i, p_i ) \Bigg\rbrace \\
 - \Big[\Phi_G (X^i) - (D-3)  \Psi_G(X^i)\Big]  \sin \Big[  \frac{\hbar}{2}  \overleftarrow{\partial_X} \cdot \overrightarrow{\partial_p}\Big]  \Bigg\lbrace   \frac{p}{\hbar }\cdot \partial_X   \Big[ a^{D-2} F_{00} \Big](X^i, p_i ) \Bigg\rbrace\\
  -  \frac{\partial}{\partial X^k}  \Big[ \Phi_G (X^i) - (D-3)  \Psi_G(X^i) \Big]  \sin \Big[ \frac{\hbar}{2} \overleftarrow{\partial_X} \cdot \overrightarrow{\partial_p}\Big]  \cdot  \Bigg\lbrace  \frac{p_k}{\hbar}  \Big[ a^{D-2} F_{00} \Big](X^i, p_i )\Bigg\rbrace   \\ 
 - \frac{ m^2}{\hbar^2} a^2 \Big[1+ \Phi_G(X^i) - (D-1) \Psi_G(X^i) \Big] \cos \Big[ \frac{\hbar}{2} \, \overleftarrow{\partial_X} \cdot \overrightarrow{\partial_p} \Big]\Big[ a^{D-2} F_{00} \Big](X^i,p_i ) \\      +  \Big[1+ \Phi_G(X^i) + (D-1) \Psi_G(X^i) \Big] \cos \Big[ \frac{\hbar}{2} \, \overleftarrow{\partial_X} \cdot \overrightarrow{\partial_p} \Big]\Big[ a^{-(D-2)} F_{11} \Big] (X^i,p_i)    \, .\label{F+Fin}
\end{multline}
\begin{multline}
 F_{-}^{\prime}(X^i, p_i )  =
-\frac{\partial}{\partial X^k}   \Big[ \Phi_G (X^i) - (D-3)  \Psi_G(X^i) \Big]   \sin \Big[  \frac{\hbar}{2}  \overleftarrow{\partial_X} \cdot \overrightarrow{\partial_p}\Big]    \frac{1}{2}\frac{\partial}{\partial X^k}  \Big[ a^{D-2} F_{00} \Big](X^i, p_i )    \\
-   \Big[  \Phi_G (X^i) - (D-3)  \Psi_G(X^i) \Big]   \sin\Big[  \frac{\hbar}{2}  \overleftarrow{\partial_X} \cdot \overrightarrow{\partial_p}\Big]  \Bigg\lbrace \Big[  \frac{\Delta_X}{4}- \frac{p^2}{\hbar^2} \Big]  \Big[ a^{D-2} F_{00} \Big](X^i, p_i ) \Bigg\rbrace \\
 - \Big[1+\Phi_G (X^i) - (D-3)  \Psi_G(X^i)\Big]  \cos \Big[  \frac{\hbar}{2}  \overleftarrow{\partial_X} \cdot \overrightarrow{\partial_p}\Big]  \Bigg\lbrace   \frac{p}{\hbar }\cdot \partial_X   \Big[ a^{D-2} F_{00} \Big](X^i, p_i ) \Bigg\rbrace\\
  -  \frac{\partial}{\partial X^k}  \Big[ \Phi_G (X^i) - (D-3)  \Psi_G(X^i) \Big]  \cos \Big[ \frac{\hbar}{2} \overleftarrow{\partial_X} \cdot \overrightarrow{\partial_p}\Big]  \cdot  \Bigg[  \frac{p_k}{\hbar}  \Big[ a^{D-2} F_{00} \Big](X^i, p_i )\Bigg]   \\ 
     + \frac{ m^2}{\hbar^2} a^2 \Big[\Phi_G(X^i) - (D-1) \Psi_G(X^i) \Big] \sin \Big[ \frac{\hbar}{2} \, \overleftarrow{\partial_X} \cdot \overrightarrow{\partial_p} \Big]\Big[ a^{D-2} F_{00} \Big](X^i,p_i ) \\      +  \Big[\Phi_G(X^i) + (D-1) \Psi_G(X^i) \Big] \sin \Big[ \frac{\hbar}{2} \, \overleftarrow{\partial_X} \cdot \overrightarrow{\partial_p} \Big]\Big[ a^{-(D-2)} F_{11} \Big] (X^i,p_i)   \, .\label{F-Fin}
\end{multline}
\begin{multline}
 a^{-(D-2)} F_{11}^{\prime}(X^i, p_i ) = 
- 2 \frac{ m^2}{\hbar^2} a^2  \Big[1+ \Phi_G(X^i) - (D-1) \Psi_G(X^i) \Big] \cos \Big[  \frac{\hbar}{2}  \overleftarrow{\partial_X} \cdot \overrightarrow{\partial_p} \Big] F_{+}(X^i, p_i )\\
+ 2 \frac{ m^2}{\hbar^2} a^2  \Big[1+ \Phi_G(X^i) - (D-1) \Psi_G(X^i) \Big] \sin \Big[  \frac{\hbar}{2}  \overleftarrow{\partial_X} \cdot \overrightarrow{\partial_p} \Big] F_{-}(X^i, p_i )\\
+2 \Bigg\lbrace 1+ \Phi_G (X^i) - (D-3)  \Psi_G(X^i)\Bigg\rbrace  \cos \Big[  \frac{\hbar}{2}  \overleftarrow{\partial_X} \cdot \overrightarrow{\partial_p} \Big] \Bigg\lbrace   \Big[\frac{\Delta_X}{4}  - \frac{p^2}{\hbar^2} \Big]  F_{+}(X^i, p_i )\Bigg\rbrace    \\
-2 \Bigg\lbrace  \Phi_G (X^i) - (D-3)  \Psi_G(X^i)\Bigg\rbrace  \sin \Big[  \frac{\hbar}{2}  \overleftarrow{\partial_X} \cdot \overrightarrow{\partial_p} \Big] \Bigg\lbrace   \Big[\frac{\Delta_X}{4}  - \frac{p^2}{\hbar^2} \Big] F_{-}(X^i, p_i )\Bigg\rbrace    \\
- 2 \Bigg\lbrace  \Phi_G (X^i) - (D-3)  \Psi_G(X^i)\Bigg\rbrace  \sin \Big[  \frac{\hbar}{2}  \overleftarrow{\partial_X} \cdot \overrightarrow{\partial_p} \Big] \Bigg\lbrace    \frac{p}{\hbar }\cdot \partial_X    F_{+}(X^i, p_i )\Bigg\rbrace \\
- 2 \Bigg\lbrace 1+ \Phi_G (X^i) - (D-3)  \Psi_G(X^i)\Bigg\rbrace  \cos \Big[  \frac{\hbar}{2}  \overleftarrow{\partial_X} \cdot \overrightarrow{\partial_p} \Big] \Bigg\lbrace    \frac{p}{\hbar }\cdot \partial_X   F_{-}(X^i, p_i )\Bigg\rbrace \\
+ \Bigg\lbrace  \frac{\partial}{\partial X^k}  \Big[ \Phi_G (X^i) - (D-3)  \Psi_G(X^i) \Big] \Bigg\rbrace  \cos \Big[  \frac{\hbar}{2}  \overleftarrow{\partial_X} \cdot \overrightarrow{\partial_p} \Big] \cdot\Bigg\lbrace   \frac{\partial}{\partial X^k}   F_{+}(X^i, p_i )\Bigg\rbrace\\
- \Bigg\lbrace  \frac{\partial}{\partial X^k} \Big[ \Phi_G (X^i) - (D-3)  \Psi_G(X^i) \Big] \Bigg\rbrace  \sin \Big[  \frac{\hbar}{2}  \overleftarrow{\partial_X} \cdot \overrightarrow{\partial_p} \Big] \cdot\Bigg\lbrace   \frac{\partial}{\partial X^k}   F_{-}(X^i, p_i )\Bigg\rbrace\\
- 2\Bigg\lbrace  \frac{\partial}{\partial X^k} \Big[ \Phi_G (X^i) - (D-3)  \Psi_G(X^i) \Big] \Bigg\rbrace  \sin \Big[  \frac{\hbar}{2}  \overleftarrow{\partial_X} \cdot \overrightarrow{\partial_p} \Big] \cdot\Bigg\lbrace   \frac{p_k}{\hbar}   F_{+}(X^i, p_i )\Bigg\rbrace\\
- 2\Bigg\lbrace  \frac{\partial}{\partial X^k} \Big[ \Phi_G (X^i) - (D-3)  \Psi_G(X^i) \Big] \Bigg\rbrace  \cos \Big[  \frac{\hbar}{2}  \overleftarrow{\partial_X} \cdot \overrightarrow{\partial_p} \Big] \cdot\Bigg\lbrace    \frac{p_k}{\hbar}  F_{-}(X^i, p_i )  \Bigg\rbrace \, .\label{F11Fin}
\end{multline}
For later discussion, we remark the following important equal time properties
\begin{multline}
  F_{00} ( X^i , p_i) = \frac{1}{2}  \int {d^{D-1} r}\, e^{- \frac{i}{\hbar}p_i  r^i}   \vac{ \Big\lbrace \hat{\phi} \Big( X^i + \frac{r^i}{2}\Big)\, , \hat{\phi} \Big( X^i - \frac{r^i}{2}\Big) \Big\rbrace}  =
 F_{00}  ( X^i , - p_i) \, , \label{F00IsEven}
\end{multline}
\begin{multline}
 F_{-} ( X^i , p_i)  = \frac{i}{4} \int {d^{D-1} r}e^{- \frac{i}{\hbar}p_i  r^i}  \Bigg[   \vac{  \Big\lbrace \hat{\Pi}\Big( X^i + \frac{r^i}{2}\Big)\, , {\hat{\phi}\Big( X^i - \frac{r^i}{2}\Big)} \Big\rbrace }  \\- \vac{ \Big\lbrace \hat{\phi}\Big( X^i + \frac{r^i}{2}\Big)\, , {\hat{\Pi}\Big( X^i - \frac{r^i}{2}\Big)}\Big\rbrace } \Bigg]  =  -  F_{-} ( X^i , - p_i)\, , \label{FMinusIsOdd}
\end{multline}
\bea
\int \frac{ d^{D-1}p}{(2\pi \hbar)^{D-1}} \Big[ p_{k_1} ... p_{k_{2n+1}} \Big]  F_{00} ( X^i , p_i) &=& 0 \, , \\
\int \frac{ d^{D-1}p}{(2\pi \hbar)^{D-1}} \Big[ p_{k_1} ... p_{k_{2n}} \Big] F_{-} ( X^i , p_i)  &=& 0 \, .
\eea
These equations tell us that $ F_{00}$ is even in $p_i$ and that $F_{-}$ is odd in $p_i$. After an appropriate rescaling that also accounts for the right dimensions, these two quantities will play the role of even and odd phase-space densities.
\paragraph{Phase-space dynamics including $\varepsilon_{\hbar}^2$ and $\varepsilon_{\hbar}^2 \cdot \varepsilon_{g}^2$ corrections.}
In this section we want to manipulate the unintegrated dynamical equations for these two-point functions to see whether we can reproduce equations that mimic the Vlasov-equation. First, write down the phase-space dynamics perturbatively including  $\varepsilon_{\hbar}^2$ that correspond to the next order in the gradient expansion we described in subsection \ref{whyQuantum}. We also include for illustration  $\varepsilon_{\hbar}^2 \cdot \varepsilon_{g}^2$ corrections that result from multiplying terms of the gradient expansion with the gravitational potentials. 
\begin{multline}
a^{D-2} F_{00}^{\prime}
=  2 \Big[ 1 + \Phi_G  + (D-1)  \Psi_G \Big] F_+ \\ 
- \frac{\hbar^2}{4} \frac{\partial^2}{\partial X^i \partial X^j } \Big[ \Phi_G  + (D-1)  \Psi_G \Big]\frac{\partial^2}{\partial p_i \partial p_j } F_+ \\+ \hbar \frac{\partial}{\partial X^i} \Big[ \Phi_G  + (D-1)  \Psi_G \Big] \frac{\partial}{\partial p_i}  F_- \\ -\hbar^3
\frac{1}{24} \frac{\partial^3}{\partial X^i \partial X^j \partial X^k}  \Big[ \Phi_G  + (D-1)  \Psi_G \Big] \frac{\partial^3}{\partial p_i\partial p_j\partial p_k}  F_- \, , 
\end{multline}
\begin{multline}
F_{+}^{\prime}  = 
  \frac{1}{2} \frac{\partial}{\partial X^k} \Big[ \Phi_G - (D-3) \Psi_G \Big] \frac{\partial}{\partial X^k}\Big[ a^{D-2} F_{00} \Big]
  +\Big[1+ \Phi_G - (D-3) \Psi_G \Big] \Big[  \frac{\Delta_X}{4}- \frac{p^2}{\hbar^2} \Big]  \Big[ a^{D-2} F_{00} \Big]  \\
    + \frac{1}{8}\frac{\partial^2}{\partial X^i\partial X^j} \Big[\Phi_G - (D-3) \Psi_G \Big] \frac{\partial^2}{\partial p_i\partial p_j}\Big[ {p^2}  \Big[ a^{D-2} F_{00} \Big] \Big]  \\
  - \frac{1}{2} \frac{\partial}{\partial X^k} \Big[ \Phi_G - (D-3) \Psi_G \Big] \frac{\partial}{\partial p_k} \Big[p^i \frac{\partial}{\partial X^i} \Big[ a^{D-2} F_{00} \Big]\Big] \\
  - \frac{1}{2} \frac{\partial^2}{\partial X^k\partial X^i} \Big[ \Phi_G - (D-3) \Psi_G \Big] \frac{\partial}{\partial p_i} \Big[p^k \Big[ a^{D-2} F_{00} \Big]\Big] \\
 - \frac{ m^2}{\hbar^2} a^2 \Big[1+ \Phi_G - (D-1) \Psi_G \Big] \Big[ a^{D-2} F_{00} \Big]
  + \frac{ m^2 a^2}{8} \frac{\partial^2}{\partial X^i\partial X^j}\Big[ \Phi_G - (D-1) \Psi_G \Big]\frac{\partial^2}{\partial p_i\partial p_j}  \Big[ a^{D-2} F_{00} \Big] \\      +  \Big[1+ \Phi_G + (D-1) \Psi_G \Big] \Big[ a^{-(D-2)} F_{11} \Big] 
  - \frac{ \hbar^2 }{8}  \frac{\partial^2}{\partial X^i\partial X^j}\Big[\Phi_G + (D-1) \Psi_G \Big]\frac{\partial^2}{\partial p_i\partial p_j} \Big[ a^{-(D-2)} F_{11} \Big]  \, ,
\end{multline}
\begin{multline}
 F_{-}^{\prime}  = - \frac{\hbar}{4}\frac{\partial^2}{\partial X^k\partial X^i} \Big[ \Phi_G - (D-3) \Psi_G \Big] \frac{\partial^2}{\partial p_i \partial X^k} \Big[ a^{D-2} F_{00} \Big] \\
- \frac{\hbar}{2}  \frac{\partial}{\partial X^k} \Big[ \Phi_G - (D-3) \Psi_G \Big] \frac{\partial}{\partial p_k}
 \Bigg[ \Big[ \frac{\Delta_X}{4} - \frac{p^2}{\hbar^2}  \Big]\Big[ a^{D-2} F_{00} \Big] \Bigg]\\
 - \frac{\hbar}{48}  \frac{\partial^3}{\partial X^k \partial X^i \partial X^j} \Big[ \Phi_G - (D-3) \Psi_G \Big] \frac{\partial^3}{\partial p_k\partial p_i\partial p_j}
 \Bigg[   {p^2} \Big[ a^{D-2} F_{00} \Big] \Bigg]
\\
 - \Big[1+ \Phi_G - (D-3) \Psi_G \Big] \frac{p_k }{\hbar} \frac{\partial}{\partial X^k}   \Big[ a^{D-2} F_{00} \Big] \\
  + \frac{\hbar}{8} \frac{\partial^2}{\partial X^i \partial X^j} \Big[\Phi_G - (D-3) \Psi_G \Big] \frac{\partial^2}{\partial p_i \partial p_j} \Big[{p_k } \frac{\partial}{\partial X^k}   \Big[ a^{D-2} F_{00} \Big]  \Big]\\
 +  \frac{1}{2} \frac{ m^2}{\hbar} a^2   \frac{\partial}{\partial X^i} \Big[\Phi_G - (D-1) \Psi_G \Big] \frac{\partial}{\partial p_i} \Big[ a^{D-2} F_{00} \Big]  \\  
 -  m^2 a^2 \hbar
\frac{1}{48} \frac{\partial^3}{\partial X^i \partial X^j \partial X^k}  \Big[\Phi_G - (D-1) \Psi_G \Big]\frac{\partial^3}{\partial p_i\partial p_j\partial p_k}  \Big[ a^{D-2} F_{00} \Big]  \\    +  \hbar \frac{1}{2} \frac{\partial}{\partial X^i} \Big[\Phi_G + (D-1) \Psi_G \Big] \frac{\partial}{\partial p_i}  \Big[ a^{-(D-2)} F_{11} \Big] \\
 -\hbar^3
\frac{1}{48} \frac{\partial^3}{\partial X^i \partial X^j \partial X^k} \Big[\Phi_G + (D-1) \Psi_G \Big] \frac{\partial^3}{\partial p_i\partial p_j\partial p_k}   \Big[ a^{-(D-2)} F_{11} \Big]\, ,
\end{multline}
\begin{multline}
a^{-(D-2)} F_{11}^{\prime} =   - 2 \frac{ m^2}{\hbar^2} a^2  \Big[1+ \Phi_G - (D-1) \Psi_G \Big] F_{+}\\
+ \frac{1}{4} { m^2} a^2 \frac{\partial^2}{\partial X^i \partial X^j}  \Big[\Phi_G - (D-1) \Psi_G \Big]\frac{\partial^2}{\partial p_i\partial p_j} F_{+}
\\+2   \Big[1+ \Phi_G - (D-3) \Psi_G \Big] \Big[\frac{\Delta_X}{4}  - \frac{p^2}{\hbar^2} \Big]  F_{+} 
+\frac{1}{4}\frac{\partial^2}{\partial X^i \partial X^j}\Big[ \Phi_G - (D-3) \Psi_G \Big]    \frac{\partial^2}{\partial p_i\partial p_j} \Big[  p^2  F_{+} \Big]  \\
- \hbar \frac{\partial}{\partial X^k}   \Big[ \Phi_G - (D-3) \Psi_G \Big] \frac{\partial}{\partial p_k} \Bigg[ \Big[\frac{\Delta_X}{4}  - \frac{p^2}{\hbar^2} \Big]  F_{-} \Bigg]
-\frac{\hbar}{24}\frac{\partial^3}{\partial X^i \partial X^j \partial X^k }\Big[ \Phi_G - (D-3) \Psi_G \Big]    \frac{\partial^3}{\partial p_i\partial p_j\partial p_k} \Big[  p^2  F_{-} \Big]  \\
-  \frac{\partial}{\partial X^k}   \Big[ \Phi_G - (D-3) \Psi_G \Big]\frac{\partial}{\partial p_k}  \Big[{p}\cdot \partial_X   F_{+}\Big] \\ - 2 \Big[1+ \Phi_G - (D-3) \Psi_G \Big] \frac{p}{\hbar }\cdot \partial_X   F_{-} + \frac{\hbar}{4} \frac{\partial^2}{\partial X^i \partial X^j }\Big[\Phi_G - (D-3) \Psi_G \Big]  \frac{\partial^2}{\partial p_i \partial p_j }\Bigg[ {p}\cdot \partial_X   F_{-}\Bigg] \\
+\frac{\partial}{\partial X^k}\Big[ \Phi_G - (D-3) \Psi_G \Big] \frac{\partial}{\partial X^k}  F_{+} 
- \frac{\hbar}{2}\frac{\partial^2}{\partial X^k\partial X^i}\Big[ \Phi_G - (D-3) \Psi_G \Big] \frac{\partial^2}{\partial X^k\partial p_i}  F_{-} \\
-\frac{\partial^2}{\partial X^k \partial X^i}\Big[ \Phi_G - (D-3) \Psi_G \Big] \frac{\partial}{\partial p_i} \Big[ p_k F_{+}  \Big]
- 2 \frac{p}{\hbar} \cdot \partial_X \Big[ \Phi_G - (D-3) \Psi_G \Big] F_{-}  \\
+ \frac{\hbar}{4}  \frac{\partial^3}{\partial X^i\partial X^j\partial X^k}\Big[ \Phi_G - (D-3) \Psi_G \Big] \frac{\partial^2}{\partial p_i \partial p_j} \Bigg[ p_k F_{-} \Bigg] \\
+ \frac{ m^2a^2}{\hbar}    \frac{\partial}{\partial X^i} \Big[ \Phi_G  - (D-1)  \Psi_G \Big] \frac{\partial}{\partial p_i}  F_- \\ -\hbar  m^2 a^2 
\frac{1}{24} \frac{\partial^3}{\partial X^i \partial X^j \partial X^k}  \Big[ \Phi_G  - (D-1)  \Psi_G \Big] \frac{\partial^3}{\partial p_i\partial p_j\partial p_k}  F_-   \, .
\end{multline}
\paragraph{Phase-space dynamics including $\varepsilon_{\hbar}^2$ but dropping $\varepsilon_{\hbar}^2 \cdot \varepsilon_{g}^2$ corrections.}
Let us further drop the  $\varepsilon_{\hbar}^2 \cdot \varepsilon_{g}^2$ corrections which is at least naively consistent with our linearization in gravity, i.e. consistent with not keeping $\varepsilon_{g}^4$ terms as we did from the very beginning of this paper. We do not expect the corrections $\varepsilon_{\hbar}^2$ from the gradient expansion to become important unless we have a very light scalar field ($m \approx 10^{-22} \text{eV}$) as pointed out in section \ref{whyQuantum} which is considered an extreme case.  Thus, for typical masses at scales $\gtrsim\text{eV}$, the following equations are perfectly accurate and we will even drop the $\varepsilon_{\hbar}^2$ corrections when discussing them further in the main text.
\begin{multline}
a^{D-2} F_{00}^{\prime}
=  2 \Big[ 1 + \Phi_G  + (D-1)  \Psi_G \Big] F_+  
+ \hbar \frac{\partial}{\partial X^i} \Big[ \Phi_G  + (D-1)  \Psi_G \Big] \frac{\partial}{\partial p_i}  F_- \, , 
\end{multline}
\begin{multline}
 F_{+}^{\prime}  =  \frac{\Delta_X}{4} \Big[ a^{D-2} F_{00} \Big] 
   -\frac{p^2}{\hbar^2}  \Big[1+ \Phi_G - (D-3) \Psi_G \Big] \Big[ a^{D-2} F_{00} \Big]  \\
 - \frac{ m^2}{\hbar^2} a^2 \Big[1+ \Phi_G - (D-1) \Psi_G \Big] \Big[ a^{D-2} F_{00} \Big]\\      +  \Big[1+ \Phi_G + (D-1) \Psi_G \Big] \Big[ a^{-(D-2)} F_{11} \Big]  \, ,
\end{multline}
\begin{multline}
 F_{-}^{\prime}  = 
 \frac{1}{2 \hbar}  \frac{\partial}{\partial X^k} \Big[ \Phi_G - (D-3) \Psi_G \Big] \frac{\partial}{\partial p_k}
 \Bigg[ {p^2} \Big[ a^{D-2} F_{00} \Big] \Bigg]\\
 - \frac{p}{ \hbar} \cdot \partial_X \Big[ \Phi_G - (D-3) \Psi_G \Big] \Big[ a^{D-2} F_{00} \Big]
 - \Big[1+ \Phi_G - (D-3) \Psi_G \Big] \frac{p_k }{\hbar} \frac{\partial}{\partial X^k}   \Big[ a^{D-2} F_{00} \Big] 
  \\+  \frac{1}{2} \frac{ m^2}{\hbar} a^2   \frac{\partial}{\partial X^i} \Big[\Phi_G - (D-1) \Psi_G \Big] \frac{\partial}{\partial p_i} \Big[ a^{D-2} F_{00} \Big]  \\  
   +  \hbar \frac{1}{2} \frac{\partial}{\partial X^i} \Big[\Phi_G + (D-1) \Psi_G \Big] \frac{\partial}{\partial p_i}  \Big[ a^{-(D-2)} F_{11} \Big]\, ,
\end{multline}
\begin{multline}
 a^{-(D-2)} F_{11}^{\prime} =  - 2 \frac{ m^2}{\hbar^2} a^2  \Big[1+ \Phi_G - (D-1) \Psi_G \Big] F_{+}\\
+  \frac{\Delta_X}{2}  F_{+} 
- 2  \frac{p^2}{\hbar^2} \Big[1+ \Phi_G - (D-3) \Psi_G \Big]  F_{+}  \\
+ \frac{1}{\hbar} \frac{\partial}{\partial X^k}   \Big[ \Phi_G - (D-3) \Psi_G \Big] \frac{\partial}{\partial p_k} \Bigg[  {p^2}   F_{-} \Bigg] 
 - 2 \Big[1+ \Phi_G - (D-3) \Psi_G \Big] \frac{p}{\hbar }\cdot \partial_X   F_{-} \\
- 2 \frac{p}{\hbar} \cdot \partial_X \Big[ \Phi_G - (D-3) \Psi_G \Big] F_{-}  
+ \frac{ m^2a^2}{\hbar}    \frac{\partial}{\partial X^i} \Big[ \Phi_G  - (D-1)  \Psi_G \Big] \frac{\partial}{\partial p_i}  F_-  \, .
\end{multline}
\pagebreak
\subsection{Einstein Equations in Longitudinal Gauge with Scalar Perturbations \label{appEinstein}}
The dynamical equations of the previous section are supplemented by the Einstein equations. Since we neglected gravitons with the choice of our metric, the Einstein equations are constraint equations that will determine the gravitational potentials in terms of two-point functions of the scalar field and thereby induce non-linear interactions. 
We write down the metric in this gauge
\bea
g_{00} (\eta, x^i ) &=& - a^{2}(\eta ) \left[ 1 + 2 \Phi_G (\eta, x^i ) \right] \,, \,  g_{ij}(\eta, x^i ) = a^2 (\eta ) \delta_{ij} \left[ 1- 2 \Psi_G(\eta, x^i ) \right] \, , \\
g^{00} (\eta, x^i ) &=& - a^{-2}(\eta ) \left[ 1 - 2 \Phi_G (\eta, x^i )\right] \,, \,  g^{ij}(\eta, x^i ) = a^{-2} (\eta ) \delta^{ij} \left[ 1 + 2 \Psi_G(\eta, x^i ) \right]\, , \\
\sqrt{-g}(\eta, x^i ) &=& a^D (\eta)  \left[ 1 +  \Phi_G(\eta, x^i )   - (D-1) \Psi_G(\eta, x^i )  \right] \, .
\eea
Let us collect the linearized connection coefficients in longitudinal gauge
\bea
\Gamma^0_{00} &=&  \mathcal{H} + \Phi_G ^{\prime} \, ,  \\
\Gamma^0_{0i} &=& \partial_i \Phi_G  \, , \\
\Gamma^i_{00} &=& \delta^{ij} \partial_j \Phi_G  \, , \\
\Gamma^0_{ij} &=& \mathcal{H}\delta_{ij}  - \Big[ 2 \mathcal{H} (\Phi_G + \Psi_G) + \Psi_G^{\prime} \Big] \delta_{ij}  \, , \\
\Gamma^i_{j0} &=& \mathcal{H}\delta^{i}_{\; j}  - \Psi_G^{\prime}  \delta^{i}_{\; j}   \, , \\
\Gamma^i_{jk} &=& -  \partial_j \Psi_G \delta^i_{\; k} - \partial_k \Psi_G \delta^i_{\; j} + \partial_l \Psi_G \delta^{i l} \delta_{j k}  \, .
\eea
We have the temporal Ricci tensor components
\bea
R_{00} &=&- (D-1)\mathcal{H}^{\prime} +  \Delta \Phi_G  + (D-1) \Psi_G^{\prime \prime} + (D-1) \mathcal{H}\Big[ \Phi_G^{\prime} + \Psi_G^{\prime} \Big] \, , \\
R_{0i} &=& (D-2)\partial_i \Psi_G^{\prime} + (D-2) \mathcal{H} \partial_i \Phi_G\, .
\eea
as well as the purely spatial part
\begin{multline}
R_{ij} = \Big[ \mathcal{H}^{\prime} +(D-2) \mathcal{H}^2 \Big] \delta_{ij}  + (D-3) \partial_i \partial_j \Psi_G - \partial_i \partial_j \Phi_G + \Delta \Psi_G\delta_{ij} \\
   - \Big [ \Psi_G^{\prime \prime} + 2(D-2) \mathcal{H}^2 (\Phi_G + \Psi_G) + 2 \mathcal{H}^{\prime} (\Phi_G + \Psi_G) + \mathcal{H} \Phi_G^{\prime} + (2D-3) \mathcal{H}\Psi_G^{\prime} \Big] \delta_{ij}  \, .
\end{multline}
This leaves us with
\begin{multline}
a^2 R = (D-1) \Big[ 2 \mathcal{H}^{\prime} + (D-2) \mathcal{H}^2 \Big] - 2 \Delta \Phi_G + (2D - 4) \Delta \Psi_G 
-2 (D-1) \Psi_G^{\prime \prime}\\ -2(D-1) \Phi_G \Big[ 2 \mathcal{H}^{\prime} +(D-2) \mathcal{H}^2 \Big] 
-2(D-1)\mathcal{H}\Phi_G^{\prime} -2(D-1)^2 \mathcal{H}\Psi_G^{\prime} \, .
\end{multline}
The $00$- and $0i$-components of the linearized Einstein tensor then read
\bea
G_{00} &=& \frac{1}{2}(D-1)(D-2) \mathcal{H}^2 +(D-2)\Delta \Psi_G - (D-1)(D-2) \mathcal{H}\Psi_G^{\prime} \,, \\
G_{0i} &=& (D-2)\partial_i \Psi_G^{\prime} + (D-2) \mathcal{H} \partial_i \Phi_G\,.
\eea
The $ij$-components of the linearized Einstein tensor read 
\begin{multline}
G_{ij} = - \Big[(D-2) \mathcal{H}^{\prime} +\frac{1}{2}(D-2)(D-3) \mathcal{H}^2 \Big] \delta_{ij} \\+ (D-2) \Psi_G^{\prime \prime}\delta_{ij} + \Delta \Big[ \Phi_G  - (D - 3)  \Psi_G \Big]\delta_{ij}
 -  \partial_i \partial_j \Big[ \Phi_G-  (D-3) \Psi_G \Big]  \\
   + \Bigg[ (D-2) \Big[ 2\mathcal{H}^{\prime} + (D-3) \mathcal{H}^2 \Big]  ( \Phi_G +\Psi_G)  +(D-2) \mathcal{H} \Big[ \Phi_G^{\prime} +(D-2)\Psi_G^{\prime} \Big] \Bigg] \delta_{ij} \,. 
\end{multline}
The energy-momentum tensor operator of the scalar field theory is given by
\be
\hat{T}_{\mu \nu} = \partial_{\mu} \hat{\phi} \partial_{\nu} \hat{\phi} - \frac{g_{\mu \nu}}{2} \Big[ g^{\alpha \beta} \partial_{\alpha} \hat{\phi} \partial_{\beta} \hat{\phi} + \frac{m^2}{ \hbar^2} \hat{\phi}^2 \ \Big]\, \label{TmunuOp}.
\ee
The composite operator \eqref{TmunuOp} needs to be renormalized by introducing on the gravity side higher order geometrical counterterms ($R^2$, square of the Weyl tensor, Gauss-Bonnet term) as well as lower order geometrical counterterms containing a bare Newton constant and a bare cosmological constant such that after substraction of UV-divergencies we are left with the observable values of the renormalized Newton constant and the renormalized cosmological constant \cite{Birrell:1982ix}. Contributions of the renormalized higher order geometrical terms completely irrelevant for the studies of large scale structures. Regularizing the two-point functions contained in the energy-momentum tensor \eqref{TmunuOp} related to the conditions we spelled out in \eqref{classicality condition}. In order to regularize the two-point functions we will split them into infrared and ultraviolet parts by introducing a cut-off in such a way that the conditions \eqref{classicality condition} can be satisfied up to that cut-off. However, for the scope of this paper we will not have to worry more about renormalization issues.
\par The semi-classical Einstein equation relates the gravitational potentials to two-point functions of the scalar field in the coincidence limit
\be
G_{\mu \nu} = \frac{\hbar}{M_P^2} {T}_{\mu \nu}   = \frac{\hbar}{M_P^2} \vac{ \hat{T}_{\mu \nu}  }\, .
\ee
Lets see how this works by looking at the purely temporal equation. We find (suppressing the time dependence in the argument) 
\be
G_{00} (X^i) = \frac{1}{2}(D-1)(D-2) \mathcal{H}^2 +(D-2)\Delta \Psi_G(X^i) - (D-1)(D-2) \mathcal{H}\Psi_G^{\prime}(X^i)  = \frac{\hbar}{M_P^2} \vac{ \hat{T}_{00} (X^i) }\, ,
\ee
where in our scalar field model
\bea
\nonumber \hat{T}_{00}(X^i) &=& \partial_0 \hat{\phi}(X^i) \partial_0 \hat{\phi} (X^i) - \frac{g_{00}}{2} \Big[ \partial^0 \hat{\phi} (X^i) \partial_0 \hat{\phi} (X^i)+ \partial^k \hat{\phi} (X^i) \partial_k \hat{\phi} (X^i) + \frac{ m^2}{\hbar^2} \hat{\phi}(X^i) \hat{\phi}(X^i) \Big] \, , \\ 
\nonumber &=&  \frac{1}{2} \hat{\phi}^{\prime}(X^i) \hat{\phi}^{\prime} (X^i) + \frac{1}{2}\Big[1+ 2 \Phi_G(X^i) +2 \Psi_G (X^i)  \Big]\partial_k \hat{\phi} (X^i) \partial_k \hat{\phi} (X^i) \\ &&  \qquad \qquad  +  \frac{a^2}{2}\Big[1+ 2 \Phi_G(X^i)  \Big] \frac{ m^2}{\hbar^2} \hat{\phi}(X^i) \hat{\phi}(X^i) \,,
\eea
and thus taking expectation values yields
\begin{multline}
\vac{\nonumber \hat{T}_{00}(X^i)} = \frac{1}{2} a^{-2(D-2)} F_{11} (X^i) \Big[1+2 \Phi_G(X^i) + 2(D-1) \Psi_G(X^i) \Big]
\\+ \frac{1}{2} \Big[1 + 2 \Phi_G(X^i) + 2 \Psi_G(X^i) \Big] \Bigg\lbrace \Bigg(\frac{\partial_X^2}{4} - \partial_r^2 \Bigg)  F_{00} (X^i,r^i)\Bigg\rbrace \Bigg|_{r^i =0} \\
+ \frac{1}{2} \frac{m^2 a^2}{\hbar^2} \Big[ 1 + 2 \Phi_G(X^i) \Big] \overline{F}_{00} (X^i) \, ,
\end{multline}
where 
\be
\overline{F}_{00} (X^i) = {F}_{00} (X^i,r^i=0)\, .
\ee
For the $0i$-equations we get
\be
 (D-2)\partial_i \Psi_G^{\prime} + (D-2) \mathcal{H} \partial_i \Phi_G =\frac{\hbar}{M_P^2}\vac{\hat{\phi}^{\prime} \partial_i \hat{\phi}}  \,.
\ee
For the $ij$-components of the energy-momentum tensor we get
\begin{multline}
 \hat{T}_{ij} = \partial_i \hat{\phi}   \partial_j \hat{\phi} + \frac{1}{2} \Big[1- 2(\Phi_G + \Psi_G) \Big] (\hat{\phi}^{\prime})^2 \delta_{ij}
 -\frac{1}{2}  \partial_k \hat{\phi} \partial_k \hat{\phi}  \delta_{ij} -  \frac{m^2 a^2}{2} \Big[1-2 \Psi_G \Big] \hat{\phi}^2 \delta_{ij}  \,.
\end{multline}
Projecting on the trace-free part we get
\be
\Bigg[ \frac{\Delta}{D-1} \delta_{ij} - \partial_i \partial_j \Bigg] \Big[ \Phi_G - (D-3) \Psi_G \Big] = \frac{\hbar}{M_P^2} \vac{ \partial_i \hat{\phi} \partial_j \hat{\phi} - \frac{\partial_k \hat{\phi} \partial_k \hat{\phi}}{D-1} \delta_{ij} }
\,.
\ee
The equation for the trace reads
\begin{multline}
\frac{G_{ii}}{D-1} = - (D-2) \mathcal{H}^{\prime} -\frac{1}{2}(D-2)(D-3) \mathcal{H}^2  + (D-2) \Psi_G^{\prime \prime} + \frac{D-2}{D-1} \Delta \Big[ \Phi_G  - (D - 3)  \Psi_G \Big]\\
   +  (D-2) \Big[ 2\mathcal{H}^{\prime} + (D-3) \mathcal{H}^2 \Big]  ( \Phi_G +\Psi_G)  +(D-2) \mathcal{H} \Big[ \Phi_G^{\prime} +(D-2)\Psi_G^{\prime} \Big] \\
   =  \frac{\hbar}{2 M_P^2} \Bigg\lbrace \Big[1- 2(\Phi_G + \Psi_G) \Big] \vac{ (\hat{\phi}^{\prime})^2 }
 -\frac{D-3}{D-1} \vac{\partial_k \hat{\phi} \partial_k \hat{\phi}}  - {m^2 a^2} \Big[1-2 \Psi_G \Big] \vac{ \hat{\phi}^2 }\Bigg\rbrace \,. 
\end{multline}
We rewrite all Einstein equations in terms of the two-point functions $F_{00}$, $F_+$, $F_-$ and $  F_{11} $: 
\begin{multline}
- (D-2) \mathcal{H}^{\prime} -\frac{1}{2}(D-2)(D-3) \mathcal{H}^2  + (D-2) \Psi_G^{\prime \prime} + \frac{D-2}{D-1} \Delta \Big[ \Phi_G  - (D - 3)  \Psi_G \Big]\\
   +  (D-2) \Big[ 2\mathcal{H}^{\prime} + (D-3) \mathcal{H}^2 \Big]  ( \Phi_G +\Psi_G)  +(D-2) \mathcal{H} \Big[ \Phi_G^{\prime} +(D-2)\Psi_G^{\prime} \Big] \\
   =  \frac{\hbar}{2  M_P^2} \Bigg\lbrace \Big[1 + 2(D-2) \Psi_G) \Big] \Big[ a^{-2(D-2)} F_{11} \Big]
 -\frac{D-3}{D-1}  \Bigg[ \Bigg(\frac{ \Delta_X}{4} -  \Delta_r \Bigg)  F_{00}\Bigg]_{r=0} \\  - \frac{ m^2 a^2}{\hbar^2} \Big[1-2 \Psi_G \Big]  F_{00}  \Bigg\rbrace \,,  \label{Einstein1}
\end{multline}
\begin{multline}
\frac{1}{2}(D-1)(D-2) \mathcal{H}^2 +(D-2)\Delta \Psi_G  - (D-1)(D-2) \mathcal{H}\Psi_G^{\prime} \\ = \frac{\hbar}{2 M_P^2} \Bigg\lbrace    a^{-2(D-2)} F_{11} \Big[1+2 \Phi_G + 2(D-1) \Psi_G \Big]
\\+  \Big[1 + 2 \Phi_G(X) + 2 \Psi_G(X) \Big] \Bigg[ \Bigg(\frac{ \Delta_X}{4} -  \Delta_r \Bigg) F_{00}\Bigg]_{r=0} 
+ {a^2} \frac{m^2}{\hbar^2} \Big[ 1 + 2 \Phi_G \Big]  F_{00}  \Bigg\rbrace  \,, \label{Einstein2}
\end{multline}
\begin{multline}
(D-2)\partial_X \Psi_G^{\prime} +  (D-2) \mathcal{H} \partial_X \Phi_G \\=  \frac{\hbar}{2 a^{(D-2)} M_P^2}    \Big[1 + \Phi_G +  (D-1) \Psi_G \Big] \Bigg\lbrace \partial_X F_+ +2i \partial_r F_-  \Bigg|_{r=0}   \Bigg\rbrace \,, \label{Einstein3}
\end{multline}
\begin{multline}
\Bigg[ \frac{\Delta_X}{D-1} \delta_{ij} - \partial_i^X \partial_j^X \Bigg] \Big[ \Phi_G - (D-3) \Psi_G \Big] 
\\ = \frac{\hbar}{ M_P^2}\Bigg[ \frac{\partial_i^X \partial_j^X}{4} +\frac{1}{2} \partial_i^X \partial_j^r -\frac{1}{2} \partial_j^X \partial_i^r - \partial_i^r \partial_j^r\\ - \frac{\delta_{ij}}{4(D-1)}  \Delta_X + \frac{\delta_{ij}}{D-1} \Delta_r \Bigg]  F_{00}  \Bigg|_{r=0}\,. \label{Einstein4}
\end{multline}
\paragraph{Eliminating the two-point functions $F_+$ and $F_{11}$ from energy-momentum tensor.}
Let us set
\be
T_{\mu \nu} := \vac{\hat{T}_{\mu\nu}}\, .
\ee
By integrating the dynamical matter equations over the momenta we can rewrite the energy-momentum tensor in such a way that it only depends on the two-point functions $F_{00}$ and $F_{-}$. We first get
\be
 a^{D-2} \overline{F}_{00}^{\prime}(X^i) =   2 \Big[1+ \Phi_G(X^i) + (D-1) \Psi_G(X^i) \Big] \overline{F}_+ (X^i) \, ,
\ee
\begin{multline}
 \overline{F}_{+}^{\prime}(X^i)  = \frac{\partial}{\partial X^k}  \Big[ \Phi_G (X^i) - (D-3)  \Psi_G(X^i) \Big]    \frac{1}{2}\frac{\partial}{\partial X^k}  \Big[ a^{D-2} \overline{F}_{00} \Big](X^i )  \\
+   \Big[  1+ \Phi_G (X^i) - (D-3)  \Psi_G(X^i) \Big]    \int \frac{ d^{D-1}p}{(2\pi \hbar)^{D-1}} \Bigg\lbrace \Big[  \frac{\Delta_X}{4}- \frac{p^2}{\hbar^2} \Big]  \Big[ a^{D-2} F_{00} \Big](X^i, p_i ) \Bigg\rbrace \\
 - \frac{ m^2}{\hbar^2} a^2 \Big[1+ \Phi_G(X^i) - (D-1) \Psi_G(X^i) \Big] \Big[ a^{D-2}  \overline{F}_{00} \Big](X^i ) \\      +  \Big[1+ \Phi_G(X^i) + (D-1) \Psi_G(X^i) \Big] \Big[ a^{-(D-2)}  \overline{F}_{11} \Big] (X^i)    \, ,
\end{multline}
and thus
\begin{multline}
 \overline{F}_{11}  (X^i) = \frac{ m^2}{\hbar^2} a^{2(D-1)} \Big[1 - 2(D-1) \Psi_G(X^i) \Big]   \overline{F}_{00} (X^i )\\+ \frac{1}{2}a^{D-2}  \Big[1- \Phi_G(X^i) - (D-1) \Psi_G(X^i) \Big] \Bigg[a^{D-2}\Big[1- \Phi_G(X^i) - (D-1) \Psi_G(X^i) \Big] \overline{F}_{00}^{\prime}(X^i) \Bigg]^{\prime} \\  - a^{2(D-2)}  \frac{\partial}{\partial X^k}  \Big[ \Phi_G (X^i) - (D-3)  \Psi_G(X^i) \Big]    \frac{1}{2}\frac{\partial}{\partial X^k}  \overline{F}_{00} (X^i )  \\
-  a^{2(D-2)}  \Big[  1 - 2(D-2)  \Psi_G(X^i) \Big]    \int \frac{ d^{D-1}p}{(2\pi \hbar)^{D-1}} \Bigg\lbrace \Big[  \frac{\Delta_X}{4}- \frac{p^2}{\hbar^2} \Big]  F_{00} (X^i, p_i ) \Bigg\rbrace  \, .
\end{multline}
We plug this expression into the energy-momentum tensor and get
\begin{multline}
T_{00}(X^i) =  \frac{m^2 a^2}{\hbar^2} \Big[1+2 \Phi_G(X^i) \Big] \overline{F}_{00} (X^i)  \\
+  a^{-(D-2)}\Big[1+\Phi_G(X^i) +(D-1) \Psi_G(X^i)\Big] \Bigg[\frac{a^{D-2}}{4} \Big[1-\Phi_G(X^i) -(D-1)\Psi_G(X^i)\Big]  \overline{F}_{00}^{\prime}  (X^i) \Bigg]^{\prime}\\
-\frac{\partial_X}{4} \Big[\Phi_G(X^i) -(D-3) \Psi_G(X^i)\Big] \partial_X \overline{F}_{00}  (X^i)
 \\+\Bigg[1+2 \Phi_G(X^i) +2 \Psi_G(X^i)  \Bigg] \int \frac{ d^{D-1}p}{(2\pi \hbar)^{D-1}} \frac{p^2}{\hbar^2}  F_{00}(X^i,p_i) \ \label{T00PureF00} \,. 
\end{multline}
\begin{multline}
T_{0i}(X^i)=  a^{-(D-2)}    \Big[1 + \Phi_G(X^i) +  (D-1) \Psi_G(X^i) \Big]\\ \times \Bigg\lbrace \frac{\partial}{\partial X^i} \Bigg[\frac{a^{D-2}}{4} \Big[1-\Phi_G(X^i) -(D-1)\Psi_G(X^i) \Big] \overline{F}_{00}^{\prime} (X^i)  \Bigg] \\ - \int \frac{ d^{D-1}p}{(2\pi \hbar)^{D-1}} \frac{p_i}{\hbar} F_- (X^i,p_i)   \Bigg\rbrace \,. \label{T0iPureF-}
\end{multline}
\begin{multline}
\frac{T_{ii}(X^i)}{D-1}   = 
   a^{-(D-2)} \Bigg[1-\Big[\Phi_G(X^i) -(D-3) \Psi_G(X^i)\Big] \Bigg] \\\times \Bigg[\frac{a^{D-2}}{4} \Big[1-\Phi_G(X^i) -(D-1)\Psi_G(X^i) \Big] \overline{F}_{00}^{\prime}(X^i) \Big] \Bigg]^{\prime}\\
-\frac{\partial_X}{4} \Big[\Phi_G(X^i) -(D-3) \Psi_G(X^i) \Big] \partial_X \overline{F}_{00} (X^i)
- \frac{D-2}{D-1} \frac{\Delta_X}{4} \overline{F}_{00} (X^i) \\
 +\frac{1}{D-1} \int \frac{ d^{D-1}p}{(2\pi \hbar)^{D-1}} \frac{p^2}{\hbar^2}  F_{00}(X^i,p_i)   \,. \label{TijTrace}
\end{multline}
\begin{multline}
T_{ij}(X^i) - \frac{\delta_{ij}}{D-1} T_{kk}(X^i)
\\ = \int \frac{ d^{D-1}p}{(2\pi \hbar)^{D-1}}  \Bigg[ \frac{\partial_i^X \partial_j^X}{4}+\frac{p_i p_j}{\hbar^2} - \frac{\delta_{ij} }{D-1} \Big( \frac{\Delta_X}{4}     +  \frac{p^2}{\hbar^2} \Big) \Bigg] F_{00}(X^i,p_i) \,. \label{TijTraceless}
\end{multline}
\begin{multline}
T_{ij}(X^i)
   =  {\delta_{ij} } \Bigg\lbrace
  a^{-(D-2)}\Bigg[1-\Big[\Phi_G(X^i) -(D-3) \Psi_G(X^i)\Big] \Bigg] \\ \times \Bigg[\frac{a^{D-2}}{4} \Big[1-\Phi_G(X^i) -(D-1)\Psi_G(X^i) \Big]  \overline{F}_{00}^{\prime}(X^i)  \Bigg]^{\prime}\\
-\frac{\partial_X}{4} \Big[\Phi_G(X^i) -(D-3) \Psi_G(X^i)\Big] \partial_X \overline{F}_{00} (X^i)
- \frac{\Delta_X}{4} \overline{F}_{00}(X^i)
\Bigg\rbrace \\+\int \frac{ d^{D-1}p}{(2\pi \hbar)^{D-1}}  \Bigg[ \frac{\partial_i^X \partial_j^X}{4}+\frac{p_i p_j}{\hbar^2} \Bigg]  F_{00} (X^i,p_i)  \,. \label{Tij}
\end{multline}
In the equation for the traceless spatial energy-momentum-tensor we used the fact that $ F_{00} $ is even in $p_i$ to eliminate two terms. 
By using the Bianchi identities, one can show that the purely spatial constraint equations follow from the constrained equations involving time-components. We thus have two independent constraint equations that relate the gravitational potentials $\Phi_G$ and $\Psi_G$ to momentum integrals over the two-point functions $ F_{00}$ and $F_{-}$.
\subsection{Energy-Momentum Conservation for Composite Fluids \label{appGenCompPerfFluid}}
In this section we want to derive non-relativistic fluid equations by assuming an energy-momentum tensor in a composite form that is for example due to taking expectation values of two-point functions. We also assume a linearized scalar metric in longitudinal gauge. We have
\be
\widetilde{T}_{\mu \nu}^{\textsc{npf}}:=    \Big[ (\widetilde{e}+\widetilde{P}) \widetilde{u}_{\mu} \widetilde{u}_{\nu} \Big]_{\text{com}} +g_{\mu \nu} \widetilde{P}\, , \label{NPFEMT}
\ee
where the composite term is basically a placeholder for any symmetric two-tensor whose indices can be raised and lowered with the metric. Thus, the quantity $\widetilde{P}$ is only an apparent pressure and the real pressure might get contributions from the composite term. We use a tilde to make a distinction between a general case and the real scalar field case we discuss throughout the paper. The superscript $\text{NPF}$ refers to the idea that although the energy-momentum tensor \eqref{NPFEMT} has the apparent form of a perfect fluid, it does not need to be the energy-momentum tensor of such due to its composite nature. Since the composite term ought to be a generalization of the non-composite perfect fluid term we impose
\be
\Big[ (\widetilde{e}+\widetilde{P}) \widetilde{u}^{\mu} \widetilde{u}_{\mu} \Big]_{\text{com}} =: - (\widetilde{e}+\widetilde{P}) \, ,
\ee 
which acts also as a definition for the energy $\widetilde{e}$.
Local energy conservation then gives 
\begin{multline}
0=\nabla_{\mu}  \Big[\widetilde{T}^{\textsc{npf}}\Big] ^{\mu}_{\;\; 0} = \partial_{\eta} \Bigg[\Big[ (\widetilde{e}+\widetilde{P}) \widetilde{u}^{0} \widetilde{u}_0 \Big]_{\text{com}} +\widetilde{P} \Bigg] + \partial_i \Big[ (\widetilde{e}+\widetilde{P}) \widetilde{u}^{i} \widetilde{u}_0 \Big]_{\text{com}}\\+ \Gamma^i_{i0} \Big[ (\widetilde{e}+\widetilde{P}) \widetilde{u}^{0} \widetilde{u}_0 \Big]_{\text{com}}-\Gamma^i_{k0}\Big[ (\widetilde{e}+\widetilde{P}) \widetilde{u}^{k} \widetilde{u}_i \Big]_{\text{com}} + \Gamma^i_{ik} \Big[ (\widetilde{e}+\widetilde{P}) \widetilde{u}^{k} \widetilde{u}_0 \Big]_{\text{com}} - \Gamma^i_{00} \Big[ (\widetilde{e}+\widetilde{P}) \widetilde{u}^{0} \widetilde{u}_i \Big]_{\text{com}} \,.
\end{multline}
We raise and lower indices and commute $u_0 \, u_i$,
\begin{multline}
 \partial_{\eta} \Bigg[\Big[ (\widetilde{e}+\widetilde{P})  \widetilde{u}^{0} \widetilde{u}_0 \Big]_{\text{com}} +\widetilde{P} \Bigg] + \partial_i \Big[(\widetilde{e}+\widetilde{P}) \widetilde{u}^{i} \widetilde{u}_0 \Big]_{\text{com}}\\+ \Gamma^i_{i0} \Big[ (\widetilde{e}+\widetilde{P})  u^{0} u_0 \Big]_{\text{com}}-\Gamma^i_{k0}\Big[ (\widetilde{e}+\widetilde{P})  \widetilde{u}^{k} \widetilde{u}_i \Big]_{\text{com}}  + \Big[\Gamma^i_{ik}  - g^{00}  \Gamma^i_{00} g_{ik} \Big] \Big[ (\widetilde{e}+\widetilde{P})  \widetilde{u}^{k}  \widetilde{u}_0 \Big]_{\text{com}} =0 \,.
\end{multline}
Next we plug in the perturbed metric in longitudinal gauge and get
\begin{multline}
 \partial_{\eta} \Bigg[\Big[ (\widetilde{e}+\widetilde{P})  \widetilde{u}^{0} \widetilde{u}_0  \Big]_{\text{com}} +\widetilde{P} \Bigg] + \Big[1- \Phi_G +(D-1)  \Psi_G  \Big]\partial_k \Bigg[ \Big[1+ \Phi_G -(D-1)  \Psi_G  \Big]  \Big[ (\widetilde{e}+\widetilde{P})   \widetilde{u}^{k}  \widetilde{u}_0 \Big]_{\text{com}}\Bigg]\\+ (D-1)\Big[\mathcal{H} - \Psi_G^{\prime}\Big] \Big[ (\widetilde{e}+\widetilde{P})   \widetilde{u}^{0}  \widetilde{u}_0 \Big]_{\text{com}} - \Big[\mathcal{H} - \Psi_G^{\prime}\Big]  \Big[ (\widetilde{e}+\widetilde{P})   \widetilde{u}^{k}  \widetilde{u}_k \Big]_{\text{com}} =0 \,. \label{ContRel}
\end{multline}
To find the non-relativistic limit of this composite equation we have to define a quantity that should capture the rest-mass in the non-relativistic limit 
\be
\widetilde{\rho} := \widetilde{e} - \widetilde{P}\, ,
\ee as well as a proper fluid velocity
\be
  \widetilde{v}^{i}  :=  - \widetilde{\rho}^{-1} \Big[1+\Psi_G + \Phi_G \Big] \Big[ (\widetilde{e}+\widetilde{P})  \widetilde{u}^{i} \widetilde{u}_0\Big]_{\text{com}} \Bigg[1 +  (\widetilde{e}+\widetilde{P}) ^{-1}\Big[ (\widetilde{e}+\widetilde{P})   \widetilde{u}^{k}  \widetilde{u}_k \Big]_{\text{com}} \Bigg]^{1/2}\,.
\ee
We then have the non-relativistic continuity equation in an FLRW-universe
\be
  \Bigg[ \partial_{\eta} \widetilde{\rho}  + (D-1)\mathcal{H}  \widetilde{\rho} + \partial_i \Big[ \widetilde{\rho} \cdot \widetilde{v}^{i}  \Big] \Bigg] \Bigg[1+ \mathcal{O}\Big( \Phi_G \, , \Psi_G \Big)+ \mathcal{O}\Big( \widetilde{v}^2 ,\widetilde{P} \Big)  \Bigg] =0 \,. \label{contFRLW}
\ee
The local momentum conservation reads
\begin{multline}
0= \nabla_{\mu} \Big[\widetilde{T}^{\textsc{npf}} \Big]^{\mu}_{\; \:i} =
\partial_0 \Big[  (\widetilde{e}+\widetilde{P}) \widetilde{u}^{0} \widetilde{u}_i \Big]_{\text{com}} + \partial_k \Big[  (\widetilde{e}+\widetilde{P}) \widetilde{u}^{k} \widetilde{u}_i \Big]_{\text{com}} 
+\partial_i P \\+ \Big[D\mathcal{H}+\Phi_G^{\prime}-(D-1)\Psi_G^{\prime}  \Big]\Big[  (\widetilde{e}+\widetilde{P}) \widetilde{u}^{0} \widetilde{u}_i \Big]_{\text{com}}
+\Big[\partial_k \Phi_G -(D-1)\partial_k \Psi_G \Big] \Big[  (\widetilde{e}+\widetilde{P}) \widetilde{u}^{k} \widetilde{u}_i \Big]_{\text{com}} \\
- \partial_i \Phi_G \Big[  (\widetilde{e}+\widetilde{P}) \widetilde{u}^{0} \widetilde{u}_0 \Big]_{\text{com}} + \partial_i \Psi_G \Big[  (\widetilde{e}+\widetilde{P}) \widetilde{u}^{k} \widetilde{u}_k \Big]_{\text{com}} \, . \label{EulerPF}
\end{multline}
Defining the proper composite velocity term
\be
\Big[  \widetilde{v}^{i} \widetilde{v}^k \Big]_{\text{com}} :=   \widetilde{\rho}^{-1} \Big[1+\Psi_G - \Phi_G \Big] \Big[ (\widetilde{e}+\widetilde{P}) \widetilde{u}^{i}\widetilde{u}_k\Big]_{\text{com}} \, , \label{compVel2}
\ee
we find the Euler equation in an FLRW-space-time
\begin{multline}
0= \Bigg[ \partial_{\eta} \Big[ \widetilde{\rho} \cdot \widetilde{v}^{i} \Big] + D\mathcal{H}\Big[ \widetilde{\rho} \cdot \widetilde{v}^{i} \Big] + \partial_k \Big( \widetilde{\rho} \cdot \Big[  \widetilde{v}^{i} \widetilde{v}^k \Big]_{\text{com}} \Big) + \widetilde{\rho} \, \delta^{ij} \partial_j \Phi_G  +  \delta^{ij} \partial_j \widetilde{P} \Bigg]\\ \times \Bigg[1+ \mathcal{O}\Big( \Phi_G \, , \Psi_G \Big)+  \mathcal{O}\Big( \widetilde{v}^2 ,\widetilde{P} \Big) \Bigg]  \, . \label{EulerPFNONrel}
\end{multline}
We can now discuss two options. \\
Option one is simply given by declaring the composite term of the energy-momentum tensor \eqref{NPFEMT} and similarly its non-relativistic descendant \eqref{compVel2} to be non-composite as it would be the case for a perfect scalar field fluid based classical field theory including non-linear terms in the fluid quantities but keeping gravitational potentials linear. Equations \eqref{contFRLW} and \eqref{EulerPFNONrel} were for example derived in the scalar field context by by \cite{Marsh:2015daa} and \cite{Suarez:2015fga} for real and complex fields, respectively.
\\
The second option is of course to take into account the composite nature of the term in \eqref{compVel2} which happens for a real scalar field fluid based on non-vanishing connected two-point functions as we discuss it in section \ref{perfOrNot}. 
However, we stress once more that the above derivation made no reference to the real scalar field theory and we only assumed local conservation of a fluid energy-momentum tensor of the form \eqref{NPFEMT}. 
We will now give an independent derivation of the the fluid equations \eqref{ContRel} and \eqref{EulerPF} in the next section as a cross-check and will explicitly use the scalar field matter equations.
\subsection{Continuity and Euler Equation for Real Scalar Field Fluid \label{appConEul}}
\paragraph{Local energy conservation.}
Let us start with the local energy conservation by integrating the dynamical equations \eqref{F00Fin} and \eqref{F+Fin} as well as \eqref{F11Fin}. We find
\be
\overline{F}_+ =  \frac{\hbar^2}{2 m^2} a^{D-2} \Big[ 1 - \Phi_G - (D-1) \Psi_G \Big] \rho^{\prime} \, ,
\ee
\be
 \overline{F}_{11}  = - a^{2(D-1)} \Big[1 -2(D-1) \Psi_G \Big] \Big[  (e+P) u^0 u_0 \Big]_{\text{com}} \, ,
\ee
\begin{multline}
\int \frac{ d^{D-1}p}{(2\pi \hbar)^{D-1}}  \frac{p^2}{\hbar^2} F_{+} -
\frac{\partial}{\partial X^k } \Big[ \Phi_G  + (D-1) \Psi_G \Big] \int \frac{ d^{D-1}p}{(2\pi \hbar)^{D-1}}  \frac{p_k}{\hbar} F_{-} \\ = \frac{1}{2} a^{D-2} \Big[1- \Phi_G -(D-1) \Psi_G \Big] \Bigg[ a^2 \Big[1-2 \Psi_G \Big] \Big[  (e+P) u_i u^i \Big]_{\text{com}}  \Bigg]^{\prime} \\ - \frac{\hbar^2}{8 m^2} a^{D-2} \Big[1- \Phi_G -(D-1) \Psi_G \Big]  \Delta_X   \rho^{\prime}  -  \frac{\hbar^2}{8 m^2} a^{D-2} \rho^{\prime}  \Delta_X \Big[ \Phi_G +(D-1) \Psi_G \Big]  \, .
\end{multline}
We also have
\begin{multline}
 a^{-(D-2)} \overline{F}_{11}^{\prime} + 2 \frac{m^2a^2}{\hbar^2} \Big[1+ \Phi_G -(D-1) \Psi_G \Big] \overline{F}_+
\\=  \frac{\Delta_X}{2} \Bigg[ \Big[1+ \Phi_G -(D-3) \Psi_G \Big]  \overline{F}_+ \Bigg] 
-\overline{F}_+ \frac{\Delta_X}{2}  \Big[\Phi_G -(D-3) \Psi_G \Big] \\ -2 \Big[1+ \Phi_G -(D-3) \Psi_G \Big]\int \frac{ d^{D-1}p}{(2\pi \hbar)^{D-1}}  \frac{p^2}{\hbar^2} F_{+} \\
-2 \frac{\partial}{\partial X^k} \Bigg[\Big[1 +\Phi_G -(D-3) \Psi_G \Big] \int \frac{ d^{D-1}p}{(2\pi \hbar)^{D-1}}  \frac{p_k}{\hbar} F_{-} \Bigg]\, .
\end{multline}
We also write down the following relation for the fluid-four velocity for convenience
\begin{multline}
\Big[1 -\Phi_G -(D-1) \Psi_G \Big] \frac{\partial}{\partial X^k} \Bigg[\Big[1 +2\Phi_G +2 \Psi_G \Big]  \int \frac{ d^{D-1}p}{(2\pi \hbar)^{D-1}} \frac{p_k}{\hbar} F_- \Bigg] \\= -a^{D}\Big[1 -\Phi_G -(D-1) \Psi_G \Big] \frac{\partial}{\partial X^k} \Bigg[\Big[ 1+ \Phi_G - (D-1) \Psi_G \Big]\Big[ (e+P)u^k u_0 \Big]_{\text{com}} \Bigg] \\ + a^{(D-2)}\frac{\hbar^2}{4 m^2}\Big[1 -\Phi_G -(D-1) \Psi_G \Big] \frac{\partial}{\partial X^k} \Bigg[ \Big[1 +2\Phi_G +2 \Psi_G \Big] \frac{\partial}{\partial X^k} \Bigg[ \Big[1- \Phi_G - (D-1) \Psi_G \Big] \rho^{\prime} \Bigg]\Bigg] \, . 
\end{multline}
We plug everything in and indeed end up after a series of straightforward manipulations with,
\begin{multline}
\Bigg[ \Big[  (e+P) u^0 u_0 \Big]_{\text{com}} +P \Bigg] ^{\prime} +(D-1) \Big[ \mathcal{H}-\Psi_G^{\prime} \Big]  \Big[  (e+P) u^0 u_0 \Big]_{\text{com}} 
      \\
+ \Big[1 -\Phi_G +(D-1) \Psi_G \Big] \frac{\partial}{\partial X^k} \Bigg[\Big[ 1+ \Phi_G - (D-1) \Psi_G \Big]\Big[ (e+P)u^k u_0 \Big]_{\text{com}} \Bigg]\\ -   \Big[\mathcal{H} -\Psi_G^{\prime}  \Big] \Big[  (e+P) u_k u^k \Big]_{\text{com}}    =0 \, ,
\end{multline}
which agrees with the result \eqref{contFRLW} of the more general derivation in appendix \ref{appGenCompPerfFluid}.
The definition of the rest-mass in \eqref{restMassDef} yields
\be
\rho = e - P \, ,
\ee
such that 
\be
\Big[  (e+P) u^{\mu} u_{\mu} \Big]_{\text{com}} = - (\rho + 2 P)\, .
\ee
In order to find the non-relativistic limit of this equation we have to define proper irreducible fluid velocity
\be
  v^{i} :=  - \rho^{-1} \Big[1+\Psi_G + \Phi_G \Big] \Big[ (e+P) u^{i}u_0\Big]_{\text{com}} \Bigg[1 +  (e+P)^{-1}\Big[ (e+P) u^{k} u_k \Big]_{\text{com}} \Bigg]^{1/2}\, . \label{properIrrVel}
\ee
We then have the non-relativistic continuity equation in an FLRW-universe
\be
  \Bigg[ \partial_{\eta} \rho + (D-1)\mathcal{H}  \rho + \partial_i \Big[ \rho \cdot  v^{i} \Big] \Bigg] \Bigg[1+ \mathcal{O}\Big( \varepsilon_g^2 \Big)+ \mathcal{O}\Big(\varepsilon_{\hbar} ^2 \Big)+ \mathcal{O}\Big(\varepsilon_{p} ^2 \Big)  \Bigg] =0 \,. \label{contFRLWScalar}
\ee
\paragraph{Local momentum conservation.}
Now, we continue with the Euler equation by integrating equation \eqref{F-Fin},
\begin{multline}
\Bigg[\int \frac{ d^{D-1}p}{(2\pi \hbar)^{D-1}} \frac{p_i}{\hbar} F_- \Bigg]^{\prime}   =
\frac{1}{4}\frac{\partial^2}{\partial X^i \partial X^k}   \Big[ \Phi_G  - (D-3)  \Psi_G \Big]    \frac{\partial}{\partial X^k}  \Big[ a^{D-2} \overline{F}_{00} \Big]    \\
+\frac{1}{2} \frac{\partial}{\partial X^i} \Big[  \Phi_G - (D-3)  \Psi_G \Big] \int \frac{ d^{D-1}p}{(2\pi \hbar)^{D-1}}  \Big[  \frac{\Delta_X}{4}- \frac{p^2}{\hbar^2} \Big]  \Big[ a^{D-2} F_{00} \Big] \\
 - \Big[1+\Phi_G  - (D-3)  \Psi_G \Big] \int \frac{ d^{D-1}p}{(2\pi \hbar)^{D-1}} \frac{p_ip_k}{\hbar^2} \frac{\partial}{\partial X^k}  \Big[ a^{D-2} F_{00} \Big]\\ 
  -  \frac{\partial}{\partial X^k}  \Big[ \Phi_G - (D-3)  \Psi_G \Big]  \int \frac{ d^{D-1}p}{(2\pi \hbar)^{D-1}} \frac{p_i p_k}{\hbar^2}   \Big[ a^{D-2} F_{00} \Big]   \\ 
     - \frac{1}{2}\frac{ m^2}{\hbar^2} a^2 \frac{\partial}{\partial X^i} \Big[\Phi_G - (D-1) \Psi_G \Big] \Big[ a^{D-2} \overline{F}_{00} \Big]     
     -  \frac{1}{2}\frac{\partial}{\partial X^i}  \Big[\Phi_G + (D-1) \Psi_G \Big] \Big[ a^{-(D-2)} \overline{F}_{11} \Big]   \, .
\end{multline}
We plug in the expressions for the integrated two-point functions in terms of hydrodynamical variables
\begin{multline}
\Bigg[\frac{\hbar^2}{4 m^2} a^{D-2} \frac{\partial}{\partial X^i} \Bigg[ \Big[1- \Phi_G - (D-1) \Psi_G \Big] \rho^{\prime} \Bigg] +\Big[ 1+ \Phi_G - (D-1) \Psi_G \Big]a^{D}\Big[ (e+P) u^0 u_i\Big]_{\text{com}} \Bigg]^{\prime} \\  =
\frac{\hbar^2}{4 m^2} a^{D-2} \frac{\partial^2}{\partial X^i \partial X^k}   \Big[ \Phi_G  - (D-3)  \Psi_G \Big]    \frac{\partial}{\partial X^k}  \rho   \\
+\frac{\hbar^2}{2 m^2} a^{D-2}\frac{\partial}{\partial X^i} \Big[  \Phi_G - (D-3)  \Psi_G \Big]  \frac{\Delta_X}{4}  \rho \\
-\frac{1}{2} \frac{\partial}{\partial X^i} \Big[  \Phi_G - (D-3)  \Psi_G \Big] \Bigg[{a^{D} } \Big[1-2 \Psi_G \Big] \Big[  (e+P) u^k u_k\Big]_{\text{com}} -  {a^{D-2} }\frac{\hbar^2}{4 m^2 } \Delta_X \rho  \Bigg]\\
 - \frac{\partial}{\partial X^k} \Bigg[ \Big[1+\Phi_G  - (D-3)  \Psi_G \Big]\Bigg[ {a^{D} } \Big[1-2 \Psi_G \Big] \Big[  (e+P) u^k u_i\Big]_{\text{com}} -  {a^{D-2} }\frac{\hbar^2}{4 m^2 }\frac{\partial^2 }{\partial X^i \partial X^k} \rho \Bigg] \Bigg]  \\ 
     - \frac{1}{2} a^D \frac{\partial}{\partial X^i} \Big[\Phi_G - (D-1) \Psi_G \Big] \rho   
     + \frac{1}{2}  a^D \frac{\partial}{\partial X^i}  \Big[\Phi_G + (D-1) \Psi_G \Big]  \Big[  (e+P) u^0 u_0 \Big]_{\text{com}}  \, .
\end{multline}
Manipulating these expressions finally leads to
\begin{multline}
   \Big[ (e+P) u^0 u_i\Big]_{\text{com}}^{\prime} 
  +\Big[D \mathcal{H} +\Phi_G - (D-1) \Psi_G \Big]^{\prime} \Big[ (e+P) u^0 u_i\Big]_{\text{com}} +  \frac{\partial}{\partial X^i}   P\\   
+ \frac{\partial}{\partial X^i}   \Psi_G   \Big[  (e+P) u^k u_k\Big]_{\text{com}}      -  \frac{\partial}{\partial X^i}  \Phi_G  \Big[  (e+P) u^0 u_0 \Big]_{\text{com}} \\
 +  \Big[1-\Phi_G +(D-1)\Psi_G  \Big] \frac{\partial}{\partial X^k} \Bigg[ \Big[1+\Phi_G  - (D-1)  \Psi_G \Big]  \Big[  (e+P) u^k u_i\Big]_{\text{com}}  \Bigg]  =0\, ,
       \, .
\end{multline}
which agrees with the result \eqref{EulerPF} of the more general derivation in appendix \ref{appGenCompPerfFluid}.
Defining the proper composite velocity term
\be
\Big[ v^{i} v^k \Big]_{\text{com}} :=  \rho^{-1} \Big[1+\Psi_G - \Phi_G \Big] \Big[ (e+P) u^{i}u_k\Big]_{\text{com}}   \, ,
\ee
we find
\begin{multline}
0= \Bigg[ \partial_{\eta} \Big[ \rho \cdot v^{i} \Big]+ D\mathcal{H}\Big[ \rho \cdot v^{i} \Big]+ \partial_k \Big[ \rho \cdot \Big[ v^{i} v^k \Big]_{\text{com}}\Big]  + \rho  \, \delta^{ij} \partial_j \Phi_G  +  \delta^{ij} \partial_j P \Bigg]\\ \times \Bigg[1 + \mathcal{O}\Big( \varepsilon_g^2 \Big)+ \mathcal{O}\Big(\varepsilon_{\hbar} ^2 \Big)+ \mathcal{O}\Big(\varepsilon_{p} ^2 \Big) \Bigg]  \, . \label{EulerPFNONrelScalar}
\end{multline}
We evaluate the apparent pressure term,
\begin{multline}
0 = \Bigg\lbrace \partial_{\eta}  \Big[ \rho  \cdot v^i \Big] + D \mathcal{H}\Big[ \rho \cdot v^i \Big] + \rho \, \delta^{ik} \partial_k  \Phi_G   
 \\ + \partial_k \Big[ \rho \cdot  \Big[ v^{i} v^k \Big]_{\text{com}}\Big]  -  \frac{\hbar^2}{4m^2a^2} \Big[ \frac{\partial^2}{\partial X^i \partial X^k}  \rho + \delta^{ik} (D-2) \mathcal{H}\partial_{\eta} \rho+\delta^{ik}    \partial_{\eta}^2  \rho \Big]   \Bigg] \Bigg\rbrace \\ \times \Bigg\lbrace 1+ \mathcal{O}\Big( \varepsilon_g^2 \Big)+ \mathcal{O}\Big(\varepsilon_{\hbar} ^2 \Big)+ \mathcal{O}\Big(\varepsilon_{p} ^2 \Big) \Bigg\rbrace  \, . 
\end{multline}
As already mentioned above, the composite operators contains information about anisotropy or in other words a non-vanishing second cumulant. Using the definition in \eqref{anisoStress} together with the definition \eqref{properIrrVel} we find the stress tensor in terms of proper fluid velocities
\be
\rho \cdot \Big[ v^{i} v^k \Big]_{\text{com}} = \Bigg[  \delta^{ij}\delta^{km} \sigma_{jm}  +  \rho \cdot  v^i \cdot v^k +   \delta^{ij} \delta^{kl} \frac{\hbar^2}{4m^2a^2} \partial_j \partial_l \rho  \Bigg]\times \Bigg[ 1+ \mathcal{O}\Big( \varepsilon_g^2 \Big)+ \mathcal{O}\Big(\varepsilon_{\hbar} ^2 \Big)+ \mathcal{O}\Big(\varepsilon_{p} ^2 \Big) \Bigg]\, ,
\ee
and thus
\begin{multline}
0=\Bigg\lbrace \partial_{\eta}  \Big[ \rho  \cdot v^i \Big] + D \mathcal{H}\Big[ \rho \cdot v^i \Big] + \rho \, \delta^{ik} \partial_k  \Phi_G   
 \\ + \frac{\partial}{\partial X^k}\Bigg[\delta^{ij}\delta^{km} \sigma_{jm}  +  \rho \cdot v^i \cdot  v^k  -  \frac{\hbar^2}{4m^2a^2} \Big[ \delta^{ik} (D-2) \mathcal{H}\partial_{\eta} \rho+\delta^{ik}    \partial_{\eta}^2  \rho \Big]   \Bigg] \Bigg\rbrace \\ \times \Bigg\lbrace 1+ \mathcal{O}\Big( \varepsilon_g^2 \Big)+ \mathcal{O}\Big(\varepsilon_{\hbar} ^2 \Big)+ \mathcal{O}\Big(\varepsilon_{p} ^2 \Big) \Bigg\rbrace=0  \, . \label{EulerWithQuantum}
\end{multline}
The equations we find here are identical to the equations \eqref{contFRLW} and \eqref{EulerPFNONrel}, except that we  identify the perturbation parameters on a more fundamental level and plugged in a concrete expression for the composite velocity term and the apparent pressure.
\par Let us once more  point out that a local rest-mass-density conservation due to some internal symmetry is not necessary to derive typical fluid equations in an FLRW-universe once we impose a large mass limit. This is the reason why a minimally-coupled real scalar field is well-suited to model the fluid dynamics of cold dark matter.
It is of course true that  the real scalar field model has no number current that is locally conserved on all scales (in contrast to a complex scalar field for example) and thus the probability associated to a single-particle wave function of a real scalar field in Minkowski space is a priori not conserved.
However, the energy-momentum tensor is another locally conserved quantity in the minimally-coupled real scalar field theory. Imposing that the mass $m$ dominates all other scales, the energy density current coincides with the rest-mass density current with corrections given by terms that go like momentum over mass, Hubble rate over mass and so on which can be tuned to be small by imposing a certain initial momentum distribution for instance. It is thus no surprise that the system we discuss in this paper has a classical limit for hydrodynamic quantities that mimicks fluid equations of massive particles whose number is locally conserved (cf. the discussion in \cite{Berges:2014xea}). 

\bibliographystyle{JHEP}
\bibliography{Biblio}
\nocite{*}
\end{document}